\documentclass[12pt]{article}

\usepackage{amsmath,amssymb,amsthm,latexsym,undertilde}
\newcommand\im{{\rm i}}

\newcommand\be{\begin{eqnarray}}
\newcommand\ee{\end{eqnarray}}

\title{Gravity-Yang-Mills-Higgs unification \\ by enlarging the gauge group}
\author{Alexander Torres-Gomez and Kirill Krasnov\\
\textit{\small School of Mathematical Sciences}\\
\textit{\small University of Nottingham}}
\date{November 2009}
\begin{document}
\maketitle
\begin{abstract}
We revisit an old idea that gravity can be unified with Yang-Mills 
theory by enlarging the gauge group of  gravity formulated as gauge theory. 
Our starting point is an action that describes a generally covariant gauge 
theory for a group $G$. The Minkowski background breaks the gauge group by
selecting in it a preferred gravitational ${\rm SU}(2)$ subgroup. We expand the action around this
background and find the spectrum of linearized theory to consist of the usual gravitons plus
Yang-Mills fields charged under the centralizer of the ${\rm SU}(2)$ in G. In addition, there
is a set of Higgs fields that are charged both under the gravitational and Yang-Mills
subgroups. These fields are generically massive and interact with both gravity and
Yang-Mills sector in the standard way. The arising interaction of the Yang-Mills sector with 
gravity is also standard. Parameters such as the Yang-Mills coupling constant and
Higgs mass arise from the potential function defining the theory. Both are realistic
in the sense explained in the paper.
\end{abstract}

\section{Introduction}

There have been numerous attempts to unify Einstein's theory of gravity with gauge fields
describing other interactions. One such unification proposal is that of Kaluza-Klein, where the 
metric and gauge fields arise from a higher-dimensional metric tensor upon compactification of 
extra dimensions. This scenario has become an indispensable
part of string theory, which also provides another unifying perspective by viewing
gravity and Yang-Mills as excitations of closed and open strings respectively.  
For more details on string-inspired unification schemes see a recent exposition
\cite{Vafa:2009se}.

There have also been attempts to unify gravity with gauge theory without introducing
extra dimensions. There is, however, a very strong no-go theorem \cite{Coleman:1967ad} 
that shows that at least one type of such unification is impossible. The theorem
states that the symmetry group of the S-matrix of a consistent quantum field theory
(in Minkowski spacetime) is the product of Poincare and internal gauge group. In other
words, the spacetime and internal symmetries do not mix. The only way to go around 
this statement is via supersymmetric extensions of Poincare group \cite{Haag:1974qh}.

Now, since gravity can be (at least loosely) viewed 
as a gauge theory for the diffeomorphism group, and the later contains Poincare
group as that of rigid global transformations, the Coleman-Mandula theorem
\cite{Coleman:1967ad} is sometimes interpreted as saying that no unification
of gravity and gauge theory that puts together diffeomorphisms and gauge 
transformations is possible. In this discussion, however, one must be careful to
distinguish between local gauge invariances of a theory and global
symmetries whose presence or absence depends on a particular state one works 
with, see \cite{Percacci:2008zf} that emphasizes this point. 

While it may be difficult or impossible to ``unify'' diffeomorphisms and
gauge transformations into a single gauge group, this is not the only
possible way to approach the unification problem. To understand how
a different type of unification might be possible, let us recall 
that in the so-called first-order formalism
gravity becomes a theory of metrics as well as Lorentz group spin connections. 
The  ``internal'' Lorentz group acts by rotating the frame and has
no effect on the metric. Thus, the physical dynamical variable is still the metric, one simply
added some gauge variables and enlarged the gauge group, which in this formulation 
is a (semi-) direct product of the diffeomorphism group and ${\rm SO}(1,3)$. Further,
in the Hamiltonian formulation this theory can be easily cast into one on the Yang-Mills
phase space. This is done by adding to the action a term that vanishes on-shell \cite{Holst:1995pc}.
The phase space is then that of pairs ${\rm SU}(2)$ connection plus the canonically conjugate
``electric'' field. Thus, after the trick of adding an on-shell unimportant term, gravity
becomes a generally covariant theory of an ${\rm SU}(2)$ connection. The spacetime metric 
(tetrad) is still a dynamical variable but in this formulation it receives the interpretation of
the momentum canonically conjugate to the connection. 

Yang-Mills theory, on the other hand, after it is written for a general spacetime metric,
also becomes a generally covariant theory of a connection and spacetime metric. 
One could then attempt to put the
two generally covariant gauge theories together in some way that combines the ``internal''
gauge groups, while leaving the total gauge group to
be a (semi-) direct product of diffeomorphisms and ``internal'' symmetries. This would not
be in any conflict with the no-go theorem \cite{Coleman:1967ad} for what is
unified is not the Poincare and internal symmetry groups. This might not be what
can be legitimately called a unification, for the end gauge group is not simple,
but this idea does lead to some interesting ``unified'' theories, as we
hope to be able to demonstrate in this paper. 

As far as we are aware, the first proposal of this type was 
put forward in \cite{Percacci:1984ai,Percacci:1990wy},
with the idea being precisely to extend the gauge group of gravity formulated 
in tetrad first-order formalism as a theory of the Lorentz connection. This proposal 
was later pushed forward in \cite{Nesti:2007jz,Nesti:2007ka}, see also \cite{Nesti:2009kk}
for the most recent development. The key point of this proposal is that it is
a non-degenerate metric that breaks the gauge symmetry of the
unified theory down to a smaller group consisting of ${\rm SO}(1,3)$ for gravity
and some "internal" group for Yang-Mills fields. 

A similar in spirit, but very different in the
realization idea was proposed in  \cite{Peldan:1992iw}, and further
developed in \cite{Peldan:1992mp,Chakraborty:1994vx,Chakraborty:1994yr}. This approach
stems from the fact that Einstein's  general relativity (GR) can be reformulated as a theory on
the Yang-Mills phase space. At the time of writing \cite{Peldan:1992iw} it was 
achieved in Ashtekar's Hamiltonian formulation of
GR \cite{Ashtekar:1987gu} that interprets gravity as a special generally covariant
(complexified) ${\rm SU}(2)$ gauge theory. The fact that gravity in this formulation becomes 
a theory of connection suggests that a gauge group larger than
${\rm SU}(2)$ can be considered. This is what was attempted in 
\cite{Peldan:1992iw,Peldan:1992mp,Chakraborty:1994vx,Chakraborty:1994yr},
with the main result of \cite{Chakraborty:1994yr} being that Yang-Mills theory
arises in an expansion of the theory around the de Sitter background. 

The idea to put together the ``internal'' gauge groups of gravity and gauge theory is 
an interesting one. However, its particular realizations available in the literature are not without
problems. Thus, the approach reviewed and further developed in \cite{Nesti:2009kk}
does a very good job in describing the fermionic content of the theory. Bosons, on the
other hand, are described less convincingly in that many new propagating degrees
of freedom (DOF) are introduced. The other approach \cite{Chakraborty:1994yr} is also not
very convincing since it works at the phase space level, and it is 
generally very difficult to approach a theory if 
no action principle is prescribed. Another aspect of the particular realization
given in \cite{Chakraborty:1994yr} is that it naturally describes a complexified GR
put together with complexified Yang-Mills. No natural reality conditions that would
convert this into a physical theory were given. 

The unification by enlarging the internal gauge group proposal was recently revisited 
in \cite{Smolin:2007rx}, where the new action principle \cite{Krasnov:2006du} for a 
class of modified gravity theories \cite{Bengtsson:1990qg}, extended to a larger gauge group
was used. This work also avoided the reality conditions problem by
extending the gauge group of an explicitly real formulation of gravity that works 
with the Lorentz, not with the complexified rotation group. 
Specifically, it was suggested in \cite{Smolin:2007rx} that the action of the type proposed in 
\cite{Krasnov:2006du} considered  for a general Lie
group $G$ describes gravity in its ${\rm SO}(4)$ part plus Yang-Mills fields
in the remaining quotient $G/{\rm SO}(4)$. As in \cite{Chakraborty:1994yr},
the Yang-Mills coupling constant  is related in \cite{Smolin:2007rx} to the cosmological constant.
As in the approach \cite{Percacci:1984ai,Percacci:1990wy}, in \cite{Smolin:2007rx} it is a 
non-degenerate metric that breaks the symmetry down to a smaller gauge group. 
The approach of \cite{Smolin:2007rx} is also similar to that of 
\cite{Percacci:1984ai,Percacci:1990wy} in that many new bosonic degrees of freedom
are introduced. Thus, it was shown in \cite{Alexandrov:2008fs} that the BF-type action 
of \cite{Krasnov:2006du} for $G={\rm SO}(4)$ does not anymore describe pure gravity
theory in that it describes six new DOF. 

In this paper we take the described unification idea one step further. 
Our approach is similar in spirit to \cite{Smolin:2007rx} in that we start
from an action principle of the type first proposed in \cite{Krasnov:2006du}. However,
unlike in \cite{Smolin:2007rx}, we interpret only a (complexified) ${\rm SU}(2)$ subgroup
of the gauge group $G$ as that corresponding to gravity. The part of the
gauge group that commutes with this gravitational ${\rm SU}(2)$ is then seen to describe 
Yang-Mills fields, and the part that does not commute with ${\rm SU}(2)$ describes 
charged scalar, i.e. Higgs, fields. We note that the suggestion that in unifications of this type the 
"off-diagonal" part of the Lie algebra corresponds to Higgs fields is contained already
in \cite{Smolin:2007rx}.
 
Our approach is also similar to the original proposal \cite{Chakraborty:1994yr} that
enlarged the ${\rm SU}(2)$ gravitational gauge group. However, in contrast to
\cite{Chakraborty:1994yr} that worked at the phase space level our starting point
is an action principle that makes a much more systematic analysis possible. Also
the details of our proposal differ significantly from that of \cite{Chakraborty:1994yr}
in that a semi-realistic (more on this below) unification is achieved without the need
for a cosmological constant. Thus, the Yang-Mills coupling constant in our scheme 
is related not to the cosmological constant, which we set to zero, 
but to a certain other parameter of the theory. This features of our proposal also makes 
it different from that of \cite{Smolin:2007rx}. 

More specifically, we start from a generally covariant gauge theory for a (complex)
semi-simple Lie group $G$, with certain reality conditions later imposed to select real physical 
configurations. A particularly simple solution of the theory describes Minkowski 
spacetime. This solution breaks $G$ down to a (complexified) ${\rm SU}(2)$
times the centralizer of ${\rm SU}(2)$ in $G$. The spectrum of linearized
theory around the Minkowski background is then shown to consist of
the usual gravitons with their two propagating DOF, gauge bosons charged
under the centralizer of ${\rm SU}(2)$ in $G$, and a set of scalar Higgs fields. 
The Higgs fields are in general massive, with the mass being related to a certain
parameter of the potential defining the theory. After the reality conditions are imposed
all sectors of the theory have a positive-definite Hamiltonian. We also work out interactions
to cubic order and show that all interactions are precisely as expected. That is,
all non-gravitational fields interact with gravity via their stress-energy tensor,
and the interactions in the non-gravitational sector are also standard and
are as expected for Higgs fields. Thus, our unification scheme passes the zeroth 
order test of being not in any obvious contradiction with observations. However, to obtain a
truly realistic unification model many problems have to be solved. In particular,
fermionic DOF are not considered in this paper at all. Thus, our results provide 
only one of the first steps along this potentially interesting research direction.
We return to open questions of our approach in the discussion section.

In this paper we have illustrated the general $G$ case by considering the simplest
non-trivial example of $G={\rm SU}(3)$. This example is rather generic, and
the same technology that we develop for $G={\rm SU}(3)$ can be used for
any Lie group. We could have presented a general semi-simple case treatment phrased
in terms of the root basis in the Lie algebra. However, at this stage of the development
of the theory it is not clear if there is any added value in doing things in full generality.
We thus decided to keep our discussion as simple as possible and treat one
example that, if necessary, is easily extendible to the general situation.

Another general remark on this paper is as follows. As the reader will undoubtedly notice,
a sizable part of our paper is occupied by the Hamiltonian analysis of various
sectors, or of the full theory. We also always give the Lagrangian treatment in which 
things are much more transparent. Thus, it might at first sight seem that the
Hamiltonian formulation only clatters the exposition. We, however, believe
that some aspects of the theory are much clearer precisely in the
Hamiltonian formulation. For instance, our treatment of the reality conditions
heavily uses the Hamiltonian analysis and it would be very hard to arrive
at the correct conditions without it. This is our main reason for carrying out
such an analysis in all cases that are discussed. 

The organization of the paper is as follows. In section \ref{sec:pure-con} we define
the class of generally-covariant gauge theories that is the subject of this paper.
Section \ref{sec:two-form} performs a Legendre transformation that introduces the two-form
field as the main dynamical variable and rewrites the
action of our theory in a form most useful for practical computations. In
section \ref{sec:ham} we sketch the Hamiltonian analysis and count the
number of propagating DOF. Section \ref{sec:lin} contains a general discussion
on the problem of linearization. In section \ref{sec:grav} we warm up by
considering the case of pure gravity corresponding to $G={\rm SU}(2)$.
The Minkowski space background that we expand about is described here. 
Section \ref{sec:su3} deals with an example of a non-trivial group
for which we take $G={\rm SU}(3)$. It is here that we obtain a Lagrangian
describing the YM and Higgs sectors of our model. In section \ref{sec:inter}
we deduce interactions between various sectors of our model and show
that they are the standard interactions expected from such fields. 
In section \ref{sec:mass} we consider a more general set of defining potentials 
and show how Higgs masses are generated. We conclude with a summary and discussion.

\section{A Class of Generally Covariant Gauge Theories}
\label{sec:pure-con}

We start by giving the most compact formulation of our class of theories. This is not the 
formulation that is most suited for practical computations, but it is conceptually
the simplest. 

According to our proposal, a theory that unifies gravity with gauge fields is simply
the most general generally covariant group $G$ gauge theory. Thus, consider 
a connection $A^I$ in the principal $G$-bundle over the spacetime manifold $M$.
As is usual in physics literature, the bundle is assumed to be trivial, so the connection
can be viewed as a Lie-algebra-valued one-form on $M$. The group $G$ that we
consider is a general semi-simple complex Lie group. Reality conditions will later
need to be imposed to select a sector of the theory that corresponds to a particular
metric signature. Note, however, that at this point there is no metric, the only dynamical
variable of our theory is the connection $A^I$. 

As we have said, the idea is to consider the most general gauge and diffeomorphism 
invariant action that can be constructed from $A^I$. The following simple construction, 
generalizing verbatim considerations \cite{Krasnov:2009ip}
for the case of pure gravity, provides a Lagrangian with the
required properties. Being gauge-invariant, it must only involve the
curvature two-form $F^I=dA^I+(1/2)[A,A]^I$, where $[\cdot,\cdot]^I$ is the Lie-bracket
and the wedge product of forms is assumed. Consider the 4-form $F^I\wedge F^J$.
This is a 4-form valued in the space of symmetric bilinear forms in $\mathfrak g$,
the Lie-algebra of $G$. Choosing an arbitrary volume 4-form $(vol)$ we
can write $F^I\wedge F^J = (vol) \Omega^{IJ}$, where now $\Omega^{IJ}$
is a symmetric $n\times n$ matrix, where $n={\rm dim}({\mathfrak g})$. Since
$(vol)$ is defined only modulo rescalings $(vol)\to \alpha (vol)$, so is the matrix
$\Omega^{IJ}$ that under such rescalings transforms as $\Omega^{IJ}\to (1/\alpha)\Omega^{IJ}$.
Let us now introduce a function $f(X)$ of symmetric $n\times n$ matrices $X^{IJ}$ 
with the following properties. First, the function has to be gauge-invariant:
$f({\rm ad}_g X)=f(X)$, where ${\rm ad}_g$ is the adjoint action of the gauge group on
the space of symmetric bilinear forms on the Lie algebra. Second, the function
must be holomorphic (we work with complex-valued quantities). Third, and most
important, the function must be homogeneous of degree one $f(\alpha X)=\alpha f(X)$.
This property allows it to be applied to the 4-form $F^I\wedge F^J$, with the
result being again a 4-form. Indeed, we have $f(F^I\wedge F^J)=(vol)f(\Omega^{IJ})$,
and it is easy to see that due to the homogeneity of $f(\cdot)$, the resulting 4-form does 
not depend on which particular volume form $(vol)$ is chosen. Thus, the
quantity $f(F^I\wedge F^J)$ is an invariantly-defined 4-form, and it can be integrated 
over the spacetime manifold to produce an action:
\be\label{action-A}
S[A]=\int_M f(F^I\wedge F^J).
\ee
As we have already said, the action is complex, so later certain reality conditions will
be imposed. 

The presented formulation (\ref{action-A}) is conceptually nice, but it is very difficult
to deal with in practice. One of the main reasons for this is that there is no natural
background around which the theory can be expanded to produce a physically
meaningful perturbation theory. This can be seen as follows. The first variation of
the action (\ref{action-A}) is given by:
\be
\delta S=\int \frac{\partial f}{\partial F^I} \wedge D_A \delta A^I,
\ee
where the derivative of $f(\cdot)$ with respect to $F^I$ can be shown to make sense 
and is a certain ${\mathfrak g}$-valued 2-form. The second variation is given by:
\be\label{A-sec-var}
\delta^2 S= \int \frac{1}{2}\frac{\partial f}{\partial F^I} \wedge 
[\delta A,\delta A]^I + \frac{\partial^2 f}{\partial F^I\partial F^J} 
D_A \delta A^I \wedge D_A \delta A^J,
\ee
where the second derivative of $f(\cdot)$ is a zero-form.
Now, the most natural "vacuum" of the theory seems to be 
\be\label{A-vac}
F^I = 0, \qquad \frac{\partial f}{\partial F^I} = 0, \qquad 
\frac{\partial^2 f}{\partial F^I\partial F^J} \not =0.
\ee
Indeed, this would indeed be a "vacuum" of the theory in the sense that the first derivative of 
the "potential" function vanishes, which then automatically satisfies the field equations
$D_A (\partial f/\partial F^I)=0$, and only the second derivative is non-trivial. From (\ref{A-sec-var})
we see that in this case the first "mass" term is absent, and there is only the "kinetic" term
for the connection. Thus, it seems like the perfect vacuum to expand about.
However, an immediate problem with this vacuum is that in the absence of any 
background structure the second derivative in (\ref{A-vac}) can only be proportional to 
the Killing form $g^{IJ}$, which then gives a degenerate kinetic term. 
So, there does not seem to be any way to build a meaningful perturbation theory around 
(\ref{A-vac}). 

As an aside remark we mention that the fact that the ``kinetic'' form in (\ref{A-sec-var})
is necessarily degenerate is very important for the possibility to describe gravity
as a gauge theory. Indeed, as work \cite{Capovilla:1989ac} showed, general relativity
can be put in the form (\ref{action-A}) for $G={\rm SU}(2)$ and a very special
choice of the function $f(\cdot)$. At the same time, it is known to be impossible
to describe gravity that is mediated by a spin two particle in terms of a gauge
field that corresponds to a spin one particle. The resulution of this seeming
paradox lies in the fact that the pure connection formulation 
(\ref{action-A}) of gravity does not allow for a well defined perturbation theory
around Minkowski background, and so the particles that it describs are not 
spin one as would be the case in any other gauge theory. Below we shall see
how the usual spin two graviton arises via certain ``duality'' trick.

A conventional perturbative treatment for theory (\ref{action-A}) is possible, but requires
a rather strange, at least from the pure connection point of view, choice of
vacuum. Thus, as we shall see in details in the following sections, the usual 
perturbative expansion around a flat metric corresponds in the pure connection language 
to an expansion around the following point:
\be
F^I = 0, \qquad \frac{\partial f}{\partial F^I} \not = 0.
\ee
This is a very strange point to expand the theory about, for one seems to be sitting at
a point that is not a minimum of the "potential". However, the non-zero right-hand-side
of the first derivative of the potential receives the interpretation of essentially the Minkowski
metric, and a usual expansion then results. It might seem that this choice introduces a "mass"
term for the connection, but this is not so. In fact, the second "kinetic" term is still
a total derivative and plays no role, and there is only the "mass" term. However, as
we shall see, the connection is no longer a natural variable in this case, and one works with
a certain new two-form field $B^I$ via which the connection is expressed as $A^I\sim \partial B^I$,
so what appears as a mass term is in fact the usual kinetic one but for the two-form field. 

This discussion motivates introduction of a new set of dynamical fields. These are
originally introduced via the standard "Legendre transform" trick so that integrating them
out one gets an original action (\ref{action-A}). However, one can then also integrate out
the original connection field and obtain a theory for the new fields only. This point
of view turns out to be very profitable, and we develop it in the next section.

\section{Two-form field formulation}
\label{sec:two-form}

There are at least two different ways to arrive at the new formulation.
One of them is via a Legendre transform from (\ref{action-A}), the other one
by thinking about generalizations of BF theory.

\subsection{Legendre transform}

As we have already explained, we introduce a new set of fields, given by
a $\mathfrak g$-valued two-form $B^I$. The action that we would like to consider
is then of BF-type and is given by:
\be\label{action-AB}
S[A,B]=\int_M g_{IJ} B^I\wedge F^J - \frac{1}{2} V(B^I\wedge B^J).
\ee
Here $V(\cdot)$ is again a $G$-invariant, holomorphic and homogeneous order one
function of symmetric $n\times n$ matrices, and as such it can be applied to the
4-form $B^I\wedge B^J$, with the result being again a 4-form. The quantity $g_{IJ}$
is the Killing-Cartan form on $\mathfrak g$. 

Integrating out $B^I$ by solving its field equation:
\be
F^I = \frac{1}{2} \frac{\partial V}{\partial B^I},
\ee
which is algebraic in $B^I$, we get back the formulation (\ref{action-A}) with
$f(\cdot)$ being an appropriate Legendre transform of $V(\cdot)$. However,
the formulation (\ref{action-AB}) is much more powerful in that we can now choose
a constant $B^I$ background and obtain a well-defined perturbation theory.
We will later see how both gravity and Yang-Mills theory appear in such a
perturbative expansion. 

An alternative viewpoint on the "Legendre transform" described is as follows. 
As we shall see below, the new two-form field that we have introduced is
essentially the momentum canonically conjugate to the connection $A^I$.
Thus, a meaningful analogy for the relation between (\ref{action-A}) and
(\ref{action-AB}) is the relation between Lagrangian and Hamiltonian
formulation of mechanics. The former one uses only position variables
as dynamical variables, but leads to second-derivative equations of motion.
The later contains an independent variable - momentum, and leads to
first order equations of motion. Thus, loosely speaking, the action
(\ref{action-AB}) can be referred to as (\ref{action-A}) written in the "Hamiltonian
form" in which the momentum variable becomes an independent dynamical field. 

Before we proceed with an analysis of properties of the theory (\ref{action-AB}),
we would like to present an alternative derivation of this action.

\subsection{Generalization of BF theory}

An alternative way to arrive at (\ref{action-AB}) is to consider possible ways
to generalize the topological BF theory. For the case of $G={\rm SU}(2)$ this
was done in \cite{Krasnov:2008fm}, and here we generalize this analysis to a 
semi-simple Lie group. Following this reference we begin with the action
\begin{equation}
S[A,B]=\int g_{IJ}\, B^I \wedge F^J - \frac{1}{2} \Phi_{IJ} \, B^{I} \wedge B^{J} \, \label{abfpafG},
\end{equation}
where $B^I$ is a two-form valued in $\mathfrak g$, $F^I$ is the curvature 
$F^I=dA^I+\frac{1}{2} f^{I}_{JK} A^J \wedge A^K$ of $A^I$, $f^I_{JK}$ are
the structure constants, and $\Phi^{IJ}$ is a function 
(zero-form) valued in the symmetric product of two copies of $\mathfrak g$. At this stage
this quantity is undetermined. But we should say already now that it is not to be
thought of as an independent field to be varied with respect to, for it will later 
be fixed by Bianchi identities. Note that
only the symmetric part of $\Phi^{IJ}$ enters the action, this is why it is assumed
symmetric from the beginning. Our conventions are that we raise and lower indices with 
the Killing-Cartan metric $g_{IJ}$ and its inverse $g^{IJ}$. We also note that
for a semi-simple Lie algebra we can always find a basis in which the metric is diagonal, 
i.e. $g_{IJ}=\delta_{IJ}$, where $\delta_{IJ}$ is the Kronecker delta. 

Varying this action with respect to the connection $A^L$ and the field $B^L$ we get, respectively, 
\begin{align}
D_AB^I &\equiv dB^I+f^I_{JK}\, A^J \wedge B^K=0  \label{eqmA} \, ,\\ 
F^I &= \Phi^I_J \, B^J  \label{eqmB} \, .
\end{align}
We see that the idea of the above action ansatz is to generalize BF theory in such a way that
the equation (\ref{eqmA}) relating $B$ and $A$ is unchanged, while we now allow
for a non-zero curvature. As we have already said, we do not consider a variation with 
respect to $\Phi^{IJ}$ because we will later show that the Bianchi identities fix this quantity 
in terms of certain components of the two-form field $B^I$.

Let us now take the covariant exterior derivative of (\ref{eqmB}) and use (\ref{eqmA}) 
together with the Bianchi identity $D_AF^I=0$. We obtain
\begin{equation}
D_A\Phi^I_J \wedge B^J=0 \label{eqsmsbi}\, .
\end{equation}
Now, the covariant exterior derivative of $D_AB^I$ is 
\begin{equation}
D_A(D_AB^I)
=f^{I}_{JK} dA^J \wedge B^K + f^{I}_{JK} f^{K}_{LM}\, A^{J} \wedge A^{L} \wedge B^{M} \, . 
\end{equation}
Using the Jacobi identity $f^{N}_{IJ} f^{L}_{NK}+f^{N}_{JK} f^{L}_{NI}+ f^{N}_{KI}f^{L}_{NJ}=0$, the 
equation above can be rewritten as
\begin{equation}
D_A(D_AB^I)=f^{I}_{JL} \, F^J \wedge B^L  \, ,
\end{equation} 
and using equation (\ref{eqmA}) and  equation (\ref{eqmB}) we get
\begin{equation}
f^I_{JL} \, \Phi^J_K \, B^K \wedge B^L =0 \label{scpbb}\, . 
\end{equation}

Let us now compute the wedge product between (\ref{eqsmsbi}) and the one-form 
$\iota_{\xi}B^I$, which has components $(\iota_{\xi}B^I)_{\mu}=\xi^{\alpha} B^I_{\alpha \mu}$, where $\xi$ is an arbitrary vector field. We get:
\begin{equation}
D\Phi_{IJ} \wedge \iota_{\xi}B^{(I}\wedge B^{J)}=0 \, .
\end{equation}
But using $\iota_{\xi}B^{(I}\wedge B^{J)}=\frac{1}{2}\iota_{\xi}(B^{I}\wedge B^{J})$, we can rewrite
this as:
\begin{equation}\label{alex-1}
D\Phi_{IJ} \wedge \iota_{\xi}(B^{I}\wedge B^{J})=0 \, .
\end{equation}

Let us now define the "internal" metric $h^{IJ}$ by means of the following relation 
\be
B^I \wedge B^J= h^{IJ}\, (vol),
\ee
where $(vol)$ is an arbitrary volume 4-form. We can then rewrite (\ref{alex-1}) as:
\begin{equation}
h_{IJ}\, D\Phi^{IJ} \wedge \iota_{\xi}(vol)=0 \label{hdp3f}\, .
\end{equation}
Using the definition of $h^{IJ}$, we can also rewrite (\ref{scpbb}) as
\begin{equation}
f^I_{JK} \, \Phi^J_L \, h^{LK}=0 \label{scph} \, . 
\end{equation}
Now, computing $h_{IJ}\, D\Phi^{IJ}$
\begin{equation}
h_{IL}\, D\Phi^{IL}=h_{IL}\, (d\Phi^{IL}+2\, f^I_{JK}\, A^J\, \Phi^{KL}) \, ,
\end{equation}
we can see that the second term in the right hand side vanishes because of (\ref{scph}) 
and the condition that the Lie algebra is semi-simple. The later is used because for a semi-simple 
Lie algebra it is possible to define a Killing-Cartan metric, in our case $\delta_{IJ}$, 
with respect to which the object $f_{IJK}=\delta_{IL}\, f^L_{JK}$ is completely anti-symmetric. 
Our final result is:
\begin{equation}
h_{IJ}\, \partial_{\mu}\Phi^{IJ}\, \xi^{\mu}=0 \, ,
\end{equation}
which implies
\begin{equation}
h_{IJ}\, \partial_{\mu}\Phi^{IJ}=0 \label{ebhP}\, ,
\end{equation}
since $\xi$ is an arbitrary vector.

The above equation implies that the quantities $h^{IJ}$ and $\Phi^{IJ}$ are not independent. Let 
us define the ``potential function'' $V:=h^{IJ}\,\Phi_{IJ}$. Then, 
\begin{equation}
dV=\Phi_{IJ}\, dh^{IJ}+h_{IJ}\, d\Phi^{IJ}=\Phi_{IJ}\, dh^{IJ} \, ,
\end{equation}  
where we have used (\ref{ebhP}). This means that: a) the potential $V$ is only a function of 
$h^{IJ}$, i.e., $V=V(h^{IJ})$ and; b) the quantities $\Phi^{IJ}$ are given 
\be
\Phi_{IJ}=\frac{\partial V}{\partial h^{IJ}}
\ee
and; c) the potential $V$ is a homogeneous function of order one in $h^{IJ}$ since
\be
V=h^{IJ} \frac{\partial V}{\partial h^{IJ}}.
\ee
Thus, using the above definition of $h^{IJ}$, and the fact that $V(\cdot)$ is homogeneous, 
we can rewrite the action (\ref{abfpafG}) as 
\begin{equation}
S=\int g_{IJ} \, B^I \wedge F^J - \frac{1}{2} \, V(B^I\wedge B^J) \, ,
\end{equation}
which is exactly the action (\ref{action-AB}) we have obtained in the previous subsection.

\subsection{Parameterizations of the potential}

As defined so far, the theory is specified by the potential function $V(\cdot)$. In the
action (\ref{action-AB}) it is applied to a 4-form, which makes things rather inconvenient
in practice, since we do not have much experience with functions of forms. Thus, it is
desirable to rewrite it as a usual function of a matrix. We have already discussed
how to do it by introducing an auxiliary volume form, but it would be nice if we
could avoid any arbitrariness such as that of rescalings of $(vol)$. 
A possible way to do this is as follows. With our choice of conventions 
$dx^\mu\wedge dx^\nu \wedge dx^\rho\wedge dx^\sigma = -\tilde{\epsilon}^{\mu\nu\rho\sigma}
d^4x$ and we have:
\be
B^I\wedge B^J = \frac{1}{4} B^I_{\mu\nu} B^J_{\rho\sigma} 
dx^\mu \wedge dx^\nu \wedge dx^\rho\wedge dx^\sigma = -\frac{1}{4}
\tilde{\epsilon}^{\mu\nu\rho\sigma} B^I_{\mu\nu} B^J_{\rho\sigma}  d^4x,
\ee
where $\tilde{\epsilon}^{\mu\nu\rho\sigma}$ is a density weight one object that
does not require a metric for its definition. Thus, if we now define a densitized
"internal metric"
\begin{equation}
\tilde{h}^{IJ}=\frac{1}{4} B^I_{\mu\nu}B^J_{\rho\sigma} \tilde{\epsilon}^{\mu\nu\rho\sigma}\, ,
\end{equation} 
we can write the action as 
\begin{equation}
S[B,A]=\int g_{IJ}\, B^I \wedge F^J + \frac{1}{2} \, V(\tilde{h}) \, d^4 x\,  \label{abfppG} \, .
\end{equation}
Thus, the potential function is now applied to an $n\times n$ matrix (densitized),
and its derivatives can be computed via the usual partial differentiation. For example,
the first variation of this action can be seen to be given by
\begin{equation}
\delta S= \int \delta B^I \wedge \left( g_{IJ} F^J - 
\frac{\partial V(\tilde{h})}{\partial \tilde{h}^{IJ}} B^J \right)- 
g_{IJ} D_A B^I \wedge \delta A^J \, \label{fvabfppG}.
\end{equation}
Indeed, the variation of the last, potential term is given by:
\be
\frac{1}{2}\int \frac{\partial V(\tilde{h})}{\partial \tilde{h}^{IJ}} \frac{1}{2} \delta B^I_{\mu\nu}
B^J_{\rho\sigma} \tilde{\epsilon}^{\mu\nu\rho\sigma} d^4 x = 
-\int \frac{\partial V(\tilde{h})}{\partial \tilde{h}^{IJ}} \delta B^I\wedge B^J,
\ee
where the matrix of first derivatives $(\partial V(\tilde{h})/\partial \tilde{h}^{IJ})$ is an object of
density weight zero. Then, the field equations of our theory can be written as: 
\begin{align}
F_I &= \frac{\partial V(\tilde{h})}{\partial \tilde{h}^{IJ}} B^J  \label{feqBG} \, \\
DB^I &\equiv dB^I+f^I_{JK}\, A^J \wedge B^K=0  \label{feqAG} \, .
\end{align}

In the literature on this class of theories a different parameterization of the potential is 
sometimes used, see e.g. the original paper \cite{Krasnov:2006du}, and also 
the unification paper \cite{Smolin:2007rx}. Thus, to avoid having to take a function
of forms, and/or having to work with a homogeneous function, one can parameterize
the potential so that an ordinary function of one less variable arises. This can be done
via a Legendre transform trick. Thus, we introduce a new variable $\Psi^{IJ}$ that
is required to be tracefree $g_{IJ} \Psi^{IJ}=0$. The idea is that the matrix $\Psi^{IJ}$
is the tracefree part of the matrix of first derivatives $\Phi^{IJ}=(\partial V/\partial \tilde{h}^{IJ})$.
In other words, let us write
\begin{equation}
\Phi_{IJ}=\Psi_{IJ} -\frac{\Lambda}{n} g_{IJ} \, ,
\end{equation}
where $\Psi_{IJ}$ is  traceless. With $\Phi^{IJ}$ being a function of $\tilde{h}^{IJ}$, so is the 
trace part $\Lambda$. However, we can also declare $\Lambda$ to be a function of
$\Psi^{IJ}$, make $\Psi^{IJ}$ and independent variable and write the action in the form:
\be\label{action-AB-Psi}
S[B,A,\Psi]=\int g_{IJ}\, B^I \wedge F^J - \frac{1}{2} 
\left( \Psi_{IJ} -\frac{\Lambda(\Psi)}{n} g_{IJ}\right) B^I\wedge B^J .
\ee
Varying the action with respect to $\Psi^{IJ}$ one gets an equation for this matrix, which,
after being solved and substituted into the action gives back (\ref{abfppG}) with
$V(\cdot)$ being an appropriate Legendre transform of $\Lambda(\Psi)$. In the
formulation (\ref{action-AB-Psi}) the function $\Lambda(\Psi)$ is an arbitrary
function of a tracefree matrix $\Psi^{IJ}$, so there is no complication of having to
require $V(\cdot)$ to be homogeneous. This formulation was used in the first
papers on this class of theories, but it was later realized that the formulation that
works solely with the two-form field $B^I$ is more convenient. Thus, we do not
use (\ref{action-AB-Psi}) in this paper.

\section{Hamiltonian Analysis}
\label{sec:ham}

To exhibit the physical content of the above theory it is useful to perform the
canonical analysis. After the 3+1 decomposition the action reads, up to an unimportant
overall numerical factor:
\be
S = \int dt \int_\Sigma d^3 x \left( \tilde{P}^{aI} \dot{A}_a^I - H\right),
\ee
where 
\be
\tilde{P}^{aI} := \tilde{\epsilon}^{abc} B_{bc}^I,
\ee
and the Hamiltonian $H$ is:
\be\label{enm-h}
-\tilde{H} = A_0^I D_a \tilde{P}^{aI} + B_{0a}^I 
\tilde{\epsilon}^{abc} F_{bc}^I - V(B_{0a}^{(I} \tilde{P}^{aJ)}).
\ee
If we dealt with the pure BF theory the last ``potential'' term 
would be absent and all the quantities $B_{0a}^I$ would be Lagrange
multipliers. However, now the Lagrangian is not linear in $B_{0a}^I$,
and, as we shall see, all but 4 of these quantities are no longer
Lagrange multipliers and should be solved for. The equations one
obtains by varying the Lagrangian with respect to $B_{0a}^I$ are:
\be\label{B0-eqs}
\tilde{\epsilon}^{abc} F_{bc}^I = V_{(1)}^{IJ} \tilde{P}^{aJ},
\ee
where $V_{(1)}^{IJ}$ denotes the matrix of first partial derivatives of the
function $V(\cdot)$ with respect to its arguments:
\be
V_{(1)}^{IJ}:= \frac{\partial V(\tilde{h})}{\partial \tilde{h}^{IJ}}.
\ee

The equations (\ref{B0-eqs}) can be solved in quite a generality by 
finding a convenient basis in the Lie algebra. Thus, consider the momenta
$\tilde{P}^{aI}$. There are at least $n-3$ vectors $N^I_\alpha, \alpha=1,\ldots,n-3$
that are orthogonal to the momenta:
\be
\tilde{P}^{aI} N^I_\alpha = 0, \quad \forall a,\alpha.
\ee
These vectors can be chosen (uniquely up to ${\rm SO}(n-3)$ rotations) by requiring:
\be
N^I_\alpha N^I_\beta = \delta_{\alpha\beta}.
\ee
We can then use the qauntities $\tilde{P}^{aI}, a=1,2,3, N^I_\alpha, \alpha=1,\ldots,n-3$
as a basis in the Lie algebra. 

We can now decompose the quantity $B_{0a}^I$ as:
\be
B_{0a}^I = \tilde{P}^{bI} \utilde{B}_{ab} + N^I_\alpha B_a^\alpha,
\ee
where $\utilde{B}_{ab}, B_a^\alpha$ are components of $B_{0a}^I$ in this basis.
There are in total $3n$ components of $B_{0a}^I$ and they are represented
here as $9$ quantities $\utilde{B}_{ab}$ as well as $3(n-3)$ quantities $B_a^\alpha$.
The argument of the function $V(\cdot)$ is now given by:
\be
B_{0a}^{(I} \tilde{P}^{aJ)} = \tilde{P}^{b(I} \tilde{P}^{aJ)} \utilde{B}_{ab}
+ N^{(I}_\alpha B_a^\alpha \tilde{P}^{aJ)}.
\ee
It is clear that this depends only on the symmetric part $\utilde{B}_{ab}$
of the components $\utilde{B}_{ab}$. Thus, the anti-symmetric part
of this $3\times 3$ matrix cannot be determined from the equations (\ref{B0-eqs})
and thus $N^a$ in $\utilde{B}_{[ab]}:=(1/2)\epsilon_{abc} N^c$ remain
Lagrange multipliers. It is also clear that due to the homogeneity of
$V(\cdot)$ one more component of $B_{0a}^I$ cannot be solved for. This
can be chosen for example to be the trace part $B_{0a}^{I} \tilde{P}^{aI}$,
which will then play the role of the lapse function. All other $6 + 3(n-3) - 1$
components of $B_{0a}^I$ can be solved for for a generic function $V(\cdot)$,
i.e. under the condition that the matrix of second derivatives of $V(\cdot)$
is non-degenerate. We are not going to demonstrate this in full generality,
but will verify it in the linearized theory below.

After the quantities $B_{0a}^I$ are solved for we substitute them into
(\ref{enm-h}) and obtain the following Hamiltonian:
\be
-\tilde{H} = A_0^I D_a \tilde{P}^{aI} + N^a \tilde{P}^{bI} F_{ab}^I + 
\tilde{N}\Lambda(F,P),
\ee
where $\tilde{N}$ is the lapse function and 
$\Lambda(F,P)$ is an approprite Legendre transform of $V(\cdot)$ that
now becomes a function of the curvature $F_{ab}^I$ and momentum $\tilde{P}^{aI}$.
Thus, there are $n$ Gauss as well as 4 diffeomorphism constraints 
in the theory. It should be possible to check by an explicit computation that they 
are first class, as was done, for example for the case of $G={\rm SU}(2)$ in 
\cite{Krasnov:2007cq}, but we shall not attempt this here, postponing such an 
analysis till the linearized case considerations. The above arguments 
allow a simple count of the degrees of freedom described by the theory:
we have $3n$ configurational degrees of freedom minus $n$ Gauss constraints
minus $4$ diffeomoprhisms, thus leading to $2n-4$ DOF. Thus, when $G=K\times {\rm SU}(2)$ 
the above count of DOF gives the right number for a gravity plus $K$ Yang-Mills theory. 
For a general $G$ one might suspect that the centralizer of the gravitational
${\rm SU}(2)$ describes Yang-Mills, while other part of the Lie algebra corresponds
to some new kind of fields. Below we will unravel their nature by considering
the linearized theory. We also note that the above count of degrees of
freedom agrees with the one presented in \cite{Alexandrov:2008fs} for
the case $G={\rm SO}(4)$. Thus, it was seen there that the theory
describes in total $2\cdot 6-4=8$ DOF, which were interpreted
as those corresponding to 2 graviton polarizations plus six
new DOF.

\section{The Linearized Theory: General considerations}
\label{sec:lin}

As we have seen in the previous section, the mechanism that selects the
gravitational ${\rm SU}(2)$ in $G$ is that the momentum variable $\tilde{P}^{aI}$
provides a map from the (co-) tangent space to the spatial slice into
$\mathfrak g$. This selects a 3-dimensional subspace in $\mathfrak g$ that
plays the role of the gravitational gauge group. Below we are going to
see this mechanism at play at the level of the Lagrangian formulation,
by studying the linearization of the action (\ref{action-AB}). In this section
it will be convenient to introduce a certain numerical prefactor in
front of this action so that the normalization of the graviton
kinetic term in the case of gravity will come out right. Thus, we
shall from now on consider the following action 
\be\label{action-AB*}
S[A,B]=4\im \int_M g_{IJ} B^I\wedge F^J - \frac{1}{2} V(B^I\wedge B^J),
\ee
where $\im=\sqrt{-1}$.

\subsection{Kinetic term}

In this section we present some general considerations that apply to any
background. We specialize to the Minkowski spacetime background in the
next section. Let us call the first term in (\ref{action-AB*}) $S_{BF}$ and the second 
``potential'' term $S_{BB}$. Then, the second variation of $S_{BF}$ is given by:
\begin{equation}\label{bf-lin}
\delta^2 S_{BF}= 4\im \int 2\delta B^I \wedge D_A \delta A^I + B^I \wedge [\delta A,\delta A]^I,
\end{equation} 
and the action linearized around $B_0, A_0$ is obtained by evaluating this on $B_0, A_0$.

As we have already mentioned, we are to view our theory as that of the two-form field
$B^I$, with the connection $A^I$ to be eliminated (whenever possible, see below) by solving its
field equations. Thus, let us assume that we are given a background two-form $B_0^I$.
The linearized connection is then to be determined from the linearized equation 
(\ref{eqmA}) that reads:
\be\label{comp-lin}
D_0 \delta B^I + [\delta A, B_0]^I = 0,
\ee
where $D_0$ is the covariant derivative with respect to the background connection $A_0^I$.
Now the background two-form $B_0^I$ is a map from the six-dimensional space of bivectors
to $\mathfrak g$, and thus selects in $\mathfrak g$ at most a 6-dimensional preferred
subspace. Let us denote this subspace by $\mathfrak k$. This subspace may or may not be
closed under Lie brackets, but for simplicity, in this paper we shall assume
that our background $B_0^I$ is such that $\mathfrak k$ is a Lie subalgebra (below
we shall make an even stronger assumption about $\mathfrak k$). It is then clear that
the part of $\delta A^I$ that lies in the centralizer of $\mathfrak k$ in $\mathfrak g$
drops from the equation (\ref{comp-lin}) and cannot be solved for. As we shall later
see, this will be the part of the group that is to describe Yang-Mills fields. The other
part of $\delta A^I$ can in general be found. For this part of the connection 
both terms in (\ref{bf-lin}) are of the same form due to (\ref{comp-lin}), and
the linearized action can be written compactly as:
\be\label{bf-lin-higgs}
\delta^2 S_{BF}= 4\im \int \delta B^I \wedge D_0 \delta A^I ,
\ee
where $\delta A^I$ has to be solved for from (\ref{comp-lin}). On the other hand, for
the subgroup of $\mathfrak g$ that centralizes $\mathfrak k$ the last term in (\ref{bf-lin})
is absent and we have:
\be\label{bf-lin-ym}
\delta^2 S_{BF}= 8\im \int \delta B^I \wedge D_0 \delta A^I.
\ee
Thus, our analysis of the "kinetic" term is going to be different for different parts of the Lie algebra.

\subsection{Potential term}

In this subsection we compute the second variation of the potential term 
$S_{BB}$ and discuss how it can be evaluated on a given background. We have:
\begin{equation}
\label{bb-lin}
\delta^2 S_{BB}=4\im\int  
2 \frac{\partial^2 V(\tilde{h})}{\partial \tilde{h}^{KL} \partial \tilde{h}^{IJ}} \, 
(B_0 \delta B)^{IJ} (B_0 \delta B)^{KL}+\frac{\partial V(\tilde{h})}{\partial \tilde{h}^{IJ}}\, 
(\delta B \delta B)^{IJ} \, ,
\end{equation}
where the integration measure $d^4x$ is implied, and we have introduced notations
\be
(B_0 \delta B)^{IJ} = \frac{1}{4}\tilde{\epsilon}^{\mu\nu\rho\sigma} B_{0\,\mu\nu}^{(I} 
\delta B_{\rho\sigma}^{J)}, \qquad
(\delta B \delta B)^{IJ} = \frac{1}{4}\tilde{\epsilon}^{\mu\nu\rho\sigma} \delta B_{\mu\nu}^{I} 
\delta B_{\rho\sigma}^{J},
\ee
and where the matrix of second derivatives is of density weight minus one. 

Let us now discuss how the derivatives of the potential can be computed. 
In general, with the potential function $V(\tilde{h})$ being homogeneous order one
function of an $n\times n$ matrix, it can be reduced to a function of ratios of its
invariants. A subset of invariants is obtained by considering traces of powers of
$\tilde{h}^{IJ}$.  However, in general these are not all invariants, and other invariants
will be introduced and discussed below in section \ref{sec:mass}. But for now,
to simplify the discussion, let us consider a special class of potentials that only
depend on the invariants obtained as the traces of powers of $\tilde{h}^{IJ}$.
Many aspects of our theory can be seen already for this special choice. 
Thus, consider the potential of the form:
\begin{equation}\label{V-f}
V=\frac{Tr \,\tilde{h}}{n}\, f\left(\frac{Tr \,\tilde{h}^2}{(Tr\, \tilde{h})^2}\, , \dots, 
\frac{Tr\, \tilde{h}^n}{(Tr\, \tilde{h})^n}\right) \, .
\end{equation}
where $f$ is now an arbitrary function of its $n-1$ arguments, 
$Tr\, \tilde{h}=g_{IJ}\, \tilde{h}^{IJ}$ and 
\begin{equation}
Tr \,\tilde{h}^p=\tilde{h}^{M_1}_{\;\;\;\,M_2}\,\tilde{h}^{M_2}_{\;\;\;\,M_3}\cdots \cdots 
\tilde{h}^{M_p}_{\;\;\;\,M_1} \, ,
\end{equation}
for $p\ge 2$. In fact, in view of the fact that the rank of $\tilde{h}^{IJ}$ is at most
six, not all the invariants are independent, so we could consider only 5 first 
arguments of $f(\cdot)$. Note that $f(\cdot)$ here is distinct from the function
used in the action (\ref{action-A}) in the pure connection formulation of our theory:
it is now an arbitrary function of its arguments, while this symbol in (\ref{action-A})
stands for a homogeneous order one function. 

The parameterization given allows derivatives to be computed.
Thus, the first derivative of the potential function with respect to $\tilde{h}^{IJ}$ is
\begin{equation}
\frac{\partial V(\tilde{h})}{\partial \tilde{h}^{IJ}}
=\frac{g_{IJ}}{n} \, f + \frac{Tr\, \tilde{h}}{n}\, 
\frac{\partial f}{\partial \tilde{h}^{IJ}} \label{fdVG}\, , 
\end{equation}
with $(\partial f/\partial \tilde{h}^{IJ})$ given by 
\begin{align}
\frac{\partial f}{\partial \tilde{h}^{IJ}}
=&\sum_{p=2}^{n} f'_p \,\, \frac{\partial}{\partial \tilde{h}^{IJ}}
\left( \frac{Tr\, \tilde{h}^p}{(Tr\, \tilde{h})^p}\right) 
\notag \\
=&\sum_{p=2}^{n} p f'_p \,\, \left( \frac{\tilde{h}^{p-1}_{IJ}}{(Tr\, \tilde{h})^p}
-\frac{Tr\, \tilde{h}^p}{(Tr\, \tilde{h})^{p+1}}\,g_{IJ} \right)
\end{align}
where $f'_p$ is the derivative of $f$ with respect to its argument 
$(Tr\, \tilde{h}^p/(Tr\, \tilde{h})^p)$ and $\tilde{h}^p_{IJ}$ is
\begin{equation}
\tilde{h}^p_{IJ}=\tilde{h}_{I M_1}\,\tilde{h}^{M_1}_{\;\;\;\,M_2}\cdots \cdots 
\tilde{h}^{M_{p-1}}_{\;\;\;\;\;\;\;\;J} \, .
\end{equation}

The second derivative of $V(\tilde{h})$ is given by:
\begin{equation}
\frac{\partial^2 V(\tilde{h})}{\partial \tilde{h}^{KL}\partial \tilde{h}^{IJ}}=
\frac{g_{IJ}}{n} \, 
\frac{\partial f}{\partial \tilde{h}^{KL}}+\frac{g_{KL}}{n} \, 
\frac{\partial f}{\partial \tilde{h}^{IJ}} + \frac{Tr\, \tilde{h}}{n}\, 
\frac{\partial^2 f}{\partial \tilde{h}^{KL}\partial \tilde{h}^{IJ}} \label{sdVG}\, ,
\end{equation}
with $(\partial^2 f/\partial \tilde{h}^{KL}\partial \tilde{h}^{IJ})$ given by
\begin{align}
\frac{\partial^2 f}{\partial \tilde{h}^{KL}\partial \tilde{h}^{IJ}}
=&\sum_{p=2}^{n}\sum_{q=2}^{n} f''_{pq} \,\, 
\frac{\partial}{\partial \tilde{h}^{IJ}}
\left( \frac{Tr\, \tilde{h}^p}{(Tr\, \tilde{h})^p}\right) \,\, 
\frac{\partial}{\partial \tilde{h}^{KL}}
\left( \frac{Tr\, \tilde{h}^q}{(Tr\, \tilde{h})^q}\right) \notag \\
&+ \sum_{p=2}^{n} f'_p \,\, \frac{\partial^2}{\partial \tilde{h}^{KL}\partial \tilde{h}^{IJ}}
\left( \frac{Tr\, \tilde{h}^p}{(Tr\, \tilde{h})^p}\right) \, ,
\end{align}
where $f''_{pq}$ stands for the derivative of $f'_p$ with respect to its $q$ argument and 
\begin{align}
\frac{\partial^2}{\partial \tilde{h}^{KL}\partial \tilde{h}^{IJ}}
\left( \frac{Tr\, \tilde{h}^p}{(Tr\, \tilde{h})^p}\right)
=&\frac{p}{(Tr\, \tilde{h})^p}\, \frac{\partial \tilde{h}^{p-1}_{IJ}}{\partial \tilde{h}^{KL}}
-\frac{p^2\,\,\tilde{h}^{p-1}_{IJ}}{(Tr\, \tilde{h})^{p+1}}\,\, g_{KL}
-\frac{p^2\,\,\tilde{h}^{p-1}_{KL}}{(Tr\, \tilde{h})^{p+1}}\,\, g_{IJ}\notag \\
&+\frac{p (p+1)\,\, Tr\,\tilde{h}^p}{(Tr\, \tilde{h})^{p+2}}\,\, g_{IJ} g_{KL} \, ,
\end{align}
with
\begin{equation}\label{h-der-h}
\frac{\partial \tilde{h}^{p-1}_{IJ}}{\partial \tilde{h}^{KL}}=g_{I(K} \, \tilde{h}_{L)M_1}\cdots \cdots \tilde{h}^{M_{p-3}}_{\;\;\;\;\;\;\;\;J}+\tilde{h}_{I(K}\tilde{h}_{L)M_1}\cdot\cdots\tilde{h}^{M_{p-4}}_{\;\;\;\;\;\;\;\;J}+\cdots  \cdots+\tilde{h}_{IM_1}\cdots \cdots \tilde{h}^{M_{p-3}}_{\;\;\;\;\;\;\;\;(K} \, g_{L)J} \, . 
\end{equation}
With the above formulas for the fist and second derivative of the potential it 
is relatively easy to find the linearized action for any semi-simple Lie group. 

\section{The $G={\rm SU}(2)$ Case: Gravity}
\label{sec:grav}

As we have already mentioned, the case $G={\rm SU}(2)$ describes (complexified)
gravity theory. A particular choice of the potential function, see below, gives general
relativity, while a general potential corresponds to a family of deformations of GR.
In this section, as a warm-up to the general $G$ case, we shall study the corresponding
linearized theory. Such an analysis has already appeared in \cite{Freidel:2008ku}. However,
our method and goals here differ significantly from that reference.

\subsection{The metric}

To understand how $G={\rm SU}(2)$ case can describe gravity we need to see how
the spacetime metric described by the theory is encoded. The answer to this is very simple: there is a
unique (conformal) metric that makes the triple $B^i$, where $i$ is the ${\mathfrak su}(2)$
index, into a set of self-dual two-forms. This is the so-called Urbantke metric \cite{Urbantke:1984eb}
\be\label{Urb}
\sqrt{-g} g_{\mu\nu} \sim \epsilon^{ijk} B^i_{\mu\alpha} B^j_{\nu\beta} B^k_{\rho\sigma} 
\tilde{\epsilon}^{\alpha\beta\rho\sigma}
\ee
that is defined modulo an overall factor. We remind the reader that at this stage
all our fields are complex, and later reality conditions will be imposed to
select physical real Lorentzian signature metrics. 

Alternatively, given a metric $g_{\mu\nu}$ one can easily construct a ``canonical'' triple of self-dual
two-forms that encode all information about $g_{\mu\nu}$. This proceeds via introducing 
tetrad one-forms $\theta^I$, with $I=0,1,2,3$ here. One then constructs the two-forms
$\Sigma^{IJ}:=\theta^I\wedge \theta^J$ and takes the self-dual part of $\Sigma^{IJ}$ with
respect to $IJ$. The resulting two-forms are automatically self-dual. They can be explicitly
constructed by decomposing $I=(0,a)$ and then writing:
\be\label{Sigma}
\Sigma^a = \im \theta^0 \wedge \theta^a - \frac{1}{2} \epsilon^{abc} \theta^b\wedge \theta^c.
\ee
Here $\im=\sqrt{-1}$ is the imaginary unit. Its presence in this formula has to do with
the fact that self-dual quantities in a spacetime of Lorentzian signature are necessarily 
complex. Thus, even though at this stage there is no well defined signature (all
quantities are complex), it is convenient to introduce $\im$ here so that later
appropriate reality conditions are easily imposed. We note that ``internal'' Lorentz
rotations of the tetrad $\theta^I$ at the level of $\Sigma^a$ boil down to (complexified)
${\rm SU}(2)$ rotations of $\Sigma^a$. 

A general ${\mathfrak su}(2)$-valued two-form field $B^i$ carries more information than
just that about a metric. Indeed, one needs $3\times 6$ numbers to specify it, while only
10 are necessary to specify a metric. A very convenient description of the other components
is obtained by introducing a metric defined by $B^i$ via (\ref{Urb}) and then using
the ``metric'' self-dual two-forms (\ref{Sigma}) as a basis and decomposing:
\be\label{B-Sigma}
B^i = b^i_a \Sigma^a.
\ee
The quantities $b^i_a$ give 9 components, the metric gives 10, and the choice of ``internal''
frame for $\Sigma^a$ adds 3 more components. There is also a freedom
of rescalings $b^i_a\to \Omega^{-2} b^i_a, \Sigma^a \to \Omega^2 \Sigma^a$, 
as well as freedom of ${\rm SO}(3)$ rotations acting simultaneously on $\Sigma^a$
and $b^i_a$, overall producing 18 independent components of $B^i$.

When one substitutes the parameterization (\ref{B-Sigma}) into the action (\ref{action-AB})
one finds that the fields $b^i_a$ are non-propagating and should be integrated out. Once
this is done one obtains an ``effective'' Lagrangian for the metric described by $\Sigma^a$.
Below we shall see how this works in the linearized theory. However, we first need to choose
a background.

\subsection{Minkowski background}

The Minkowski background is described in our framework by a collection of metric
two-forms (\ref{Sigma}) constructed from the Minkowski metric. Thus, we choose an
arbitrary time plus space split and write:
\be\label{s}
\Sigma_0^a = \im dt \wedge dx^a - \frac{1}{2} \epsilon^{abc} dx^b\wedge dx^c,
\ee
where $dt, dx^a, a=1,2,3$ form a tetrad for the Minkowski metric $ds^2=-dt^2+\sum_a (dx^a)^2$.
Our two-form field background is then chosen to be
\be\label{backgr}
B_0^i = \delta^i_a \Sigma^a_0,
\ee
where $\delta^i_a$ is an arbitrary ${\rm SO}(3)$ matrix that for simplicity can be chosen
to be the identity matrix. 

In what follows we will also need a triple of anti-self dual metric forms that, together with
(\ref{Sigma}) form a basis in the space of two-forms. A convenient choice is given by:
\be\label{sb}
\bar{\Sigma}_0^a = \im dt \wedge dx^a + \frac{1}{2} \epsilon^{abc} dx^b\wedge dx^c.
\ee

The following formulas, which can be shown to follow directly from definitions 
(\ref{s}) and (\ref{sb}), are going to be very useful  
\begin{align}
\Sigma^a_{0\, \mu\sigma} \Sigma^{b\sigma}_{0 \; \:\nu}&
=-\delta^{ab}\, \eta_{\mu\nu} + \epsilon^{abc}\,\Sigma^c_{0\, \mu\nu} \, , \label{asub}\\
\Sigma^{a\mu\nu}_0 \Sigma^b_{0\, \mu\nu}&=4\, \delta^{ab} \label{as1} \, ,\\
\epsilon^{abc}\, \Sigma^a_{0\, \mu\sigma} \Sigma^{b\sigma}_{0 \; \:\lambda} 
\Sigma^c_{0\, \lambda\mu}&=-4! \label{as2} \, ,\\
\epsilon^{abc}\, \Sigma^a_{0\, \mu\nu} \Sigma^{b}_{0\, \rho\sigma} 
\Sigma^{d\nu\sigma}_0&=-2\delta^{cd}\, \eta_{\mu\rho}  \label{as3} \, ,\\
\Sigma^a_{0\,\mu\nu} \Sigma^a_{0\,\rho\sigma} &= \eta_{\mu\rho}\eta_{\nu\sigma}-
\eta_{\mu\sigma}\eta_{\nu\rho}-\im \epsilon_{\mu\nu\rho\sigma}\, \label{ss-proj},
\end{align} 
where $\eta_{\mu\nu}$ is the Minkowski metric.
We are going to refer to them as the algebra of $\Sigma$'s.

The first of the relations above, namely (\ref{asub}), is central, for all others
(apart from (\ref{ss-proj})) can be derived from it. It is useful to develop some basis-independent
understanding of this relation. We are working with the Lie algebra ${\mathfrak su}(2)$
and are considering a basis $X^a$ in it in which the structure constants read
$[X^a,X^b]=\epsilon^{abc} X^c$. This is the basis given by $X^a=-(\im/2)\sigma^a$,
where $\sigma^a$ are Pauli matrices. The metric $g^{ab}=\delta^{ab}$ on the Lie algebra 
can be obtained as $g^{ab}=-2 {\rm Tr}(X^aX^b)$. Then (\ref{asub}) can be
understood as follows: the product of two $\Sigma$'s is given by minus the metric
plus the structure constants times $\Sigma$. We will see that in this form the
relations (\ref{asub}) persist to any basis in ${\mathfrak su}(2)$. 

\subsection{Linearized action}

We are now going to linearize the $G={\rm SU}(2)$ theory around the background
(\ref{backgr}). Thus, we take:
\begin{equation}
B^i=B^i_0+b^i \, .
\end{equation} 
As we have already discussed, to linearize the kinetic BF term of the action we need
to solve for the linearized connection if we can. This is certainly possible
for the case at hand, as we shall now see.

If we denote the linearized connection by $a^i$ we have to solve the following system of equations
\begin{equation}
db^i+\epsilon^i_{\,jk}\, a^j \wedge B_0^k =0 \label{cesu2}\, ,
\end{equation}
where we have used the fact that the background connection is zero. It is convenient
at this stage to replace all $i$-indices by $a$-ones, which we can do using the
background object $\delta^i_a$ that provides such an identification. We can now
use the self-duality $\epsilon^{\mu\nu\rho\sigma}\Sigma^a_{0\,\mu\nu}=2i \Sigma^{a\,\mu\nu}_0$
of the background to rewrite this equation as
\be\label{su2-comp}
\frac{1}{2\im}\epsilon^{\mu\nu\rho\sigma} \partial_\nu b^a_{\rho\sigma} +
\epsilon^{abc} a^b_{\nu} \Sigma^{c\,\mu\nu}_0 = 0.
\ee
We now multiply this equation by $\Sigma^{a\,\alpha\beta}_0\Sigma^d_{0\,\alpha\mu}$,
and use the identity (\ref{as3}) to get:
\be\label{a-1}
a^{a}_\beta= \frac{1}{2} \Sigma^{b\!\quad\alpha}_{0\,\beta} \Sigma^a_{0\,\alpha\mu}
\frac{1}{2\im}\epsilon^{\mu\nu\rho\sigma} \partial_\nu b^b_{\rho\sigma}, \qquad
\mathrm{or}\qquad 
a^{a}_{\beta} = \frac{1}{4\im} \Sigma^{b\!\quad\alpha}_{0\,\beta} \Sigma^a_{0\,\alpha\mu}
 (\partial b^b)^\mu,
\ee
where we have introduced a compact notation:
\be\label{db}
(\partial b^b)^\mu := \epsilon^{\mu\nu\rho\sigma} \partial_\nu b^b_{\rho\sigma}
\ee
for a multiple of the Hodge dual of the exterior derivative of the perturbation two-form. 

The BF part of the linearized action was obtained in (\ref{bf-lin-higgs}). We need
to divide the second variation given in this formula by 2 to get the correct
action quadratic in the perturbation. Thus, we have:
\be\label{S-2-2}
S^{(2)}_{BF} = 2\im \int b^a \wedge da^a = 
-\im \int a_\mu^a (\partial b^a)^\mu,
\ee
where we have written everything in index notations and integrated by parts to put the
derivative on $b_{\mu\nu}^a$, and used the definition (\ref{db}). Now 
substituting (\ref{a-1}) we get:
\be\label{S-bf-lin}
S^{(2)}_{BF} = \frac{1}{4}
\int \eta^{\alpha\beta} \Sigma^{a}_{0\, \alpha \mu} (\partial b^b)^\mu \Sigma^b_{0\, \beta\nu} 
(\partial b^a)^\nu.
\ee

Let us now linearize the potential term. For this we need to know the background $\tilde{h}^{ij}$
as well as the matrices of first and second derivatives for the background. Using (\ref{s}) is easy to 
see that $\tilde{h}^{ij}_0=2\im \delta^{ij}$. Since the background volume form is 
just the identity we can now safely remove the density weight symbol from the matrix $\tilde{h}^{ij}_0$. 
Also, as before, let us replace all $i$-indices by $a$-indices using $\delta^i_a$. Using 
(\ref{fdVG}) and the fact that the first derivatives $(\partial f/\partial h^{ab})$
vanish on this background we immediately get:
\begin{equation}
\left. \frac{\partial V}{\partial h^{ab}} \right|_{h_0}=\frac{\delta_{ab}}{3}\, f_0,
\end{equation}
where $f_0$ is the background value of the function $f$ in the parameterization (\ref{V-f}).
It is not hard to see that this value plays the role of the cosmological constant of
the theory, so in our Minkowski background it is necessarily zero by the background
field equations. The matrix of second derivatives of the potential is easily evaluated
using (\ref{sdVG}) and we find:
\begin{equation}
\left. \frac{\partial^2 V}{\partial h^{cd}\partial h^{ab}}\right|_{h_0}
=\frac{g}{2\im} \, 
\left( \delta_{a(c}\delta_{d)b}-\frac{1}{3}\delta_{ab}\delta_{cd} \right) \, ,
\end{equation}
where we have introduced:
\be
g:= \sum_{p=2,3} \frac{(f'_p)_0 \, p(p-1)}{3^p} \, .
\ee
This is a constant of dimensions of the cosmological constant $1/L^2$. It is going to play a
role of a parameter determining the strength of gravity modifications.

We can now write the linearized potential term (\ref{bb-lin}). We must divide it by two to
get the correct action for the perturbation. This gives:
\begin{equation}
\label{lapsu2}
S^{(2)}_{BB}=-\frac{g}{2} \int \left( \delta_{a(c}\delta_{d)b}-\frac{1}{3}\delta_{ab}\delta_{cd} \right) 
\, \left( \Sigma^{a\,\mu\nu}_0 b^b_{\mu\nu} \right)\, 
\left( \Sigma^{c\,\rho\sigma}_0 b^d_{\rho\sigma} \right) \, .
\end{equation}
Note that the tensor in brackets here is just the projector on the tracefree part.
This fact will be important in our Hamiltonian analysis below.
Our total linearized action is thus (\ref{S-bf-lin}) plus (\ref{lapsu2}).

\subsection{Symmetries}

The quadratic form obtained above is degenerate, and its degenerate directions correspond
to the symmetries of the theory. These are not hard to write down. An obvious symmetry is
that under (complexified) ${\rm SO}(3)$ rotations of the fields. Considering an infinitesimal gauge 
transformation of the background $\Sigma^a_{0\,\mu\nu}$ we find that the
action must be invariant under the following set of transformations:
\be\label{delta-su2-b}
\delta_\omega b^a_{\mu\nu} = \epsilon^{abc} \omega^b \Sigma^c_{0\,\mu\nu},
\ee
where $\omega^a$ are infinitesimal generators of the transformation. It is clear that
(\ref{lapsu2}) is invariant since it involves only the $ab$-symmetric part of 
$(\Sigma^{a\,\mu\nu}_0 b^b_{\mu\nu})$, and the transformation (\ref{delta-su2-b})
affects the anti-symmetric part. Let us check the invariance of the kinetic term
(\ref{S-bf-lin}). We have the following expression for the variation:
\be\label{gauge-1}
\frac{1}{2} \int \eta^{\alpha\beta} 
\Sigma^{a}_{0\,\alpha\mu} (\partial \delta_\omega b^b)^\mu 
\Sigma^b_{0\,\beta\nu} (\partial b^a)^\nu.
\ee
Substituting here the expression (\ref{delta-su2-b}) for the variation we find:
\be
\eta^{\alpha\beta} \Sigma^{a}_{0\,\alpha\mu} (\partial \delta_\omega b^b)^\mu 
\Sigma^b_{0\,\beta\nu}  = 2\im \eta^{\alpha\beta} \Sigma^{a}_{0\,\alpha\mu} 
\epsilon^{bcd} \partial_\rho \omega^c \Sigma^{d\,\mu\rho}_0  
\Sigma^b_{0\,\beta\nu}= 4\im \partial_\nu \omega^i,
\ee
where we have used the self-duality of $\Sigma^a_{0\,\mu\nu}$ and applied the identity
(\ref{as3}) once. Substituting this to (\ref{gauge-1}) and integrating by parts
to move the derivative from $\omega^a$ to $b^a$ we get under the integral 
$\epsilon^{\mu\nu\rho\sigma}\partial_\mu\partial_\nu b^a_{\rho\sigma}=0$,
since the partial derivatives commute. This proves the invariance under gauge transformations.

Another set of symmetries of the action is that of diffeomorphisms. These are given by:
\be\label{diff-1}
\delta_\xi b^a = d \iota_\xi \Sigma_0^a,
\ee
where $\iota_\xi$ is the operator of interior product with a vector field $\xi^\mu$. It is not hard
to compute this explicitly in terms of derivatives of the components of the vector field. However,
we do not need all the details of this two-form. Indeed, let us first note that
the first "kinetic" term of the action is in fact invariant under a larger symmetry:
\be \label{topo}
\delta_\eta b^a = d\eta^a,
\ee
where $\eta^a$ is an arbitrary Lie-algebra valued one-form. Indeed, this is obvious
given that the kinetic term is constructed from the components of the 3-form $db^a$
given by the exterior derivative of the perturbation two-form. Thus, (\ref{topo})
indeed leaves the kinetic term invariant. Then, since (\ref{diff-1}) is of the
form (\ref{topo}) with $\eta^a=\iota_\xi \Sigma^a_0$ we have the invariance of the
first term. To see that the potential term (\ref{lapsu2}) 
is invariant we should simply show that the symmetric 
tracefree  part of the matrix $(\Sigma_0 \delta_\xi b)^{ab}$ is zero. Let us compute the symmetric
part explicitly. We have:
\be
\Sigma^{(a\,\mu\nu}_0 \partial_\mu \xi^\rho \Sigma^{b)}_{0\,\rho\nu} =
\delta^{ab} \partial_\rho \xi^\rho,
\ee
where we have used (\ref{asub}). Thus, there is only the
trace symmetric part, so the part that enters into the variation of the action (\ref{gauge-1}) is zero.
This proves the invariance under diffeomorphisms. Note that the second "potential" term
is not invariant under all transformations (\ref{topo}), since for such a transformation that is
not a diffeomorphism the matrix $(\Sigma_0 \delta_\eta b)^{ab}$ contains a non-trivial symmetric 
tracefree part, as can be explicitly checked. 

We will see that these are the only symmetries when we perform the Hamiltonian analysis. 
However, before we do this, let us show how the usual linearized GR appears from
our theory.

\subsection{Relation to GR}

In this subsection we would like to describe how general relativity (linearized)
with its usual gravitons appears from the linearized Lagrangian described above.
We shall see that to get GR we must take the limit when the ``mass'' parameter $g$
for the components $(\Sigma_0 b)^{ab}_{tf}$, where $tf$ stands for the tracefree
part, is sent to infinity. Indeed, the potential part (\ref{lapsu2}) depends
precisely on these components, and when the parameter $g$ is sent to infinity
these components are effectively set to zero. We shall now see that this gives
GR. 

It is not hard to show that in general the tracefree part 
$h^{tf}_{\mu\nu}:=h_{\mu\nu}-(1/4)\eta_{\mu\nu} h^\rho_\rho$ of 
the metric perturbation $h_{\mu\nu}$ defined via $g_{\mu\nu}=\eta_{\mu\nu}+h_{\mu\nu}$ 
corresponds in our language of two forms to the 
anti-self-dual part of the two-form perturbation:
\be\label{b-h}
(b_{\mu\nu}^a)_{asd} = \Sigma^{a\!\quad\rho}_{0\,[\mu} h^{tf}_{\nu]\rho}.
\ee
The fact that this two-form is anti-self-dual
can be easily checked by contracting it with $\Sigma^{b\,\mu\nu}_0$
and using the algebra (\ref{asub}). The result is zero, as appropriate for an
anti-self-dual tow-form. In addition to (\ref{b-h}) there is
in general also the self-dual part of the two-form perturbation. However,
in the limit $g\to\infty$ all but the trace part of this gets set to
zero by the potential term. The trace part, on the other hand, is proportional
to the trace part $\eta^{\mu\nu} h_{\mu\nu}$ of the metric perturbation.
To simplify the analysis it is convenient to set this to zero
$\eta^{\mu\nu} h_{\mu\nu}=0$. This is allowed since in pure gravity the trace
of the perturbation does not propagate. Then (\ref{b-h}) is the complete
two-form perturbation, and we can drop the $tf$ symbol.

To simplify the analysis further, instead of deriving the full linearized action 
for the metric perturbation $h_{\mu\nu}$,
let us work in the gauge where the perturbation is transverse
$\partial^\mu h_{\mu\nu}=0$. Let us then compute the quantity $(\partial b^a)^\mu$ in 
this gauge. Using anti-self-duality of $b^a_{\mu\nu}$ given by (\ref{b-h}) we have:
\be
\epsilon^{\mu\nu\rho\sigma}\partial_\nu b^a_{\rho\sigma} = -2\im \partial_\nu b^{a\,\mu\nu}.
\ee
Substituting here the explicit expression (\ref{b-h}) and using the transverse gauge
condition we get:
\be
(\partial b^a)^\mu = \im \Sigma^{a\,\nu\rho}_0 \partial_\nu h_\rho^\mu.
\ee
We can now substitute this into the action (\ref{S-bf-lin}) to get:
\begin{align}
S^{(2)}=&-\frac{1}{4}\int \eta^{\alpha\beta} 
\Sigma^a_{0\,\alpha\mu} \Sigma^{b\, \rho\sigma}_0 \partial_\rho h_{\sigma}^\mu
\Sigma^b_{0\,\beta\nu} \Sigma^{a\, \gamma\delta}_0 \partial_\gamma h_\delta^\nu
\\ \nonumber =&
-\frac{1}{4} \int \eta^{\alpha\beta}
( \delta_\alpha^\gamma \delta_\mu^\delta - \delta_\alpha^\delta \delta_\mu^\gamma 
-\im \epsilon_{\alpha\mu}^{\quad\gamma\delta})
( \delta_\beta^\rho \delta_\nu^\sigma - \delta_\beta^\sigma \delta_\nu^\rho 
-\im \epsilon_{\beta\nu}^{\quad\rho\sigma}) \partial_\rho h_\sigma^\mu \partial_\gamma h_\delta^\nu,
\end{align}
where we have used (\ref{ss-proj}) to get the second line. We can now contract the indices
and take into account the tracefree as well as the transverse condition on $h_{\mu\nu}$.
We get the following simple action as the result:
\be\label{S-grav}
S^{(2)} = -\frac{1}{2} \int \partial_\mu h_{\rho\sigma} \partial^\mu h^{\rho\sigma},
\ee
which is the correctly normalized transverse traceless graviton action. Note that in the
passage to GR we have secretly assumed that $h_{\mu\nu}$ in (\ref{b-h}) is a real
metric perturbation. Below we will see how to impose the reality conditions on our
theory that this comes out. Also note that the sign in front of (\ref{S-grav})
is correct for our choice of the signature being $(-,+,+,+)$.

\subsection{Hamiltonian analysis of the linearized theory}

For a finite $g$ our theory describes a deformation of GR. Since not all components
of the two-form perturbation $b^a_{\mu\nu}$ are dynamical, the nature
of this deformation is most clearly seen in the Hamiltonian framework. 
This is what this subsection is about. 

We note that the outcome of this rather technical subsection is that at "low" energies
$E^2\ll g$ the modification can be ignored and one can safely work with the
usual linearized GR. Thus, it may be advisable to skip this subsection on the
first reading. Let us start by analyzing the kinetic BF-part.

\bigskip
\noindent \textbf{Kinetic term.}
Expanding the product of two $\Sigma$-matrices in (\ref{S-bf-lin}) using (\ref{asub}) 
we can write the linearized Lagrangian density for the BF-part as
\begin{equation}
\mathcal{L}_{BF}= \frac{1}{4}
\,(\partial b^a)^\mu (\partial b^b)^\nu
\, \left( \epsilon^{abc}\,\Sigma^c_{0\,\mu\nu}+
\delta^{ab}\, \eta_{\mu\nu}  \right) \label{llbfsu2}\, .
\end{equation} 
Let us now perform the space plus time decomposition. Thus, we split the spacetime
index as $\mu=(0,a)$, where $a=1,2,3$. Note that we have denoted the spatial index
by the same lower case Latin letter from the beginning of the alphabet that we
are already using to denote the internal ${\mathfrak su}(2)$ index. This is
allowed since we can use spatial projection of the $\Sigma^a_{0\,\mu\nu}$ two-form
to provide such an identification. Thus, from (\ref{Sigma}) we have:
\be\label{sigm-ab}
\Sigma^a_{0\,bc} = -\epsilon^a_{\,\, bc} \, ,
\ee
and
\be\label{sigm-0a}
\Sigma^a_{0\, 0b}= \im \delta^a_b \, .
\ee

Let us now use these simple relations to obtain the space plus time decomposition
of the Lagrangian. First, we need to know components of the $(\partial b^a)^\mu$
vector. The time component is given by:
\be
(\partial b^a)^0=\epsilon^{0bcd} \partial_b b_{cd}^a = - \partial_b t^{ab},
\ee
where our conventions are $\epsilon^{0abc}=-\epsilon^{abc}$ and we have
introduced:
\be\label{t-su2}
t^{ab}:=\epsilon^{bcd} b_{cd}^a.
\ee
The spatial component of $(\partial b^a)^\mu$ is given by:
\be
(\partial b^a)^b = \epsilon^{b0cd}\partial_0 b_{cd}^a + 2\epsilon^{bc0d} \partial_c b_{0d}^a=
\partial_0 t^{ab} - 2\epsilon^{bcd} \partial_c b_{0d}^a.
\ee

Now, the Lagrangian (\ref{llbfsu2}) is given by:
\be
\mathcal{L}_{BF}= -\frac{1}{4}(\partial b^a)^0 (\partial b^a)^0 + \frac{1}{2}
(\partial b^a)^0 (\partial b^b)^d \epsilon^{abc}\Sigma^c_{0d} 
+ \frac{1}{4} (\partial b^a)^e (\partial b^b)^f (\epsilon^{abc}\Sigma^c_{ef}+\delta^{ab}\delta_{ef}).
\ee
Substituting the above expressions we get:
\begin{align}
\mathcal{L}_{BF}=& -\frac{1}{4} \partial_b t^{ab} \partial_c t^{ac} 
-\frac{\im}{2} \partial_d t^{ad} (\partial_0 t^{bc} - 2\epsilon^{cef} \partial_e b_{0f}^b) \epsilon^{abc}
\\ \nonumber
-&\frac{1}{4}(\partial_0 t^{ae} - 2\epsilon^{emn} \partial_m b_{0n}^a)
(\partial_0 t^{bf} - 2\epsilon^{fpq} \partial_p b_{0q}^b)
(\epsilon^{abc}\epsilon^c_{\,\,ef} - \delta^{ab}\delta_{ef}).
\end{align}

Our fields are now therefore $b^a_{0b}$ and $t^{ab}$. There will also be another, potential
part to this Lagrangian, but it does not contain time derivatives, so the conjugate momenta
can be determined already at this stage. Thus, it is clear that the field $b^a_{0b}$
is non-dynamical since the Lagrangian does not depend on its time derivatives. The
momentum conjugate to $t^{ab}$, on the other hand, is given by:
\be
\pi^{ab}:=\frac{\partial {\cal L}_{BF}}{\partial (\partial_0 t^{ab})} =
-\frac{\im}{2} \epsilon^{abc} \partial_d t^{cd} 
-\frac{1}{2}(\partial_0 t^{ef} - 2\epsilon^{fpq} \partial_p b_{0q}^e)
(\epsilon^{aec}\epsilon^{cbf} - \delta^{ae}\delta^{bf}).
\ee
It is not hard to check that the momentum variable is simply related to the spatial
projection of the connection (\ref{a-1}) as:
\be\label{pi-su2}
\pi^a_b = -2\im a^a_b.
\ee

To rewrite the Lagrangian in the Hamiltonian form one must solve for the velocities
$\partial_0 t^{ab}$ in terms of the momenta $\pi^{ab}$. However, it is clear that not
all the velocities can be solved for - there are constraints. A subset of these constraints 
is given by the $\mu=0$ component of the (\ref{su2-comp}) equation that, when written
in terms of $\pi^{ab}$, becomes:
\be\label{Gauss-su2}
{\cal G}^a:=\epsilon^{abc} \pi^{bc}+\im \partial_b t^{ab}=0.
\ee
These are primary constraints that must be added to the Hamiltonian with 
Lagrange multipliers. 

Thus, the expression for velocities in terms of momenta will contain undetermined
functions. These functions are simply the $a^a_0$ components of the connection,
as well as (at this stage undetermined) $b^a_{0b}$ components of the two-form
field. The expression for velocities is given by the spatial components of equation
(\ref{su2-comp}). After some algebra it gives:
\be
\partial_0 t^{ab}=2\epsilon^{bef} \partial_e b^a_{0f} - 2\epsilon^{abc} a_0^c 
- \epsilon^{aed}\epsilon^{dbf}\pi^{ef}.
\ee 

Let us now obtain a slightly more convenient expression for the Lagrangian. Indeed,
recall that using the compatibility equation between the connection and the two-form
perturbation, we could have chosen to write our linearized action (\ref{S-2-2}) as
\be\label{su2-aa}
S_{BF}^{(2)}=-2\im \int \epsilon^{abc} \Sigma^a_0 \wedge a^b\wedge a^c=
-2\int \Sigma^{a\,\mu\nu} \epsilon^{abc} a^b_\mu a^c_\nu.
\ee
Introducing the time plus space split and writing the result in terms of the momentum 
variable (\ref{pi-su2}) we get the following Lagrangian:
\be
{\cal L}_{BF}=-2\epsilon^{abc} \pi^{ab} a_0^c 
- \frac{1}{2} \epsilon^{aef} \epsilon^{abc} \pi^{be} \pi^{cf}.
\ee
We can now easily find the BF-part of the Hamiltonian:
\be
{\cal H}_{BF}=\pi^{ab}\partial_0 t^{ab} - {\cal L}_{BF}=
2\pi^{ab} \epsilon^{bef}\partial_e b^a_{0f} 
- \frac{1}{2} \epsilon^{aef} \epsilon^{abc} \pi^{be} \pi^{cf}.
\ee
We need to add to this the primary constraints (\ref{Gauss-su2}) with Lagrange
multipliers. Thus, the total Hamiltonian coming from the BF part of the
action is
\be
{\cal H}_{BF}^{total}=2\pi^{ab} \epsilon^{bef}\partial_e b^a_{0f} 
- \frac{1}{2} \epsilon^{aef} \epsilon^{abc} \pi^{be} \pi^{cf}+
\omega^a {\cal G}^a.
\ee
This is, of course, the standard result for the linearized BF Hamiltonian. If not for the potential
term, the Hamiltonian would be a sum of terms generating the topological
constraint $\partial_{[b} \pi^a_{c]}=0$ and the Gauss constraint (\ref{Gauss-su2}).
Let us now consider the other BB part of the Lagrangian.

\noindent \textbf{Potential part.}
We can rewrite the linearized Lagrangian density for the BB part (\ref{lapsu2}) as
\begin{equation}
\mathcal{L}_{BB}=-\frac{g}{2}\, 
\left( b^{(a}_{\mu\nu} \Sigma^{b)\mu\nu}_0 \right)_{\text{tf}}\, 
\left( b^{(a}_{\rho\sigma} \Sigma^{b)\rho\sigma}_0 \right)_{\text{tf}} \, ,
\end{equation}
where $tf$ stands for the tracefree parts of the matrices.
Splitting the space and time indices gives:
\be
\left( b^{(a}_{\mu\nu} \Sigma^{b)\mu\nu}_0 \right)_{\text{tf}}=-
\left(2\im b^{(ab)}_{\,\,0} + t^{(ab)}\right)_{tf}, 
\ee
and so
\begin{equation}
\mathcal{L}_{BB}=-\frac{g}{2}\, 
\left(2\im b^{(ab)}_{\,\,0} + t^{(ab)}\right)_{tf}\left(2\im b^{(ab)}_{\,\,0} + t^{(ab)}\right)_{tf}\, .
\end{equation}

\noindent \textbf{Analysis of the constraints.} Thus, the total linearized Hamiltonian density 
$\mathcal{H}=\mathcal{H}_{BF}^{total}-\mathcal{L}_{BB}$ is given by
\begin{align}\nonumber
\mathcal{H}= 2\pi^{ab} \epsilon^{bef}\partial_e b^{af}_{\,\,0} 
- \frac{1}{2} \epsilon^{aef} \epsilon^{abc} \pi^{be} \pi^{cf}+
\omega^a {\cal G}^a 
+\frac{g}{2}\, 
\left(2\im b^{(ab)}_{\,\,0} + t^{(ab)}\right)_{tf}\left(2\im b^{(ab)}_{\,\,0} + t^{(ab)}\right)_{tf}\, .
\end{align}
It is now clear that only the anti-symmetric part and trace parts of $b^{ab}_{\,\,0}$
remain Lagrange multipliers in the full theory. These are the generators of the diffeomorphisms.
The other part of $b^{ab}_{\,\,0}$, namely the symmetric traceless is clearly non-dynamical
and should be solved for from its field equations. Varying the Hamiltonian with respect
to this symmetric tracefree part we get
\be\label{btf-su2}
\left(2\im b^{(ab)}_{\,\,0} + t^{(ab)}\right)_{tf}=
\frac{\im}{g} \left(\epsilon^{ef(a}\partial_e \pi^{b)}_f\right)_{tf}.
\ee
Now writing:
\be\label{b0-su2}
b^{ab}_{\,\,0}=\im N \delta^{ab} + \frac{1}{2} \epsilon^{abc} N^c + (b^{(ab)}_{\,\,0})_{tf}
\ee
and substituting the symmetric tracefree part from (\ref{btf-su2}) we get the following 
Hamiltonian
\begin{align}\label{Ham-su2}
{\cal H}&=-2N\im\epsilon^{abc}\partial_a \pi_{bc} - 2 \partial_{[a} \pi^a_{b]} N^b
+ \omega^a {\cal G}^a \\ \nonumber
&- \frac{1}{2} \epsilon^{aef} \epsilon^{abc} \pi^{be} \pi^{cf}
+ \im \left(\epsilon^{ef(a}\partial_e \pi^{b)}_f\right)_{tf} (t^{(ab)})_{tf}
+\frac{1}{2g} \left(\epsilon^{ef(a}\partial_e \pi^{b)}_f\right)_{tf}
\left(\epsilon^{pq(a}\partial_p \pi^{b)}_q\right)_{tf} .
\end{align}
The reason why we introduced a factor of $\im$ in front of the lapse function will become clear 
below. One can recognize in the first line the usual Hamiltonian, diffeomorphism
and Gauss linearized constraints of Ashtekar's Hamiltonian formulation of
general relativity \cite{Ashtekar:1987gu}. The first two terms in the second line comprise the
Hamiltonian. Finally, the last term is due to the modification and goes away
in the limit $g\to\infty$. 

It is not hard to show that the reduced phase space for the above system is obtained
by considering $\pi^{ab}, t^{ab}$ that are symmetric, traceless and transverse
$\partial_a \pi^{ab}=0, \partial_a t^{ab}=0$. On such configurations the matrix
$\epsilon^{efa}\partial^e \pi^{fb}$ is automatically symmetric traceless and transverse.
The reduced phase space Hamiltonian density is then given by:
\be\label{H-phys-su2}
{\cal H}^{phys}=\frac{1}{2}(\pi^{ab})^2 +\im \epsilon^{efa}\partial^e t^{fb} \pi^{ab}+
\frac{1}{2g}(\partial^a \pi^{bc})^2,
\ee
where we have integrated by parts and put the derivative on $t^{ab}$ in the second term.
This Hamiltonian is complex, so we need to discuss the reality conditions.

\noindent {\bf Reality conditions.}
So far our discussion was in terms of complex-valued fields. Thus, the reduced phase space
obtained above after imposing the constraints and quotienting by their action was
complex dimension $2+2$. Reality conditions need to be imposed to select the
physical phase space corresponding to Lorentzian signature gravity. 

In the case of GR that corresponds to $g\to\infty$ the reality condition could be guessed
from the form of the Hamiltonian (\ref{H-phys-su2}). Indeed, we can write it as:
\be
{\cal H}^{phys}_{GR}=\frac{1}{2}\left( \pi^{ab} +\im \epsilon^{efa} \partial^e t^{fb}\right)^2
+\frac{1}{2} (\partial^a t^{bc})^2.
\ee
Thus, it is clear that we just need to require $t^{ab}$ and 
$\pi^{ab}+\im\epsilon^{efa} \partial^e t^{fb}$ to be real. This procedure, however,
does not work for the full Hamiltonian because of the last term in (\ref{H-phys-su2}).

Let us now note that the last term in (\ref{H-phys-su2}), when written in momentum space
behaves as $E^2/M^2$, where $E$ is the energy and $M^2=g$ is the modification
parameter. Thus, for energies $E\ll M$ the modification term is much smaller than the term
$\pi^2$ and can be dropped. It is natural to expect that gravity is only modified close
to the Planck scale, so it is natural to expect $M^2\approx M^2_p$, where $M_p$ is
the Planck mass. With this assumption the last term in (\ref{H-phys-su2}) is unimportant
for "ordinary" energies and can be dropped. Thus, if we are to work at energies
much smaller than the Planck scales ones then we do not need to go beyond GR
described by the first two terms in (\ref{H-phys-su2}). 

The above discussion shows that a discussion of the reality conditions for the
full Hamiltonian (\ref{H-phys-su2}), even though possible and necessary if
one is interested in the behavior of the theory close to the Planck scale, is
not needed if one only wants to work for with much smaller energies. For
this reason, and in order not to distract the reader from the main line of the argument,
a somewhat technical reality conditions discussion for the full theory is placed in the Appendix.

Now that we understood how the simplest case $G={\rm SU}(2)$ gives rise
to gravity we can apply the same procedure to more interesting cases of
a larger gauge group. We consider the example of ${\rm SU}(3)$
that well illustrates the general pattern.

\section{The $G={\rm SU}(3)$ Case: Gravity-Maxwell system}
\label{sec:su3}

In this section we perform an analysis analogous to that in the previous section but
taking a larger gauge group. As before, we first consider the complex theory, and
only at the end impose the reality conditions. Let us start by reviewing some basic
facts about the ${\mathfrak su}(3)$ Lie algebra.

\subsection{Lie algebra of ${\rm SU}(3)$}

The standard  matrix representation of the Lie algebra of ${\rm SU}(3)$ consist of all 
traceless anti-hermitian 3 x 3 complex matrices. The standard basis for 
${\mathfrak su}(3)$ space is given by the imaginary unit times a generalization of Pauli 
matrices, known as Gell-Mann matrices. These hermitian matrices are given by:
\begin{align}
\lambda_1=&
\begin{pmatrix}
0 & 1 & 0 \\
1 & 0 & 0 \\
0 & 0 & 0 \\
\end{pmatrix} \, , &
\lambda_2=&
\begin{pmatrix}
0 & -\text{i} & 0 \\
\text{i} & 0 & 0 \\
0 & 0 & 0 \\
\end{pmatrix} \, , &
\lambda_3=&
\begin{pmatrix}
1 & 0 & 0 \\
0 & -1 & 0 \\
0 & 0 & 0 \\
\end{pmatrix} \, , \notag \\
\lambda_4=&
\begin{pmatrix}
0 & 0 & 1 \\
0 & 0 & 0 \\
1 & 0 & 0 \\
\end{pmatrix} \, , &
\lambda_5=&
\begin{pmatrix}
0 & 0 & -\text{i} \\
0 & 0 & 0 \\
\text{i} & 0 & 0 \\
\end{pmatrix} \, , &
\lambda_6=&
\begin{pmatrix}
0 & 0 & 0 \\
0 & 0 & 1 \\
0 & 1 & 0 \\
\end{pmatrix} \, , \notag \\
\lambda_7=&
\begin{pmatrix}
0 & 0 & 0 \\
0 & 0 & -\text{i} \\
0 & \text{i} & 0 \\
\end{pmatrix} \, , &
\lambda_8=\frac{1}{\sqrt{3}} &
\begin{pmatrix}
1 & 0 & 0 \\
0 & 1 & 0 \\
0 & 0 & -2 \\
\end{pmatrix} \, .
\end{align}

However, in our computations the Cartan-Weyl basis is going to be more convenient. Let us
recall that in the Cartan-Weyl formalism one starts with the maximally commuting
Cartan subalgebra, which in our case is spanned by two elements $\lambda_3,\lambda_8$.
One then selects basis vectors that are eigenstates of the elements of Cartan under 
the adjoint action. This leads to the following basis, see \cite{geor},\cite{cahn}
\begin{align}
T_{\pm}=&\frac{1}{\sqrt{2}} ( T_x \pm \text{i}\,T_y) & V_{\pm}=&\frac{1}{\sqrt{2}} 
( V_x \pm \text{i}\,V_y) & W_{\pm}=&\frac{1}{\sqrt{2}} ( W_x \pm \text{i}\,W_y) \notag \\
T_z=& \frac{1}{2} \lambda_3 & Y=& \frac{1}{2} \lambda_8 \, , \label{basis-su3}
\end{align}
where $T_x=\frac{1}{2} \lambda_1$, $T_y=\frac{1}{2} \lambda_2$, $V_x=\frac{1}{2} \lambda_4$, 
$V_y=\frac{1}{2} \lambda_5$, $W_x=\frac{1}{2} \lambda_6$ and  $W_y=\frac{1}{2} \lambda_7$. 
Then the Cartan subalgebra is $H_i={\rm Span}(T_z,Y)$,  and the commutator between any 
of the $H_i$'s and the rest of the elements of the basis $E_{\alpha}$, 
$E_{\alpha}=\{  T_{+}, T_{-}, T_z, V_{+}, V_{-}, W_{+}, W_{-} \}$, is a multiple of 
$E_{\alpha}$, i.e. $[H_i,E_{\alpha}]=\alpha_i\, E_{\alpha}$. One considers the 
$\alpha_i$'s, for $i=1,2$, as the components of a vector, called a root of the system. 
In this case we have six roots, i.e. $\{1,0\}$, $\{-1,0\}$, $\{\frac{1}{2},\frac{\sqrt{3}}{2}\}$, 
$\{-\frac{1}{2},-\frac{\sqrt{3}}{2}\}$, $\{-\frac{1}{2},\frac{\sqrt{3}}{2}\}$, 
$\{\frac{1}{2},-\frac{\sqrt{3}}{2}\}$. The Lie brackets between elements of this basis
are given in Table \ref{commutators}. We also need to know the 
metric $g_{IJ}=-2{\rm Tr}(T_I T_J)$ in this basis. It is given in 
Table {\ref{metric}}.

\begin{table}
\centering{\footnotesize
\renewcommand{\tabcolsep}{1pt}
\renewcommand{\arraystretch}{1.7}
\begin{tabular}{|c|c|c|c|c|c|c|c|c|}\hline
$[ \downarrow, \to ]$  & $T_{+}$ & $T_{-}$ & $T_z$ & $V_{+}$ & $V_{-}$ & $W_{+}$ & $W_{-}$ & $Y$ \\ \hline
$T_{+}$ & 0 & $T_z$ & $-T_{+}$ & 0 & $-\frac{1}{\sqrt{2}}W_{-}$ & $\frac{1}{\sqrt{2}}V_{+}$ & 0 & 0 \\ \hline
$T_{-}$ & $-T_z$ & 0 & $T_{-}$ & $\frac{1}{\sqrt{2}}W_{+}$ & 0 & 0 & $-\frac{1}{\sqrt{2}}V_{-}$ & 0 \\ \hline
$T_{z} $ & $T_{+}$ & $-T_{-}$ & 0 & $\frac{1}{2}V_{+}$ & $-\frac{1}{2}V_{-}$ & $-\frac{1}{2}W_{+}$ & $\frac{1}{2}W_{-}$ & 0 \\ \hline
$V_{+}$ & 0 & $-\frac{1}{\sqrt{2}}W_{+}$ & $-\frac{1}{2}V_{+}$ & 0 & $\frac{1}{2} (\sqrt{3} Y+T_z)$ & 0 & $\frac{1}{\sqrt{2}}T_+$ & $-\frac{\sqrt{3}}{2}V_{+}$    \\ \hline
$V_-$ & $\frac{1}{\sqrt{2}}W_{-}$ & 0 & $\frac{1}{2}V_{-}$ & $-\frac{1}{2}(\sqrt{3} Y+T_z)$ & 0 & $-\frac{1}{\sqrt{2}}T_-$ & 0 & $\frac{\sqrt{3}}{2}V_{-}$  \\ \hline
$W_{+}$ & $-\frac{1}{\sqrt{2}}V_{+}$ & 0 & $\frac{1}{2}W_{+}$ & 0 & $\frac{1}{\sqrt{2}}T_-$ & 0 & $\frac{1}{2}(\sqrt{3} Y - T_z)$ & $-\frac{\sqrt{3}}{2}W_{+}$  \\ \hline
$W_{-}$ & 0 & $\frac{1}{\sqrt{2}}V_{-}$  & $-\frac{1}{2}W_{-}$  & $-\frac{1}{\sqrt{2}}T_+$  & 0 & $-\frac{1}{2}(\sqrt{3} Y - T_z)$  & 0 & $\frac{\sqrt{3}}{2}W_{-}$  \\ \hline
$Y$ & 0 & 0 & 0 & $\frac{\sqrt{3}}{2}V_{+}$ & $-\frac{\sqrt{3}}{2}V_{-}$ & $\frac{\sqrt{3}}{2}W_{+}$ & $-\frac{\sqrt{3}}{2}W_{-}$ & 0 \\ \hline
\end{tabular}}
\caption{Commutators between $T_{+}, T_{-}, T_z, V_{+}, V_{-}, W_{+}, W_{-}, Y$.}\label{commutators}
\end{table}
\begin{table} 
\centering{\small
\renewcommand{\arraystretch}{1.5}
\begin{tabular}{|c|c|c|c|c|c|c|c|c|}\hline
$\langle \downarrow | \to \rangle$  & $T_{+}$ & $T_{-}$ & $T_z$ & $V_{+}$ & $V_{-}$ & $W_{+}$ & $W_{-}$ & $Y$ \\ \hline
$T_{+}$ & 0 & $-1$ & 0 & 0 & 0 & 0 & 0 & 0 \\ \hline
$T_{-}$ & $-1$ & 0 & 0 & 0 & 0 & 0 & 0 & 0 \\ \hline
$T_{z} $ & 0 & 0 & $-1$ & 0 & 0 & 0 & 0 & 0 \\ \hline
$V_{+}$ & 0 & 0 & 0 & 0 & $-1$ & 0 & 0 & 0    \\ \hline
$V_-$ & 0 & 0 & 0 & $-1$ & 0 & 0 & 0 & 0  \\ \hline
$W_{+}$ & 0 & 0 & 0 & 0 & 0 & 0 & $-1$ & 0  \\ \hline
$W_{-}$ & 0 & 0  & 0 & 0  & 0 & $-1$  & 0 & 0  \\ \hline
$Y$ & 0 & 0 & 0 & 0 & 0 & 0 & 0 & $-1$ \\ \hline
\end{tabular}}
\caption{Components for the internal metric in the base $\{T_{+}, T_{-}, T_z, V_{+}, V_{-}, W_{+}, W_{-}, Y\}$.}\label{metric}
\end{table}

\subsection{Background}

Let us now discuss how a background to expand around can be chosen. A background
two-form field $B_0^I$ is a map from the space of bivectors, which is 6-dimensional,
to the Lie algebra in question. Thus, its image is at most 6-dimensional subspace
in ${\mathfrak su}(3)$. There are many different subspaces one can consider. 
In this paper we study the simplest possibility. Thus, we choose $B_0^I$ such
that the image of the space of 2-forms that it produces in ${\mathfrak su}(3)$
is 3-dimensional. Moreover, we choose this image to be an ${\mathfrak su}(2)$
Lie sub-algebra. Even further, we choose this sub-algebra to be that spanned
by $\{ T_+, T_-, T_z\}$. Clearly, this is not the only ${\mathfrak su}(2)$ sub-algebra in
${\mathfrak su}(3)$. Other possibilities include $\{ V_+, V_-, \frac{1}{2}\left(\sqrt{3}Y + T_z\right)\}$ 
and $\{ W_+, W_-, \frac{1}{2}\left(\sqrt{3}Y - T_z\right)\}$. In this paper we do not study
these different possibilities, leaving a more thorough investigation to further
research. We believe that the example we choose to study is sufficiently 
illustrating. 

Thus, our background is essentially the same as the one we considered in the
previous section. This is motivated by our desire to have the usual gravity
theory arising as the part of the larger theory we are now considering. Since
in the general Lie algebra context it is convenient to work with the Cartan-Weyl
basis, we need to change the basis of basic two-forms (\ref{s}) as well. This can be
worked out as follows. In the previous section we were using a basis in the
Lie algebra in which the structure constants were given by $\epsilon_{abc}$.
If we denote the corresponding generators by $X_a$ then $[X_a,X_b]=\epsilon_{abc}X_c$.
On the other hand, for generators $T_a$ used in (\ref{basis-su3}) we have
$[T_a,T_b]=\im \epsilon_{abc} T_c$. The relation between these two bases is
$X_a=-\im T_a$. We can then define a new set of self-dual two-forms $\Sigma^\pm, \Sigma^z$
via:
\be
\Sigma\equiv \sum_{a=1,2,3} \Sigma^a X_a = \Sigma^+ T_+ + \Sigma^- T_- + \Sigma^z T_z.
\ee
This gives
\begin{align}
\Sigma^+&=\frac{-\im}{\sqrt{2}}\left( \Sigma^1 - \text{i}\, \Sigma^2 \right) & 
\Sigma^-&=\frac{-\im}{\sqrt{2}}\left( \Sigma^1 + \text{i}\, \Sigma^2 \right) & 
\Sigma^z &=-\im\, \Sigma^3 \label{spm} \, . 
\end{align} 
The ${\mathfrak su}(3)$-valued two-form $\Sigma$ is our background to expand about. 

\subsection{Linearization: Kinetic term}

As before, the first step of the linearization procedure is to solve for those components
of the connection for which this is possible. As we have discussed in section \ref{sec:lin},
this is in general possible for the components of the connection in the directions in
the Lie algebra that do not commute with the directions spanned by the background
two-forms. In our case these are the directions spanned by $T_\pm,T_z$ and 
$V_\pm, W_\pm$. We already know how to solve for the connection components
in the directions $T_\pm,T_z$. Indeed, the solution is given by (\ref{a-1}) which 
we just have to rewrite in the different basis. It is, however, more practical
to solve the equations once more by working in the different basis from the
very beginning. 

\noindent {\bf The ${\mathfrak su}(2)$ part.}
The ${\mathfrak su}(2)$ sector equations in the
Cartan-Weyl basis are:
\be\nonumber
db^+ + a^z \wedge \Sigma^+ - a^+\wedge \Sigma^z = 0, \\
db^- + a^-\wedge \Sigma^z - a^z\wedge \Sigma^-=0, \\ \nonumber
db^z + a^+\wedge \Sigma^- - a^-\wedge \Sigma^+=0.
\ee
We rewrite them in spacetime notations, take the Hodge dual, and use the self-duality
of the $\Sigma^\pm, \Sigma^z$ matrices to get:
\be\nonumber
\frac{1}{2\im} (\partial b^+)^\mu + a^z_\nu \Sigma^{+\,\mu\nu} - a^+_\nu \Sigma^{z\,\mu\nu}=0, \\
\label{comp-su2-part}
\frac{1}{2\im} (\partial b^-)^\mu + a^-_\nu \Sigma^{z\,\mu\nu} - a^z_\nu \Sigma^{-\,\mu\nu}=0, \\
\nonumber
\frac{1}{2\im} (\partial b^z)^\mu + a^+_\nu \Sigma^{-\,\mu\nu} - a^-_\nu \Sigma^{+\,\mu\nu}=0,
\ee
where the notation is, as before 
$(\partial b)^\mu = \epsilon^{\mu\nu\rho\sigma}\partial_\nu b_{\rho\sigma}$.
We now need the algebra of the new $\Sigma$-matrices. It can be worked out from 
the relations (\ref{spm}) and the algebra (\ref{asub}). We get:
\begin{align}
\Sigma^+_{\mu\sigma}\, \Sigma^{-\sigma}_{\;\;\: \nu}=\,\,&\eta_{\mu\nu} +\Sigma^z_{\mu\nu} \, , & 
\Sigma^z_{\mu\sigma}\, \Sigma^{+\sigma}_{\;\;\: \nu}&=\Sigma^+_{\mu\nu} \, , &
\Sigma^z_{\mu\sigma}\, \Sigma^{-\sigma}_{\;\;\: \nu}&=-\Sigma^-_{\mu\nu} \, , \notag \\
\Sigma^z_{\mu\sigma}\, \Sigma^{z\sigma}_{\;\;\: \nu}&= \eta_{\mu\nu}\, , &
\Sigma^+_{\mu\sigma}\, \Sigma^{+\sigma}_{\;\;\: \nu}&=0 \, , &
\Sigma^-_{\mu\sigma}\, \Sigma^{-\sigma}_{\;\;\: \nu}&=0 \label{aspmz}\, .
\end{align} 
For purposes of
the calculation it is very convenient to rewrite these relations in the schematic
form, by viewing them as matrix algebra. Our matrix multiplication convention for the two-forms 
is $(X Y)_\mu^{\,\,\,\,\nu} = X_{\mu}^{\,\,\rho} Y_\rho^{\,\,\nu}$. We have:
\begin{align}
\Sigma^+ \Sigma^- =\,\, & \eta + \Sigma^z \, , &
\Sigma^z \Sigma^{+} &=\Sigma^+\, , &
\Sigma^z \Sigma^{-} &=-\Sigma^-\, , \notag \\
\Sigma^z \Sigma^{z} = \,\, &\eta \, , &
\Sigma^+ \Sigma^{+} &=0 \, , &
\Sigma^-\Sigma^{-} &=0 \label{su2-rels} \, .
\end{align} 
This is precisely the relations (\ref{asub}), just written in terms of metric and the 
structure constants on ${\mathfrak su}(2)$ for a different basis.

In matrix product conventions, the equations (\ref{comp-su2-part}) take the 
following transparent form:
\be\nonumber
\frac{1}{2\im} (\partial b^+) + \Sigma^+ a^z - \Sigma^z a^+=0,
\\ \label{comp-su2*}
\frac{1}{2\im} (\partial b^-) + \Sigma^z a^- - \Sigma^- a^z=0,
\\  \nonumber
\frac{1}{2\im} (\partial b^z) + \Sigma^- a^+ - \Sigma^+ a^-=0,
\ee
where the convention is that the second spacetime index of $\Sigma$ is contracted with the 
spacetime index of $a$.

We can now solve (\ref{comp-su2*}) by using the algebra (\ref{su2-rels}).
To this end we multiply the first equation by $\Sigma^+$ and the second one
by $\Sigma^-$. This leads to two equations involving only $a^\pm$ but not $a^z$.
We can obtain another two equations of the same sort by multiplying the last
equation in (\ref{comp-su2-part}) by $\Sigma^\pm$. Then adding-subtracting
the resulting equations we get:
\be\label{apm-su2}
a^+ = -\frac{1}{4\im} \left( \Sigma^- \Sigma^+ (\partial b^+) + \Sigma^+ (\partial b^z) \right),
\qquad
a^- = -\frac{1}{4\im} \left( \Sigma^+ \Sigma^- (\partial b^-) - \Sigma^- (\partial b^z) \right).
\ee

To obtain the last component of the connection we multiply the first equation in
(\ref{comp-su2*}) by $\Sigma^-$ and second by $\Sigma^+$, and then subtract
the resulting equations. We find 
$\Sigma^- a^+-\Sigma^+ a^-=-(1/2\im) (\partial b^z)$ using (\ref{apm-su2}). We get:
\be
a^z=-\frac{1}{4\im} \left( (\partial b^z) + \Sigma^- (\partial b^+)-\Sigma^+ (\partial b^-)\right).
\ee

It is now easy to write the ${\mathfrak su}(2)$ part of the linearized BF part of the action. 
Using the metric components given in Table \ref{metric}, from (\ref{S-2-2}) we have:
\be
S^{{\mathfrak su}(2)}_{BF} = -\frac{1}{4}\int (\partial b^+)
\left( \Sigma^+ \Sigma^- (\partial b^-) - \Sigma^- (\partial b^z) \right)+
(\partial b^-)\left( \Sigma^- \Sigma^+ (\partial b^+) + \Sigma^+ (\partial b^z) \right)
\\ \nonumber 
+(\partial b^z) \left( (\partial b^z) + \Sigma^- (\partial b^+)-\Sigma^+ (\partial b^-)\right),
\ee
where again our convenient schematic form of the notation is used. This is simplified
to give:
\be
S^{{\mathfrak su}(2)}_{BF} =- \frac{1}{2}\int (\partial b^+)(\eta+\Sigma^z)(\partial b^-)
+(\partial b^-)\Sigma^+ (\partial b^z) - (\partial b^+)\Sigma^- (\partial b^z)
+\frac{1}{2} (\partial b^z)(\partial b^z) .
\ee
We could now use this as the starting point of the Hamiltonian analysis similar to 
the one in the previous section. However, it is clear that its results are basis-independent,
so we do not need to repeat it. Still, the above considerations are quite useful
as a warm-up for the more involved analysis that now follows.

\noindent {\bf The part that does not commute with ${\mathfrak su}(2)$.}
Let us denote the four directions $V_\pm,W_\pm$ collectively by index 
$\alpha=4,5,6,7$. We have to solve the following system of equations:
\begin{equation}
db^\alpha+  f^\alpha_{\;\,\beta a}\, a^\beta 
\wedge \Sigma^a =0 \, ,
\end{equation}
where the terms $f^\alpha_{\;\,ab}\, a^a \wedge \Sigma^b$
are absent since the corresponding structure constants are zero.
Explicitly, using table \ref{commutators} we have:
\begin{align}
db^4 - \frac{1}{\sqrt{2}}\, a^6 \wedge \Sigma^+ -\frac{1}{2}\, a^4 \wedge \Sigma^z &=0 \, , \\
db^5 + \frac{1}{\sqrt{2}}\, a^7 \wedge \Sigma^- + \frac{1}{2}\, a^5 \wedge \Sigma^z &=0 \, ,  \\
db^6 - \frac{1}{\sqrt{2}}\, a^4 \wedge \Sigma^- + \frac{1}{2}\, a^6 \wedge \Sigma^z &=0 \, , \\
db^7 + \frac{1}{\sqrt{2}}\, a^5 \wedge \Sigma^+ - \frac{1}{2}\, a^7 \wedge \Sigma^z &=0 \, .
\end{align}

We can solve this system using the same technology that we used above for the
${\mathfrak su}(2)$ sector. Thus, we take the Hodge dual of the above equations, use
the self-duality of the $\Sigma$'s, and rewrite everything in the schematic matrix form.
We get:
\begin{align}\nonumber
\frac{1}{2\im} (\partial b^4)- \frac{1}{\sqrt{2}}\, \Sigma^+ a^6  
-\frac{1}{2}\, \Sigma^z a^4 &=0 \, , \\ \nonumber
\frac{1}{2\im} (\partial b^5) + \frac{1}{\sqrt{2}}\, \Sigma^- a^7 
+ \frac{1}{2}\, \Sigma^z a^5  &=0 \, ,  \\ \label{higgs-compat}
\frac{1}{2\im} (\partial b^6) - \frac{1}{\sqrt{2}}\, \Sigma^- a^4 
+ \frac{1}{2}\,  \Sigma^z a^6 &=0 \, , \\ \nonumber
\frac{1}{2\im} (\partial b^7) + \frac{1}{\sqrt{2}}\, \Sigma^+ a^5 
- \frac{1}{2}\, \Sigma^z a^7 &=0 \, .
\end{align}

We can now manipulate these equations using the algebra (\ref{su2-rels}). Thus, let us
multiply the third equation by $\sqrt{2}\Sigma^+$ and subtract the result from the
first equation. This gives:
\be
\frac{1}{2\im} (\partial b^4)-\frac{\sqrt{2}}{2\im} \Sigma^+ (\partial b^6)
+ (\eta+\frac{1}{2}\Sigma^z) a^4=0.
\ee
It is now easy to find $a^4$ by noting that $(\eta+(1/2)\Sigma^z)^{-1}=(4/3)(\eta-(1/2)\Sigma^z)$.
Thus, we have:
\be\label{a4}
a^4 = \frac{1}{3i}\left( \sqrt{2} \Sigma^+ (\partial b^6) - (2\eta - \Sigma^z)(\partial b^4)\right).
\ee

Similarly, we multiply the last equation by $\sqrt{2}\Sigma^-$ and add it to the second equation.
Multiplying then by the inverse of $(\eta-(1/2)\Sigma^z)$ we get:
\be\label{a5}
a^5 = -\frac{1}{3i}\left( \sqrt{2} \Sigma^- (\partial b^7) + (2\eta + \Sigma^z)(\partial b^5)\right).
\ee

To find $a^6$ we multiply the first equation by $\sqrt{2}\Sigma^-$ and subtract the result
from the third equation. We then multiply the result by the inverse of $(\eta-(1/2)\Sigma^z)$.
We get:
\be\label{a6}
a^6 = \frac{1}{3i}\left( \sqrt{2} \Sigma^- (\partial b^4) - (2\eta + \Sigma^z)(\partial b^6)\right).
\ee

Finally, to find $a^7$ we multiply the second equation by $\sqrt{2}\Sigma^+$ and add
the result to the last equation. Multiplying the result by the inverse of $(\eta+(1/2)\Sigma^z)$
we get:
\be\label{a7}
a^7 = -\frac{1}{3i}\left( \sqrt{2} \Sigma^+ (\partial b^5) + (2\eta - \Sigma^z)(\partial b^7)\right).
\ee

We should now substitute the above results into the relevant part of the action. This
is again obtained from (\ref{S-2-2}) by taking into account the expression for the metric.
We shall refer to this part of the action as "Higgs" in view of its interpretation to be
developed later. We have:
\be
S^{Higgs}_{BF}= \im\int a^4 (\partial b^5) + a^5 (\partial b^4)+a^6 (\partial b^7)
+a^7 (\partial b^6),
\ee
where we took into account and extra minus sign that comes from the metric. 
Substituting here the above connections, we get, after some simple algebra:
\begin{align}\label{S-bf-Higgs}
S^{Higgs}_{BF}=&\frac{2}{3}\int \sqrt{2} (\partial b^5)\Sigma^+(\partial b^6)
-\sqrt{2} (\partial b^4)\Sigma^-(\partial b^7)
\\ \nonumber
-&(\partial b^4)(2\eta+\Sigma^z)(\partial b^5)
-(\partial b^6)(2\eta-\Sigma^z)(\partial b^7).
\end{align}
A more illuminating way to write this action is by introducing two two-component
fields:
\be
\left(\begin{array}{c} b^4 \\ b^6 \end{array}\right) \qquad
\left(\begin{array}{c} b^5 \\ b^7 \end{array}\right)\, .
\ee
It is not hard to see that this split of the "Higgs" sector part of the Lie algebra is just the split
into two irreducible representation spaces with respect to the action of the gravitational
${\rm SU}(2)$. In terms of these columns the above action takes the following form:
\be\label{S-bf-Higgs*}
S^{Higgs}_{BF}=\frac{2}{3} \int \left( (\partial b^5) \, (\partial b^7)\right) 
\left( \begin{array}{cc} -2\eta +\Sigma^z & \sqrt{2} \Sigma^+ \\ 
\sqrt{2} \Sigma^-  & -2\eta -\Sigma^z \end{array}\right) 
\left(\begin{array}{c} (\partial b^4) \\ (\partial b^6) \end{array}\right)\, .
\ee
Below we will use this action as the starting point for an analysis that will eventually
exhibit the physical DOF propagating in this sector. 

\noindent{\bf Centralizer ${\rm U}(1)$ part.} We cannot solve for the 
components of the connection in the part
that commutes with ${\mathfrak su}(2)$. In our case this is the direction $Y$ of the Lie
algebra. We shall refer to this part of the action as "YM". 
Thus, the action remains of BF type:
\be\label{S-bf-YM}
S^{YM}_{BF}=-4\im \int b^8 \wedge da^8,
\ee
where the extra minus sign is the one in the metric. 

\subsection{Linearization: Potential term}

As in the ${\rm SU}(2)$ case our background internal metric $\tilde{h}^{IJ}_0$ is just
$2\im g^{ab}$ in the ${\mathfrak su}(2)$ directions and zero in all other directions. 
Since the background metric is flat we shall drop the tilde from $\tilde{h}^{IJ}$ in this section.
We compute the matrix of first derivatives of the potential using (\ref{fdVG}). We get:
\begin{align}
\left. \frac{\partial V}{\partial h^{ab}} \right|_0=&\frac{f_0}{8}\,g_{ab} ,\\
\left. \frac{\partial V}{\partial h^{a\alpha}} \right|_0=& 0  \, ,\\
\left. \frac{\partial V}{\partial h^{\alpha\beta}} \right|_0=& 
\left( \frac{f_0}{8}-\frac{1}{8}\sum_{p=2}^6 (f'_p)_0 \, \frac{p}{3^{p-1}}  \right) g_{\alpha\beta}\, .
\end{align}
Here $f_0, (f'_p)_0$ are the value of the function and its derivatives at the
background, and index $\alpha$ stands for all directions in the Lie algebra
that are not in ${\mathfrak su}(2)$. The quantity $f_0$ can be identified with a
multiple of the cosmological constant. More specifically:
\be
\Lambda = -\frac{3f_0}{8}.
\ee
Let us also define another constant of dimensions $1/L^2$:
\be\label{kappa}
\kappa \equiv \frac{1}{8} \sum_{p=2}^6 (f'_p)_0 \, \frac{p}{3^{p-1}}.
\ee
Then we have:
\be
\left. \frac{\partial V}{\partial h^{\alpha\beta}} \right|_0 = - (\Lambda/3+\kappa) g_{\alpha\beta}\, .
\ee
The sum here and in the previous formula is
taken over $p=2,\ldots,6$, because the function $f$ can at most depend
on 5 ratios of 6 invariants of the matrix $h^{IJ}$. It has at most only 6 independent
invariants since it is constructed from the map $B^I_{\mu\nu}$ that has the rank at
most six. Since we want to work with the Minkowski spacetime background
we should set $\Lambda=0$, which we do in what follows.

We now need to compute the matrix of second derivatives. Let us first obtain its
${\mathfrak su}(2)$ part. Using (\ref{sdVG}) we get:
\be
\frac{\partial^2 V}{\partial h^{cd}\partial h^{ab}}=\frac{g}{2\im} \, 
\left( g_{a(c} g_{d)b}-\frac{1}{3} g_{ab} g_{cd} \right) \, ,
\ee
where we have defined:
\be
g=\frac{1}{8} \sum_{p=2}^6 (f'_p)_0 \, \frac{p (p-1)}{3^{p-1}}.
\ee
As in the ${\rm SU}(2)$ case this constant is going to measure strength of gravity modifications.
Both $\kappa$ and $g$ constants have dimensions of $1/L^2$ and are, in general, 
independent parameters of our linearized theory, related to first
derivatives $(f'_p)_0$ of the function $f$ of the ratios.

Let us now compute the matrix of second derivatives in its part not in ${\mathfrak su}(2)$.
We only need its mixed components $a\alpha$ and $b\beta$. The computation is easy 
and using (\ref{sdVG}) we get:
\be\label{g}
\left. \frac{\partial^2 V}{\partial h^{a\alpha}\partial h^{b\beta}}\right|_0=\frac{\kappa}{4\im} 
\, g_{ab} g_{\alpha\beta}.
\ee
We note that in this computation only one of the terms in (\ref{h-der-h}) survives,
and this is the reason why it is the constant $\kappa$ that appears in this formula.

We can now compute all the potential parts. We use (\ref{bb-lin}) which we have
to divide by two to get the correct quadratic action. For the ${\mathfrak su}(2)$ 
gravitational part the result is unchanged from that in the previous section and we have:
\be
S^{grav}_{BB}=-\frac{g}{2} \int 
\left( g_{a(c} g_{d)b}-\frac{1}{3}g_{ab}g_{cd} \right) 
\, \left( \Sigma^{a\,\mu\nu}_0 b^b_{\mu\nu} \right)\, 
\left( \Sigma^{c\,\rho\sigma}_0 b^d_{\rho\sigma} \right) .
\ee

The "Higgs" and "YM" parts of the potential term are both given by:
\be\label{pot-higgs-ym}
S^{Higgs-YM}_{BB}=-\frac{\kappa }{4}
\int g_{ab}g_{\alpha\beta} (\Sigma^{a\,\mu\nu} b^\alpha_{\mu\nu})
(\Sigma^{b\,\rho\sigma} b^\beta_{\rho\sigma}) + 2\im
g_{\alpha\beta} \epsilon^{\mu\nu\rho\sigma} b^\alpha_{\mu\nu} b^\beta_{\rho\sigma},
\ee
so the indices $\alpha,\beta$ here take values $4,5,6,7,8$. We can further simplify this
using (\ref{ss-proj}). We get:
\be\label{S-bb-higgs}
S^{Higgs-YM}_{BB}=-\kappa \int g_{\alpha\beta} b^{\alpha\,\mu\nu}
b^{\beta\,\rho\sigma}P^-_{\mu\nu\rho\sigma}.
\ee
where
\be
P^-=\frac{1}{2}\left(\eta_{\mu[\rho}\eta_{\sigma]\nu}+\frac{\im}{2}\epsilon_{\mu\nu\rho\sigma}
\right)
\ee
is the anti-self-dual projector. 

\subsection{Symmetries}

We have seen that the ${\mathfrak su}(2)$ sector of the theory is completely unchanged
from what we have obtained in the $G={\rm SU}(2)$ case. One can moreover see
that diffeomorphisms still act only within this sector. Indeed, the action of
a diffeomorphism in the direction of a vector field $\xi^\mu$ is still
given by (\ref{diff-1}) and only changes the ${\mathfrak su}(2)$ part of
the two-form field. Similarly, the ${\rm SU}(2)$ gauge transformations
act only on the ${\mathfrak su}(2)$ sector. Thus, the gravity
story that we have considered in the previous section is unchanged.

Let us now consider what happens in directions not in ${\mathfrak su}(2)$. 
Let us first consider the "Higgs" sector spanned by $V_\pm,W_\pm$. 
A gauge transformation with the gauge parameter $\omega$ valued in this sector
acts as $\delta_\omega b=[\omega, \Sigma]$. In components this reads:
\begin{align}\nonumber
\delta_\omega b^4 =  - \frac{1}{\sqrt{2}}\, \omega^6 \Sigma^+ 
-\frac{1}{2}\, \omega^4 \Sigma^z & \, , \\ \nonumber
\delta_\omega b^5 =  \frac{1}{\sqrt{2}}\, \omega^7 \Sigma^- 
+ \frac{1}{2}\, \omega^5 \Sigma^z & \, ,  \\ \label{higgs-sym}
\delta_\omega b^6 =  - \frac{1}{\sqrt{2}}\, \omega^4 \Sigma^- 
+ \frac{1}{2}\, \omega^6  \Sigma^z & \, , \\ \nonumber
\delta_\omega b^7 =  \frac{1}{\sqrt{2}}\, \omega^5 \Sigma^+ 
- \frac{1}{2}\, \omega^7 \Sigma^z & \, .
\end{align}

The remaining part of the Lie algebra is that spanned by $Y$. The corresponding
gauge transformation has no effect on the two-form field $b^8$ 
(nor on $b^\alpha, \alpha=4,5,6,7$) since it commutes with the background. 
However this gauge transformation does act on the connection $a^8$ by the
usual ${\rm U}(1)$ gauge transformation $a^8\to a^8+d\omega^8$. 
The kinetic part (\ref{S-bf-YM}) clearly remains invariant, and the potential
part is invariant since it only depends on $b^8$ that does not transform. 

\subsection{Low energy limit of the "Higgs" sector}

Our analysis of the "YM" sector presented below will show that the parameter $\kappa$
that appeared in the "Higgs-YM" part of the potential (\ref{pot-higgs-ym}) must be
taken to be of the order $M_p^2$, where $M_p$ is the Planck mass. This will be
follow from the fact that the YM coupling constant should be of order unity in a realistic
unification scheme, which then immediately implies $\kappa\sim M_p^2$. 
Another way to reach the same conclusion is to note that $M_p$ is the only
scale in our problem, so all dimensionful quantities must be of the Planck size,
see more on this in the last discussion section. If this
is the case then the role of the potential term (\ref{pot-higgs-ym}) for the "Higgs"
sector is to make the anti-self-dual components of the two-forms $b^\alpha_{\mu\nu}$
"infinitely massive" and thus effectively set them to zero. This is quite similar to what
happened in the gravitational sector in the limit $g\to \infty$ with the $b^{ab}_{tf}$
components. Thus, we see that in the low energy limit $E^2\ll \kappa$ the two-forms
$b^\alpha_{\mu\nu}$ can be effectively assumed to be self-dual. As such they
can be expanded in the background self-dual two-forms $\Sigma^a_{0\,\mu\nu}$. After
such an ansatz is substituted into the action (\ref{S-bf-Higgs*}) the result 
simplifies considerably. However, in order to exhibit the physical modes we need to 
introduce some convenient gauge-fixing. Inspecting (\ref{higgs-sym}) we see that 
it is possible to set to zero the following components of the $b^{\alpha a}$:
\be\label{higgs-gauge}
b^{4}_{+}=0, \qquad b^{5}_{-} =0, \qquad b^{6}_{-}=0, \qquad b^{7}_{-}=0.
\ee
This gauge turns out to be very convenient. We now write the gauge-fixed two-forms
$b^{\alpha}_{\mu\nu}$ as follows:
\be\nonumber
b^4_{\mu\nu} = \frac{1}{2} \left( \frac{1}{\sqrt{2}} b^{4}_{-} \Sigma^{-}_{\mu\nu} 
+\frac{\sqrt{3}}{2} b^{4}_z \Sigma^z_{\mu\nu} \right), 
\\ \nonumber
b^5_{\mu\nu} = \frac{1}{2} \left( \frac{1}{\sqrt{2}} b^{5}_{+} \Sigma^{+}_{\mu\nu} 
+\frac{\sqrt{3}}{2} b^{5}_z \Sigma^z_{\mu\nu} \right), 
\\  \label{b-alpha}
b^6_{\mu\nu} = \frac{1}{2} \left( \frac{1}{\sqrt{2}} b^{6}_{+} \Sigma^{+}_{\mu\nu} 
+\frac{\sqrt{3}}{2} b^{6}_z \Sigma^z_{\mu\nu} \right), \\ \nonumber
b^7_{\mu\nu} = \frac{1}{2} \left( \frac{1}{\sqrt{2}} b^{7}_{-} \Sigma^{-}_{\mu\nu} 
+\frac{\sqrt{3}}{2} b^{7}_z \Sigma^z_{\mu\nu} \right), 
\ee
where the independent fields are now $b^{4}_{-}, b^{5}_{+}, b^{6}_{+}, b^{7}_{-}$ 
and $b^{\alpha}_z$ and the "strange" normalization coefficients are chosen in 
order for the Lagrangian to be obtained to have the canonical form.

Substituting (\ref{b-alpha}) into (\ref{S-bf-Higgs*}) and using the algebra of $\Sigma$-matrices
we get the following simple effective low-energy action:
\be
S_{eff}^{Higgs}=-\int \partial^\mu b^{5}_{+} \partial_\mu b^{4}_{-} 
+ \partial^\mu b^{7}_{-} \partial_\mu b^{6}_{+}
+\partial^\mu b^{5}_z \partial_\mu b^{4}_z + \partial^\mu b^{7}_z \partial_\mu b^{6}_z.
\ee

This form of the Lagrangian makes the reality conditions necessary to get a real theory obvious.
Indeed, it is clear that the reality conditions are:
\be\label{higgs-reality-L}
(b^{5}_{+})^*=b^{4}_{-}, \qquad (b^{7}_{-})^*=b^{6}_{+}, 
\qquad (b^{5}_z)^*=b^{4}_z, \qquad (b^{7}_z)^*=b^{6}_z.
\ee
These conditions can be compactly stated by introducing the following 
${\mathfrak su}(2)\otimes {\mathfrak g}$ valued object:
\begin{align}\label{bf-b}
{\bf b} :=& \left(  b^{4}_{-} T_{+}+ b^{4}_z T_{z} \right) \otimes V_{+}
+ \left( b^{5}_{+} T_{-} + b^{5}_z T_{z} \right) \otimes V_{-}
\\ \nonumber
+&\left( b^{6}_{+} T_{-}+b^{6}_z T_{z} \right) \otimes W_{+}
+ \left( b^{7}_{-} T_{+}+ b^{7}_z T_{z}  \right) \otimes W_{-}\, 
\end{align}
and requiring it to be hermitian:
\be
{\bf b}^\dagger = {\bf b}.
\ee
The action can also be written quite compactly in terms of ${\bf b}$. Indeed, using the
pairing given by the $\langle \cdot, \cdot \rangle$ metric in the Lie algebra we get:
\be\label{L-Higgs*}
{\cal L}^{Higgs}_{eff} = -\langle \partial^\mu {\bf b}^\dagger, \partial_\mu {\bf b}\rangle 
\ee
for the low-energy $E^2\ll \kappa$ effective "Higgs" sector Lagrangian. It is thus
clear that, at least in the low energy regime, the "Higgs" sector of our theory consists
just of 4 complex massless scalar fields with the usual Lagrangian. It is not hard to 
show that in the finite $\kappa$ limit the content of this sector does not change, and
is still given by massless fields. 

\subsection{Hamiltonian formulation for the Higgs sector}

In this subsection we obtain the 
Hamiltonian formulation of the sector spanned by $V_\pm,W_\pm$. After the analysis
performed in the previous subsection such an analysis is not really necessary as
we know what the propagating DOF described by this sector are like, and we even 
know the correct reality conditions. However, 
we decided to perform such an analysis for completeness, and also to confirm the
reality conditions found from the Hamiltonian perspective. One finds the
Hamiltonian analysis to be exactly parallel to that in the gravitational case,
with even the final expression for the Hamiltonian being analogous. This subsection
is quite technical and the reader is advised to skip it on the first reading. As
in the case of gravity, we start by performing the space plus time split
of the kinetic BF part. 

\noindent \textbf{BF-Part.}
From  (\ref{S-bf-Higgs}) our Lagrangian density is:
\begin{align}\label{L-bf-Higgs}
\mathcal{L}_{BF}^{Higgs} =&  \frac{2\sqrt{2}}{3}\, \left( (\partial b^5)^{\mu} (\partial b^6)^{\nu} \, 
\Sigma^+_{\mu\nu} - (\partial b^4)^{\mu}(\partial b^7)^{\nu}\Sigma^-_{\mu\nu} \right)
\notag \\
-&\frac{2}{3} \left( (\partial b^4)^{\mu}(\partial b^5)^{\nu} ( 2\eta_{\mu\nu}+\Sigma^z_{\mu\nu})
+ (\partial b^6)^{\mu}(\partial b^7)^{\nu}( 2\eta_{\mu\nu}-\Sigma^z_{\mu\nu})  \right) \, .
\end{align}

Now, denoting the indices $4,5,6,7$ collectively by $\alpha$, we have:
\be
(\partial b^\alpha)^0 = - \partial_b t^{\alpha b}, \qquad
(\partial b^\alpha)^a = \partial_0 t^{\alpha a} - 2\epsilon^{abc} \partial_b b^\alpha_{0c},
\ee
where we have introduced the configurational variables
\be
t^{\alpha a}:=\epsilon^{abc} b^\alpha_{bc}.
\ee

We do not need an expression for the expanded Lagrangian (\ref{L-bf-Higgs}) because
a more compact expression in terms of the conjugate momenta will be obtained below. 
For now let us compute the momenta conjugate to the configurational variables
$t^{\alpha a}$. It is sufficient to compute just one of the momenta to see the pattern. We have:
\begin{align}
\pi_{4a}:=\frac{\partial {\cal L}^{Higgs}_{BF}}{\partial \partial_0 t^{4 a}} =&
-\frac{2\sqrt{2}}{3} \left( \Sigma^{-}_{ab}(\partial_0 t^{7b}-2\epsilon^{bef}\partial_e b^7_{0f})
+ \Sigma^{-}_{0a} \partial_b t^{7b}\right) \\ \nonumber
-&\frac{4}{3}(\partial_0 t^{5}_a-2\epsilon_a^{\,\,ef}\partial_e b^5_{0f})
-\frac{2}{3}\left(\Sigma^z_{ab} (\partial_0 t^{5b}-2\epsilon^{bef}\partial_e b^5_{0f})
+\Sigma^z_{0a} \partial_b t^{5b}\right).
\end{align}
Comparing it to (\ref{a5}) we see that $\pi_{4a} = 2\im a^5_a$.
This is precisely analogous to the relation (\ref{pi-su2}) we had in the case of
gravity. Indeed, the above relation can be rewritten as $\pi_{4a} = -2\im g_{4\alpha} a^\alpha_a$,
which generalizes (\ref{pi-su2}). The other momenta are obtained as follows:
\be
\pi_{\alpha a} = -2\im g_{\alpha\beta} a^\beta_a.
\ee

We now need to solve for the velocities in terms of the momenta and substitute the result
into the Lagrangian. Similarly to the case of gravity the velocities can be obtained
by taking the spatial component of the equations (\ref{higgs-compat}). We get:
\be\nonumber
\partial_0 t^4_a- 2\epsilon_a^{\,\, bc} \partial_b b^4_{0c}
= \im\sqrt{2}\, \Sigma^+_{0a} a^6_0 +\frac{1}{\sqrt{2}}\, \Sigma^{+\,\, b}_{a} \pi_{7b}  
+\im\, \Sigma^z_{0a} a^4_0 +\frac{1}{2}\, \Sigma^{z\,\, b}_{a} \pi_{5b} \, , \\ \nonumber
\partial_0 t^5_a- 2\epsilon_a^{\,\, bc} \partial_b b^5_{0c}
=- \im\sqrt{2}\, \Sigma^{-}_{0a} a^7_0 - \frac{1}{\sqrt{2}}\, \Sigma^{-\,\, b}_{a} \pi_{6b} 
- \im\, \Sigma^z_{0a} a^5_0- \frac{1}{2}\, \Sigma^{z\,\, b}_{a} \pi_{4b}  \, ,  \\ \label{higgs-vel-mom}
\partial_0 t^6_a- 2\epsilon_a^{\,\, bc} \partial_b b^6_{0c}
= \im\sqrt{2}\, \Sigma^{-}_{0a} a^4_0+ \frac{1}{\sqrt{2}}\, \Sigma^{-\,\, b}_{a} \pi_{5b} 
- \im\,  \Sigma^z_{0a} a^6_0 - \frac{1}{2}\,  \Sigma^{z\,\, b}_{a} \pi_{7b} \, , \\ \nonumber
\partial_0 t^7_a- 2\epsilon_a^{\,\, bc} \partial_b b^7_{0c}
= - \im\sqrt{2}\, \Sigma^{+}_{0a} a^5_0 - \frac{1}{\sqrt{2}}\, \Sigma^{+\,\, b}_{a} \pi_{4b}
+ \im\, \Sigma^z_{0a} a^7_0+ \frac{1}{2}\, \Sigma^{z\,\, b}_{a} \pi_{6b} \, .
\ee
The time projections of the equations (\ref{higgs-compat}) are then the Gauss
constraints. 

For the last step we start from a convenient expression for the Lagrangian. This is given
by an analog of (\ref{su2-aa}), which reads:
\begin{align}
{\cal L}^{Higgs}_{BF}=&-2g_{ab}\Sigma^{a\,\mu\nu} f^{b}_{\alpha\beta} a^\alpha_\mu a^\beta_\nu
\\ \nonumber
=& -2\sqrt{2}\Sigma^{+\,\mu\nu} a^5_\mu a^6_\nu  
+ 2\sqrt{2} \Sigma^{-\,\mu\nu} a^4_\mu a^7_\nu  
+ 2\Sigma^{z\,\mu\nu} a^4_\mu a^5_\nu  
-2\Sigma^{z\,\mu\nu} a^6_\mu a^7_\nu,
\end{align}
where $f^a_{\alpha\beta}$ are the structure constants. Expanding it, and converting
the spatial components of the connection into momenta we get:
\begin{align}\nonumber
{\cal L}^{Higgs}_{BF}=&\frac{1}{\sqrt{2}}\Sigma^{+\,ab} \pi_{4a} \pi_{7b}  
- \frac{1}{\sqrt{2}} \Sigma^{-\,ab} \pi_{5a} \pi_{6b}
- \frac{1}{2}\Sigma^{z\,ab} \pi_{5a} \pi_{4b}
+\frac{1}{2}\Sigma^{z\,ab} \pi_{7a} \pi_{6b}\\ \nonumber
-&\im\sqrt{2}\Sigma^{+\, a}_{0} (a^5_0 \pi_{7a} - \pi_{4a} a^6_0)  
+ \im\sqrt{2} \Sigma^{-\, a}_{0} (a^4_0 \pi_{6a} - \pi_{5a} a^7_0)  \\ \nonumber
+&\im\Sigma^{z\, a}_{0} (a^4_0 \pi_{4a} - \pi_{5a} a^5_0)  
-\im\Sigma^{z\, a}_{0} (a^6_0 \pi_{6a}-\pi_{7a} a^7_0)\, .
\end{align}

We can now compute the Hamiltonian:
\be
{\cal H}^{Higgs}_{BF} = \pi_{\alpha a} \partial_0 t^{\alpha a}-{\cal L}^{Higgs}_{BF}=
2\pi_{\alpha a} \epsilon^{abc} \partial_b b^\alpha_{0c} +\frac{1}{2} g_{ab} 
g^{\alpha\gamma} g^{\beta\delta} f^a_{\alpha\beta}
\Sigma^{b\, ef} \pi_{\gamma e} \pi_{\delta f}\, ,
\ee
The obtained expression is not the full Hamiltonian. To obtain the
later we need to add 4 Gauss constraints that are obtained as the time components
of the compatibility equations (\ref{higgs-compat}). We will not need an explicit form
of the Gauss constraints since we already know from (\ref{higgs-sym}) what
is generated by them. 

\noindent \textbf{The BB-Part.} Let us now consider the potential part (\ref{S-bb-higgs}).
The corresponding Lagrangian density reads:
\be
{\cal L}^{Higgs}_{BB}=-\kappa P^{-}_{\mu\nu\rho\sigma} g_{\alpha\beta} b^{\alpha\,\mu\nu}
b^{\beta\,\rho\sigma}.
\ee
Expanding the spacetime index we get:
\be
{\cal L}^{Higgs}_{BB}=\kappa g_{\alpha\beta} 
(b^{\alpha\, a}_{0} b^\beta_{0a} -\frac{1}{4} t^{\alpha a} t^\beta_{a})
+\im \kappa g_{\alpha\beta} b^\alpha_{0a} t^{\beta a}.
\ee

\noindent \textbf{Total Hamiltonian.} We now form the total Hamiltonian
${\cal H}^{Higgs}={\cal H}^{Higgs}_{BF}-{\cal L}^{Higgs}_{BB}$ and
integrate out the non-dynamical fields $b^\alpha_{0a}$. We get the
following expressions for these fields by solving their field equations:
\be\label{b0a-higgs}
b^\alpha_{0a} = \frac{1}{\kappa}  g^{\alpha\beta}\epsilon_{abc} 
\partial^b \pi_{\beta}^{c} - \frac{\im}{2}t^\alpha_{a}\, .
\ee
This should be compared with
(\ref{btf-su2}) that we have in the gravitational sector. We now substitute this back
to get the Hamiltonian with second-class constraints solved for:
\be\label{H-higgs}
{\cal H}^{Higgs}=\frac{1}{2} g_{ab} 
g^{\alpha\gamma} g^{\beta\delta} f^a_{\alpha\beta}
\Sigma^{b\, ef} \pi_{\gamma e} \pi_{\delta f}
-\im (\epsilon^{abc} \partial_b \pi_{\alpha c})  t^\alpha_{a}
+\frac{1}{\kappa} g^{\alpha\beta} (\epsilon^{abc} \partial_b \pi_{\alpha c})
(\epsilon^{aef} \partial_e \pi_{\beta f}) \, ,
\ee
plus Gauss constraints with their corresponding Lagrange multipliers. 
Note also that the Hamiltonian we have obtained is analogous to the
one in the case of gravity (\ref{Ham-su2}). Indeed, there is similarly
the $\pi^2$ term and a $(\epsilon\partial\pi)t$ term with an imaginary
unit in front. There is also a $\partial^2\pi^2$ term with a parameter
of dimensions $1/M^2$ as a coefficient. Note that for any value of the
parameter $\kappa$ this Hamiltonian describes modes that are massless.
To rewrite this Hamiltonian in terms of physical propagating modes we need to
understand the gauge-fixing. 

\noindent \textbf{Gauge-fixing.} To choose a convenient gauge-fixing that
eliminates the gauge transformation freedom, let us discuss what the
two-forms $\Sigma^\pm,\Sigma^z$ become after they get projected onto the
spatial hypersurface. Thus, let us find analogs of relations (\ref{sigm-0a}), (\ref{sigm-ab}). 
Let us introduce the following 3 spatial vectors:
\be
\Sigma^+_{0a}:= m_a, \qquad \Sigma^-_{0a} := \bar{m}_a, \qquad \Sigma^z_{0a}:= n_a.
\ee
Then, taking various projections of (\ref{su2-rels}), it is easy to check that the following relations
hold:
\be
m^a m_a = \bar{m}^a\bar{m}_a =0, \qquad m^a n_a = \bar{m}^a n_a = 0, \\ \nonumber
m^a\bar{m}_a = 1, \qquad n^a n_a=1.
\ee
Taking different projections of (\ref{su2-rels}) one finds the spatial pull-backs of the two-forms
in terms of the vectors introduced:
\be\label{sab-pm}
\Sigma^+_{ab}=n_a m_b - m_a n_b=-\im\epsilon_{abc} m^c, 
\qquad&{}& \Sigma^-_{ab}=\bar{m}_a n_b - n_a \bar{m}_b=-\im\epsilon_{abc}\bar{m}^c,
\\ \nonumber
\Sigma^z_{ab} = m_a \bar{m}_b - \bar{m}_a m_b&=&-\im\epsilon_{abc} n^c.
\ee

We now use (\ref{higgs-sym}) to fix the gauge as in (\ref{higgs-gauge}). In terms of
the configurational variables $t^{\alpha a}$ the gauge conditions read:
\be\label{gauge-t}
t^{4-}=0, \qquad t^{5+}=0, \qquad t^{6+}=0, \qquad t^{7-}=0,
\ee
where our convention is that $t^{\alpha +}:=m^a t^\alpha_a, t^{\alpha -}=\bar{m}^a t^\alpha_a,
t^{\alpha z}=n^a t^\alpha_a$. 

Let us now find the consequences of the Gauss constraints. In terms of the introduced
vectors $m^a,\bar{m}^a, n^a$ these read:
\be
\partial_a t^{4a} -\frac{1}{\sqrt{2}} m_a \pi_{7}^a - \frac{1}{2} n_a \pi_5^a=0, \qquad
\partial_a t^{5a} +\frac{1}{\sqrt{2}} \bar{m}_a \pi_{6}^a + \frac{1}{2} n_a \pi_4^a=0, 
\\ \nonumber
\partial_a t^{6a} -\frac{1}{\sqrt{2}} \bar{m}_a \pi_{5}^a + \frac{1}{2} n_a \pi_7^a=0, \qquad
\partial_a t^{7a} +\frac{1}{\sqrt{2}} m_a \pi_{4}^a - \frac{1}{2} n_a \pi_6^a=0,
\ee
Introducing more compact notations $\pi_{\alpha}^{+}:=m_a \pi_\alpha^a, 
\pi_{\alpha}^{-}=\bar{m}_a \pi_\alpha^a, \pi_{\alpha}^{z}=n_a \pi_\alpha^a$ and
passing to the momentum space we have:
\be
\im |k| t^{4z} -\frac{1}{\sqrt{2}} \pi_{7}^+ - \frac{1}{2} \pi_5^z=0, \qquad
\im |k| t^{5z} +\frac{1}{\sqrt{2}} \pi_{6}^- + \frac{1}{2} \pi_4^z=0, 
\\ \nonumber
\im |k| t^{6z} -\frac{1}{\sqrt{2}} \pi_{5}^- + \frac{1}{2} \pi_7^z=0, \qquad
\im |k| t^{7z} +\frac{1}{\sqrt{2}} \pi_{4}^+ - \frac{1}{2} \pi_6^z=0.
\ee
We now use these constraints to find the components of the momenta that are
conjugate to the gauge-fixed variables (\ref{gauge-t}). We have:
\be\label{higgs-mom}
\pi_4^+(k)=-\im\sqrt{2} |k| t^{7z} + \frac{1}{\sqrt{2}} \pi_6^z, \qquad 
\pi_5^-(k)=\im\sqrt{2} |k| t^{6z} + \frac{1}{\sqrt{2}} \pi_7^z, \\ \nonumber
\pi_6^-(k)=-\im\sqrt{2} |k| t^{5z} - \frac{1}{\sqrt{2}} \pi_4^z, \qquad
\pi_7^+(k)=\im\sqrt{2} |k| t^{4z} - \frac{1}{\sqrt{2}} \pi_5^z.
\ee
Let us now substitute these expressions into the $\pi^2$ part of the Hamiltonian. Thus, we 
have for the first term in (\ref{H-higgs}):
\be\label{part-1}
-\frac{3}{4}(\pi_4^z(-k) \pi_5^z(k) +\pi_7^z(-k) \pi_6^z(k)) 
- \frac{1}{2}(\pi_4^-(-k)\pi_5^+(k)+\pi_7^-(-k)\pi_6^+(k) )
\\ \nonumber
+\frac{\im |k|}{2}(\pi_4^z(-k) t^{4z}(k)+\pi_7^z(-k) t^{7z}(k)
-\pi_5^z(-k) t^{5z}(k)-\pi_6^z(-k) t^{6z}(k)) \\ \nonumber
-|k|^2(t^{5z}(-k)t^{4z}(k)+t^{6z}(-k)t^{7z}(k)) 
\ee

Let us now work out the second term in (\ref{H-higgs}). We use:
\be\label{eps}
-\im\epsilon^{abc}=n^a(m^b \bar{m}^c-\bar{m}^b m^c)+
m^a (\bar{m}^b n^c - n^b \bar{m}^c) + \bar{m}^a (n^b m^c - m^b n^c),
\ee
which can be easily derived from (\ref{sab-pm}) to write the second term in (\ref{H-higgs}) as:
\be\label{h-dark-2}
\im |k| (\pi_{\alpha}^{-}(-k)  t^{\alpha +}(k)-\pi_{\alpha}^{+}(-k)  t^{\alpha -}(k)).
\ee
Here we again passed to the momentum space and used
\be
\partial_a (e^{\im kx} t^\alpha_b(k)) =\im k_a e^{\im kx} t_b^\alpha(k), 
\ee 
where $k^a=|k| n^a$ is a vector in the direction of $n^a$. This makes only two of the terms 
from (\ref{eps}) survive. Expanding and using the gauge-fixing conditions 
(\ref{gauge-t}) we get for this term:
\be\label{part-2}
\im |k|(\pi_4^{-}(-k) t^{4+}(k) + \pi_7^{-}(-k) t^{7+}(k) 
- \pi_5^{+}(-k) t^{5-}(k) -\pi_6^{+}(-k) t^{6 -}(k)).
\ee
The total Hamiltonian in the $E^2\ll \kappa$ low energy limit is given by the sum of two terms,
i.e., (\ref{part-1}) and (\ref{part-2}).

\noindent {\bf Reality conditions.} Let us now discuss the reality conditions that are
appropriate in the $E^2\ll \kappa$ low energy limit. It is clear that they can be determined
by "completing the square", similar to what we have seen in the Hamiltonian formulation
of the gravitational sector (in the low energy limit). Thus, let us write the total 
Hamiltonian as:
\be
{\cal H}^{Higgs}=-\frac{3}{4}\left( \pi_4^z(-k) - \frac{2\im |k|}{3} t^{5z}(-k) \right)
\left( \pi_5^z(k) - \frac{2\im |k|}{3} t^{4z}(k)\right) - \frac{4}{3} |k|^2 t^{5z}(-k) t^{4z}(k) 
\\ \nonumber
-\frac{3}{4}\left( \pi_7^z(-k) - \frac{2\im |k|}{3} t^{6z}(-k) \right)
\left( \pi_6^z(k) - \frac{2\im |k|}{3} t^{7z}(k)\right) - \frac{4}{3} |k|^2 t^{6z}(-k) t^{7z}(k)
\\ \nonumber
-\frac{1}{2}\left( \pi_4^-(-k) - 2\im |k| t^{5-}(-k) \right)
\left( \pi_5^+(k) - 2\im |k| t^{4+}(k)\right) - 2 |k|^2 t^{5-}(-k) t^{4+}(k)
\\ \nonumber
-\frac{1}{2}\left( \pi_7^-(-k) - 2\im |k| t^{6-}(-k) \right)
\left( \pi_6^+(k) - 2\im |k| t^{7+}(k)\right) - 2 |k|^2 t^{6-}(-k) t^{7+}(k)\, .
\ee

The form of the reality conditions is now obvious. Indeed, we introduce new momenta
variables:
\be
\tilde{\pi}_4^z(k) = \pi_4^z(k) + \frac{2\im |k|}{3} t^{5z}(k), \qquad
\tilde{\pi}_5^z(k) = \pi_5^z(k) - \frac{2\im |k|}{3} t^{4z}(k), \\ \nonumber
\tilde{\pi}_6^z(k) = \pi_6^z(k) - \frac{2\im |k|}{3} t^{7z}(k), \qquad
\tilde{\pi}_7^z(k) = \pi_7^z(k) +\frac{2\im |k|}{3} t^{6z}(k), \\ \nonumber
\tilde{\pi}_4^{-}(k)=\pi_4^{-}(k)+2\im |k| t^{5-}(k), \qquad
\tilde{\pi}_5^{+}(k)=\pi_5^{+}(k)-2\im |k| t^{4+}(k), \\ \nonumber
\tilde{\pi}_6^{+}(k)=\pi_6^{+}(k)-2\im |k| t^{7+}(k), \qquad
\tilde{\pi}_7^{-}(k)=\pi_7^{-}(k)+2\im |k| t^{6-}(k),
\ee
and then require the following reality conditions:
\be\label{higgs-reality-H}
\tilde{\pi}_4^z(-k)= -(\tilde{\pi}_5^z(k))^*, \qquad \tilde{\pi}_7^z(-k)= -(\tilde{\pi}_6^z(k))^*, 
\\ \nonumber
\tilde{\pi}_4^{-}(-k)=-(\tilde{\pi}_5^{+}(k))^*, \qquad \tilde{\pi}_7^{-}(-k)=-(\tilde{\pi}_6^{+}(k))^*
\\ \nonumber
t^{5z}(-k)=-(t^{4z}(k))^*, \qquad t^{6z}(-k)=-(t^{7z}(k))^*, 
\\ \nonumber
t^{5-}(-k)=-(t^{4+}(k))^*, \qquad t^{6-}(-k)=-(t^{7+}(k))^*.
\ee
It is not hard to see that these conditions are the same as we have derived earlier in the
Lagrangian framework, see (\ref{higgs-reality-L}). Indeed, the extra minus present in
(\ref{higgs-reality-H}) is due to the following transformation properties of the basic 
two-forms:
\be
(\Sigma^+_{ab})^* = -\Sigma^-_{ab}, \qquad (\Sigma^z_{ab})^* = -\Sigma^z_{ab}
\ee
that directly follow from (\ref{sab-pm}). The obtained real positive definite Hamiltonian
is that of 4 complex massless scalar fields, so we have full agreement with our
Lagrangian analysis above. Reality conditions and the Hamiltonian for the full
finite $\kappa$ theory can be obtained via precisely the same method
as in the gravitational sector case treated in the Appendix. We refrain from
giving such an analysis in this work, as it becomes even more technical. 

\subsection{YM sector}

In this subsection we work out the Lagrangian for the remaining part of the theory,
which lives in the part of the gauge group that commutes with the background 
${\mathfrak su}(2)$. The total Lagrangian we start with is a sum of kinetic
term (\ref{S-bf-YM}) and the potential term (\ref{S-bb-higgs}), with an extra
sign in the potential term coming from the metric component $g_{88}=-1$.
This gives:
\be\label{L-YM}
{\cal L}^{YM}= 2\im \epsilon^{\mu\nu\rho\sigma} b^8_{\mu\nu} \partial_\rho a^8_\sigma
+\kappa P^{-\,\mu\nu\rho\sigma} b^8_{\mu\nu} b^8_{\rho\sigma}.
\ee
The further analysis is greatly simplified by making use of the reality condition for
the $b^8_{\mu\nu}$ two-form from the outset. Thus, as we will also confirm by
the Hamiltonian analysis in the next subsection, the two-form $b^8_{\mu\nu}$
needs to be purely imaginary:
\be
b^8_{\mu\nu}:=-\im \tilde{b}^8_{\mu\nu}, \qquad (\tilde{b}^8_{\mu\nu})^*=\tilde{b}^8_{\mu\nu}.
\ee
This immediately leads to simplifications as the real part of the Lagrangian
(\ref{L-YM}) is then given simply by:
\be
{\cal L}^{YM}_{real} =  2 \epsilon^{\mu\nu\rho\sigma} \tilde{b}^8_{\mu\nu} 
\partial_\rho a_\sigma^8 - \frac{\kappa}{2} \tilde{b}^{8\,\mu\nu} \tilde{b}^8_{\mu\nu}.
\ee
Taking a variation with respect to $\tilde{b}^8_{\mu\nu}$ we learn that:
\be\label{b8-f}
\tilde{b}^8_{\mu\nu} = \frac{1}{\kappa} \epsilon_{\mu\nu\rho\sigma} F^{\rho\sigma},
\ee
where $F_{\mu\nu}=\partial_\mu a^8_\nu -\partial_\nu a^8_\mu$ is the curvature
of our ${\rm U}(1)$ gauge field, which is therefore, for real $\kappa$, real. 
Substituting the result back into the Lagrangian we get:
\be\label{Lagr-YM}
{\cal L}^{YM}=-\frac{2}{\kappa} (F_{\mu\nu})^2.
\ee
This is the standard YM Lagrangian with the coupling constant:
\be
g_{YM}^2 = \frac{\kappa}{8}.
\ee
To convert this into a physical coupling constant we recall that we need to multiply the
Lagrangian by $32\pi G$, as this is exactly the prefactor that converts the canonically-normalized
graviton Lagrangian (\ref{S-grav}) into the Einstein-Hilbert one. Thus, the physical
coupling constant in our arising YM  theory is given by:
\be\label{YM-coupl}
g_{YM}^2 = 4\pi G \kappa.
\ee
Realistic particle physics coupling constants are of the order of unity (and smaller),
so we learn that the parameter $\kappa$ must be of the order $M_p^2$, which is
what we have been using in the previous subsections.

\subsection{Reality conditions for the YM sector}

In this subsection we perform the Hamiltoanian analysis of the YM sector with the
main aim being to obtain the reality conditions used above. As in all other
cases considered, the reality conditions become obvious once the Hamiltonian is written down.

We start from the Lagrangian (\ref{L-YM}). Expanding
\be
\epsilon^{\mu\nu\rho\sigma}b^8_{\mu\nu}\partial_\rho a^8_\sigma=
-2\epsilon^{abc} b^8_{0a} \partial_b a^8_c - t^{8a} (\partial_0 a_a^8-\partial_a a_0^8),
\ee
where $t^{8a}:=\epsilon^{abc} b^8_{bc}$, we see that
the momentum conjugate to the connection $a_a^8$ is 
\be\label{mom-YM}
\pi^{8a}:=\frac{\partial {\cal L}^{YM}}{\partial \partial_0 a_a^8}=-2\im t^{8a}.
\ee
The Hamiltonian is then:
\be\label{Ham-YM}
{\cal H}^{YM}=4\im \epsilon^{abc} b^8_{0a} \partial_b a_c^8- a_0^8 \partial_a \pi^{8a}
+\kappa(b^{8\, a}_{0} b^8_{0a} +\frac{1}{16} \pi^{8 a} \pi^8_{a})
-\frac{\kappa}{2} b^8_{0a} \pi^{8 a}.
\ee
We find the non-dynamical fields $b^8_{0a}$ via their field equations and get:
\be\label{b0-YM}
b^8_{0a}=-\frac{2\im}{\kappa} \epsilon_{abc}\partial^b a^{8c} +\frac{1}{4} \pi^8_a.
\ee
Substituting this back into (\ref{Ham-YM})  we get the "physical" Hamiltonian
\be\label{H-phys-YM}
{\cal H}_{phys}^{YM}
=\frac{4}{\kappa} \left(\epsilon^{abc} \partial_b a^8_c
+\frac{\im \kappa}{8}\pi^{8a} \right)^2
+\frac{\kappa}{16} \pi^{8a}\pi^8_a.
\ee
It is now clear that the "correct" reality conditions that give rise to a real positive definite
Hamiltonian is:
\be\label{reality-YM}
 {\rm Im}(\epsilon^{abc}\partial_b a_c^8)
+\frac{\kappa}{8}{\rm Re}(\pi^{8a}) =0,
\qquad {\rm Im}(\pi^{8a})=0.
\ee
From (\ref{b0-YM}) and (\ref{mom-YM}) it is easy to see that these reality conditions are 
equivalent to the condition that $b^8_{\mu\nu}$ two-form is purely imaginary,
which is what we have used in the previous subsection.

Passing to the real phase space and imposing the Gauss constraint $\partial_a \pi^{8a}=0$
as well as the transverse gauge condition $\partial^a a^8_a=0$ we get the following
simple expression for the real Hamiltonian:
\be
{\cal H}^{YM}_{real}= \frac{4}{\kappa} (\partial_a a^{8\, real}_b)^2 +\frac{\kappa}{16}
(\pi^{8a})^2,
\ee
which again confirms that the parameter $\kappa/8$ plays the role of $g_{YM}^2$.


\section{Interactions}
\label{sec:inter}

In this section we work out (some of the) cubic order interactions for our theory. Our main goal is
to verify that the YM and Higgs sectors interacts with the gravitational field in the usual way, 
and that the YM-Higgs interaction is also standard. We start with general considerations on the 
cubic order expansion of our theory.

\subsection{General considerations}

The third variation of the BF-term is
\begin{equation}
\delta^3 S_{BF}= 4\im \int 3\, \delta B^I \wedge [\delta A,\delta A]^I \, ,
\end{equation}
and the third variation of the BB-term is
\begin{align}\label{pot-order3}
\delta^3 S_{BB}=4\im \int d^4x &\left( 
 4 \frac{\partial^3 V(\tilde{h})}{\partial \tilde{h}^{MN} \partial \tilde{h}^{KL} \partial \tilde{h}^{IJ}} \, 
(B_0 \delta B)^{IJ} (B_0 \delta B)^{KL} (B_0 \delta B)^{MN} \right. \\ &\;\;\;+6 \left. \frac{\partial^2 
V(\tilde{h})}{\partial \tilde{h}^{KL} \partial \tilde{h}^{IJ}} \, 
(B_0 \delta B)^{IJ} (\delta B \delta B)^{KL} \right)\, .
\end{align}

As in the case of the quadratic order expansion, it is most laborious to compute the
derivatives of the potential. We have already computed the second derivative above. The third 
derivative of $V(\tilde{h})$ is given by:
\be
\frac{\partial^3 V(\tilde{h})}{\partial \tilde{h}^{MN} \partial \tilde{h}^{KL}\partial \tilde{h}^{IJ}}=
\frac{g_{IJ}}{n} \, \frac{\partial^2 f}{\partial \tilde{h}^{MN}\partial \tilde{h}^{KL}}+\frac{g_{KL}}{n} \, 
\frac{\partial^2 f}{\partial \tilde{h}^{MN}\partial \tilde{h}^{IJ}} +\frac{g_{MN}}{n} \, \frac{\partial^2 f}
{\partial \tilde{h}^{KL}\partial \tilde{h}^{IJ}}   \\ \nonumber
+ \frac{Tr\, \tilde{h}}{n}\, 
\frac{\partial^3 f}{\partial \tilde{h}^{KL}\partial \tilde{h}^{KL}\partial \tilde{h}^{IJ}} \, ,
\ee
where the third derivative of the function of the ratios is given by
\begin{align}
\frac{\partial^3 f}{\partial \tilde{h}^{MN}\partial \tilde{h}^{KL}\partial \tilde{h}^{IJ}}
=&\sum_{p=2}^{n}\sum_{q=2}^{n}\sum_{p=2}^{n} f'''_{pqr} \,\, 
\frac{\partial}{\partial \tilde{h}^{IJ}}
\left( \frac{Tr\, \tilde{h}^p}{(Tr\, \tilde{h})^p}\right) \,\, 
\frac{\partial}{\partial \tilde{h}^{KL}}
\left( \frac{Tr\, \tilde{h}^q}{(Tr\, \tilde{h})^q}\right) \,\,\frac{\partial}{\partial \tilde{h}^{MN}}
\left( \frac{Tr\, \tilde{h}^r}{(Tr\, \tilde{h})^r}\right)  \notag \\
&+\sum_{p=2}^{n}\sum_{q=2}^{n} f''_{pq} \,\, \frac{\partial}{\partial \tilde{h}^{IJ}}
\left( \frac{Tr\, \tilde{h}^p}{(Tr\, \tilde{h})^p}\right)\,\, \frac{\partial^2}{\partial \tilde{h}^{MN}\partial \tilde{h}^{KL}}
\left( \frac{Tr\, \tilde{h}^q}{(Tr\, \tilde{h})^q}\right)  \notag \\
&+\sum_{p=2}^{n}\sum_{q=2}^{n} f''_{pq} \,\, \frac{\partial}{\partial \tilde{h}^{KL}}
\left( \frac{Tr\, \tilde{h}^p}{(Tr\, \tilde{h})^p}\right)\,\, \frac{\partial^2}{\partial \tilde{h}^{MN}\partial \tilde{h}^{IJ}}
\left( \frac{Tr\, \tilde{h}^q}{(Tr\, \tilde{h})^q}\right)  \notag \\
&+\sum_{p=2}^{n}\sum_{q=2}^{n} f''_{pq} \,\, \frac{\partial}{\partial \tilde{h}^{MN}}
\left( \frac{Tr\, \tilde{h}^p}{(Tr\, \tilde{h})^p}\right)\,\, \frac{\partial^2}{\partial \tilde{h}^{KL}\partial \tilde{h}^{IJ}}
\left( \frac{Tr\, \tilde{h}^q}{(Tr\, \tilde{h})^q}\right)  \notag \\
&+ \sum_{p=2}^{n} f'_p \,\, \frac{\partial^3}{\partial \tilde{h}^{MN}\partial \tilde{h}^{KL}\partial \tilde{h}^{IJ}}
\left( \frac{Tr\, \tilde{h}^p}{(Tr\, \tilde{h})^p}\right) \, ,
\end{align}
where $f'''_{pqr}$ stands for the derivative of $f''_{pq}$ with respect to its $r$ argument and 
\begin{align}
\frac{\partial^3}{\partial \tilde{h}^{KL}\partial \tilde{h}^{KL}\partial \tilde{h}^{IJ}}\left( \frac{Tr\, \tilde{h}^p}{(Tr\, \tilde{h})^p}\right)
=&\frac{p}{(Tr\, \tilde{h})^p}\, \frac{\partial^2 \tilde{h}^{p-1}_{IJ}}{\partial \tilde{h}^{MN}\partial \tilde{h}^{KL}} \notag \\
&-\frac{p^2}{(Tr\, \tilde{h})^{p+1}}\,\left( g_{IJ}\,\frac{\partial \tilde{h}^{p-1}_{KL}}{\partial \tilde{h}^{MN}}+ g_{KL}\,\frac{\partial \tilde{h}^{p-1}_{IJ}}{\partial \tilde{h}^{MN}}+g_{MN}\,\frac{\partial \tilde{h}^{p-1}_{IJ}}{\partial \tilde{h}^{KL}} \right) \notag \\
&+\frac{p^2 (p+1)}{(Tr\, \tilde{h})^{p+2}}\left( g_{IJ}\,g_{KL}\, \tilde{h}^{p-1}_{MN}+g_{IJ}\,g_{MN}\, \tilde{h}^{p-1}_{KL}+g_{KL}\,g_{MN}\, \tilde{h}^{p-1}_{IJ} \right) \notag \\
&-\frac{p (p+1) (p+2)\,\, Tr\,\tilde{h}^p}{(Tr\, \tilde{h})^{p+3}}\,\, g_{IJ}\, g_{KL}\, g_{MN} \, .
\end{align}
The first derivative of a power of $\tilde{h}^{IJ}$ is given by (\ref{h-der-h}).
We have not found a sufficiently simple general expression for the second derivative 
of $\tilde{h}^{p-1}_{IJ}$ with respect to $\tilde{h}^{MN}\tilde{h}^{KL}$, but the expression
(\ref{h-der-h}) can be easily differentiated for any given p. The above expressions can
be used to obtain the third derivatives of the potential for our background. The results are
given in the next subsection.

\subsection{Interactions with gravity}

In this paper we shall not consider gravitational sector self-interactions. They are easily 
computable, but since the main emphasis of this work is on unification, it is of much more
interest to compute the interactions of other fields with gravity and their self-interactions.
In this subsection we consider the coupling of non-gravitational fields to gravity. 

Thus, at least one of the perturbation fields $\delta B^I$ is to be taken to lie in the gravitational
sector. It is then easy to see that this is the only interaction in the cubic order. Indeed, where
two of the three perturbation fields lie in the gravitational sector and there is only one 
non-gravitational perturbation, there is no interaction coming from the potential part since
\be
\left. \frac{\partial^2 V(\tilde{h})}{\partial \tilde{h}^{a\alpha} \partial \tilde{h}^{bc}}\right|_0=0,
\qquad 
\left. \frac{\partial^3 V(\tilde{h})}{\partial \tilde{h}^{e\alpha}\partial \tilde{h}^{cd}\partial \tilde{h}^{ab}}\right|_0=0,
\ee
where $\alpha$ stands for the non-gravitational part of the Lie algebra. There is also no 
interaction coming from the kinetic part of the action for the structure constant $f^{I}_{JK}$
is zero when two of the indices are in the ${\mathfrak su}(2)$ part and only
one index is in the non-gravitational part. Thus we only need to consider the interaction that
is linear in the graviton perturbation. It is natural to expect that this coupling is that to the
stress-energy tensor of our non-gravitational fields, and this will be confirmed below.

The interaction coming from the kinetic term is only non-trivial for the Higgs sector
fields (since the structure constant with two of its indices in the "YM" part of the Lie algebra and 
one in ${\mathfrak su}(2)$ is zero since the YM and the gravitational parts commute).
This interaction is of the schematic type $h (\partial b)^2$, which is as expected for scalar fields 
coupled to gravity. We are not going to work out this term, even though
it is not hard to do it using the explicit formulas for the connections worked out above. 

Let us concentrate on the interactions coming from the potential part of the action,
as being the most interesting one. The relevant derivatives of the potential are as follows:
\begin{align}\label{V-der-3}
\left.\frac{\partial^2 V(\tilde{h})}{\partial \tilde{h}^{\alpha\beta} \partial \tilde{h}^{ab}}
\right|_0=&0,
\\ \nonumber
\left.\frac{\partial^2 V(\tilde{h})}{\partial \tilde{h}^{b\beta} \partial \tilde{h}^{a\alpha}}
\right|_0=&\frac{\kappa}{4\im} g_{\alpha\beta} g_{ab},
\\ \nonumber
\left. \frac{\partial^3 V(\tilde{h})}{\partial \tilde{h}^{d\beta}
\partial \tilde{h}^{c\alpha}\partial \tilde{h}^{ab}}\right|_0=&\frac{g_{\alpha\beta}}{2(2\im)^2}
\left((\kappa-g)\left( g_{a(c}g_{d)b} - \frac{1}{3} g_{ab} g_{cd}\right)- \frac{\kappa}{3} g_{ab} g_{cd}
\right)\, .
\end{align}
Note that the fact that the first quantity is zero is not completely trivial, at it involves
a precise cancellation of two otherwise non-zero terms. 

We can now compute the relevant interaction terms using (\ref{pot-order3}). We need to divide 
this expression by $3!$ to remove the extra multiplicity introduced by taking the third
variation of the action. An additional simplification comes from the fact that
in the first term in the third derivative in (\ref{V-der-3}) we have a matrix projecting
onto the tracefree part of the gravitational two-form perturbation matrix 
$\Sigma^{a\,\mu\nu}_0 b^b_{\mu\nu}$. This part is zero when the parameter $g\to\infty$,
which is the limit of the usual GR that we are considering. Thus, this part drops out
and we have for the gravity-non-gravity interaction term coming from the potential:
\be\label{L3-pot}
{\cal L}^{(3)}=\frac{4\im}{3!} \Big( 4\cdot 3 \frac{1}{2(2\im)^2} \left(-\frac{\kappa}{3}\right)
\left( \frac{\im}{2}\right)^3 (\Sigma^{a\,\mu\nu}_0 b^a_{\mu\nu})
(\Sigma^{c\,\rho\sigma}_0 b^\alpha_{\rho\sigma})
 (\Sigma^{d\,\rho\sigma}_0 b^\beta_{\rho\sigma}) g_{cd} g_{\alpha\beta}
 \\ \nonumber
 + 6\cdot 2 \frac{\kappa}{4\im} \left(\frac{\im}{2}\right) (\Sigma^{a\,\mu\nu}_0 b^\alpha_{\mu\nu})
 \frac{1}{4}(\epsilon^{\rho\sigma\tau\lambda} b^b_{\rho\sigma} b^\beta_{\tau\lambda}) g_{ab}
 g_{\alpha\beta}\Big).
 \ee
Here the extra factors of $3$ in the first term and $2$ in the second come from expanding the
general Lie algebra indices in (\ref{pot-order3}), and the factors of $\im/2$ come
by using the self-duality of the background forms $\Sigma^a_{0\,\mu\nu}$. To understand
this expression it is useful to separate the coupling to the trace of the graviton perturbation,
and to the tracefree part. Let us consider the trace first. Thus, we take:
\be
b^a_{\mu\nu} = \frac{h}{3} \Sigma^a_{0\,\mu\nu},
\ee
with the field $h$ being proportional to the trace of the metric perturbation $h_{\mu\nu}$. It is then
easy to see that the expression (\ref{L3-pot}) vanishes on such gravitational
perturbations. This is, of course, as expected, for both our YM and Higgs sectors
are expected to be conformally-invariant (classically). Indeed, this is standard for
the YM fields, and for the Higgs sector this expectation follows from the fact that
the fields are (up to now) massless. Using (\ref{L3-pot}) it is not hard to check 
that there is indeed no coupling
to the trace part of the metric, which confirms our expectation. Note that this
also provides quite a non-trivial check of our scheme, for the whole scheme
would be invalidated if we had found that our YM fields couple to the trace of
the metric. 

We now confirm that the coupling to the tracefree part of the metric perturbation is also
as expected. We only need to consider the second term in (\ref{L3-pot}), as
the first term involves only the trace part of the metric perturbation. Let us consider the 
YM sector first. We now substitute:
\be
b^a_{\mu\nu}=\Sigma^{a\quad \rho}_{0\,[\mu} h_{\nu]\rho},
\ee
and use the anti-self-duality of this two-form to get:
\be
{\cal L}^{(3)}_{grav-YM} = -\frac{\kappa}{2} \Sigma^{a\,\mu\nu}_0 b^8_{\mu\nu} 
\Sigma^{a\,\rho\lambda}_0 h_\lambda^\sigma b^8_{\rho\sigma}=
-2\kappa P^{+\,\mu\nu\rho\lambda} b^8_{\mu\nu} b^8_{\rho\sigma} h_\lambda^\sigma.
\ee
Here an extra minus is due to the metric on the Lie algebra. The physical Lagrangian
is obtained from here by taking the real part. This makes only the term in the self-dual
projector $P^{+\,\mu\nu\rho\lambda}$ that contains the metric to survive. Substituting
(\ref{b8-f}) we get:
\be
{\cal L}^{(3)}_{grav-YM}=\frac{1}{\kappa} \epsilon_\rho^{\,\,\lambda\mu\nu}
F_{\mu\nu} \epsilon^{\rho\sigma\alpha\beta} F_{\alpha\beta} h_{\lambda\sigma}.
\ee
Expanding the product of two $\epsilon$'s here we get:
\be
{\cal L}^{(3)}_{grav-YM}=\frac{4}{\kappa} F_{\mu\rho} F_{\nu\sigma} h^{\mu\nu}
\eta^{\rho\sigma},
\ee
in which expression we recognize precisely the coupling to the stress-energy tensor that
arises from the YM Lagrangian (\ref{Lagr-YM}). The sign in front is different from that
in (\ref{Lagr-YM}) because the variation of the metric with two upper indices is given by minus
$h^{\mu\nu}$. Thus, the arising coupling of our YM fields
to the gravitational sector is correct.

Let us now discuss the coupling of the Higgs sector to gravity. It is easy to see that
in the low-energy approximation in which $E^2\ll \kappa$ and the two-forms
$b^\alpha_{\mu\nu}$ are self-dual there is no coupling coming from the potential term.
Indeed, we have already discussed that there is no coupling to the trace part of the
metric perturbation. Thus, there is only the second term in (\ref{L3-pot}) that
can contribute. However, it contains a factor of 
$(\epsilon^{\rho\sigma\tau\lambda} b^b_{\rho\sigma} b^\beta_{\tau\lambda})$
which is contraction of a self-dual Higgs two-form and an anti-self-dual 
gravitational one. So, it is zero and the only interaction term in the Higgs
sector comes from the kinetic term of the action. As we have already discussed,
it is of the $h (\partial b)^2$ form, which is just the coupling of the metric perturbation to 
the stress-energy tensor of our set of massless fields. We are not going to work out
the details as they are slightly messy, but we hope that the discussion given is
sufficient to show that the interaction is as expected. 

\subsection{Interactions in the non-gravitational sector}

Let us now concentrate on the interactions in the non-gravitational sector, 
most interestingly those between the YM and Higgs sectors. 

First, we note that there are no cubic interactions in the non-gravitational sector that
come from the potential term. Indeed, such an interaction term involves three
perturbation two-forms $b^\alpha_{\mu\nu}$ with the Lie algebra index outside
of ${\mathfrak su}(2)$. It is not hard to see that the corresponding derivatives of
the potential vanish:
\be 
\left. \frac{\partial^2 V(\tilde{h})}{\partial \tilde{h}^{\beta \gamma} \partial \tilde{h}^{a\alpha}} 
\right|_0= 0, \qquad 
\left. \frac{\partial^3 V(\tilde{h})}{\partial \tilde{h}^{c\gamma}\partial \tilde{h}^{b\beta}\partial \tilde{h}^{a\alpha}}\right|_0= 0 .
\ee

Thus, at cubic order we only need to consider the interactions coming from the kinetic
term. It is not hard to see that there are no self-interactions in the Higgs or YM sectors,
but there are two possible types of interaction between these sectors. One of them
comes from the term $g_{\alpha\beta} b^\alpha f^\beta_{\gamma 8} a^\gamma a^8$,
the other comes from $b^8 f^8_{\alpha\beta} a^\alpha a^\beta$, where $\alpha$
now stands for the Higgs sector index. The second of this is an interaction of the
type $(1/\kappa) F (\partial b)^2$, and is thus suppressed at low energies 
by $E^2/\kappa$. However, the first interaction is non-trivial and important even at
low energies. In fact, it is not hard to show that this is the standard interaction
of the gauge field $a^8$ with the conserved ${\rm U}(1)$ current of the Higgs
sector that is charged under the YM subgroup. We are not going to spell out
the details that are again slightly messy, but the important point is that the YM-Higgs
sectors interaction is also as expected for a set of scalar fields charged under
the YM gauge group (Higgs fields). 

\section{More general potentials: Mass generation}
\label{sec:mass}

Up to now we have for simplicity considered a very special class of potentials
that depend only on the invariants constructed from the "internal" metric $\tilde{h}^{IJ}$
using the Killing-Cartan metric $g_{IJ}$. It is not hard to show that due to the fact
that the rank of $\tilde{h}^{IJ}$ is at most six, there is at most six such independent 
invariants, and thus only at most five ratios to be considered as the arguments of
the function $f(\cdot)$ in (\ref{V-f}). However, it is clear that these are not the only
possible invariants. Indeed, the most general gauge-invariant function of 
$\tilde{h}^{IJ}$ can also involve invariants constructed using the structure constants 
$f^I_{JK}$. For instance, let us consider
\be\label{ffhhh}
ff\tilde{h}\tilde{h}\tilde{h}:=f^{PQR}f^{STU}\tilde{h}_{PS}\tilde{h}_{QT}\tilde{h}_{RU},
\ee
where the indices on the structure constants are raised using the metric on the group. 
More generally, one can construct a matrix:
\be
(ff\tilde{h}\tilde{h})^{IJ} := f^{IQR}f^{JTU}\tilde{h}_{QT}\tilde{h}_{RU}
\ee
and build more complicated invariants from traces of powers of $\tilde{h}^{IJ}$ and
$(ff\tilde{h}\tilde{h})^{IJ}$. This leads to a much more general set of gauge-invariant
functions. In this section we shall study implications of such more general potentials.
Our main point in this section is that these more general potential functions lead
naturally to Higgs fields becoming massive. This is very important for 
phenomenology, for massless Higgs fields interacting with the "visible" YM sector
in the standard way is obviously inconsistent with observations. 

\subsection{Potential with an extra invariant}

For simplicity, in this paper we shall consider only one additional invariant given by
(\ref{ffhhh}). We shall see that such a potential is sufficient to generate masses for
the Higgs sector particles. It is not hard to consider even more general potentials,
but we refrain from doing it in this already lengthy paper. 

Thus, let us consider the 
potential depending on one more invariant
\begin{equation}
V(\tilde{h})=\frac{Tr \,\tilde{h}}{n}\, F\left(\frac{Tr \,\tilde{h}^2}{(Tr\, \tilde{h})^2}\, , \dots, 
\frac{Tr\, \tilde{h}^n}{(Tr\, \tilde{h})^n}, \frac{ff\tilde{h}\tilde{h}\tilde{h}}{(Tr\, \tilde{h})^3}\right) \, ,
\end{equation}
where we have divided (\ref{ffhhh}) by $(Tr\, \tilde{h})^3$ to make the potential
 homogeneous degree one. Then, the first derivative with respect to $\tilde{h}$ is
\begin{equation}
\frac{\partial V(\tilde{h})}{\partial \tilde{h}^{IJ}}
=\frac{g_{IJ}}{n} \, F + \frac{Tr\, \tilde{h}}{n}\, 
\frac{\partial F}{\partial \tilde{h}^{IJ}} \, , 
\end{equation}
with $(\partial F/\partial \tilde{h}^{IJ})$ given by 
\begin{equation}
\frac{\partial F}{\partial \tilde{h}^{IJ}}
=\sum_{p=2}^{n} F'_p \,\, \frac{\partial}{\partial \tilde{h}^{IJ}}
\left( \frac{Tr\, \tilde{h}^p}{(Tr\, \tilde{h})^p}\right)+F'_{n+1}\, \frac{\partial}{\partial \tilde{h}^{IJ}}
\left( \frac{ff\tilde{h}\tilde{h}\tilde{h}}{(Tr\, \tilde{h})^3}  \right) 
\end{equation}
where $F'_p$ is the derivative of $F$ with respect to its argument 
$(Tr\, \tilde{h}^p/(Tr\, \tilde{h})^p)$, $F'_{n+1}$ is the derivative of $F$ 
with respect to its last argument and 
\begin{equation}
\frac{\partial}{\partial \tilde{h}^{IJ}}\left( \frac{ff\tilde{h}\tilde{h}\tilde{h}}{(Tr\, \tilde{h})^3}  \right)
=\frac{3\, f^{PQ}_{\;\;\;\;(I}f_{J)}^{\;\;RS}\, \tilde{h}_{PR}\tilde{h}_{QS}}{(Tr\, \tilde{h})^3}
-\frac{3\,\, ff\tilde{h}\tilde{h}\tilde{h}}{(Tr\, \tilde{h})^4}\,  g_{IJ} \, .  
\end{equation}

Now, let us compute the second derivative of $V$ with respect to $\tilde{h}$. We get
\begin{equation}
\frac{\partial^2 V(\tilde{h})}{\partial \tilde{h}^{KL}\partial \tilde{h}^{IJ}}=\frac{g_{IJ}}{n} \, 
\frac{\partial F}{\partial \tilde{h}^{KL}}+\frac{g_{KL}}{n} \, 
\frac{\partial F}{\partial \tilde{h}^{IJ}} + \frac{Tr\, \tilde{h}}{n}\, 
\frac{\partial^2 F}{\partial \tilde{h}^{KL}\partial \tilde{h}^{IJ}} \, ,
\end{equation}
with $(\partial^2 F/\partial \tilde{h}^{KL}\partial \tilde{h}^{IJ})$ given by
\begin{align}
\frac{\partial^2 F}{\partial \tilde{h}^{KL}\partial \tilde{h}^{IJ}}
=\sum_{p=2}^{n} F'_p \,\, \frac{\partial^2}{\partial \tilde{h}^{KL}\partial \tilde{h}^{IJ}} 
\left( \frac{Tr\, \tilde{h}^p}{(Tr\, \tilde{h})^p}\right)+ F'_{n+1}\,\, 
\frac{\partial^2}{\partial \tilde{h}^{KL}\partial \tilde{h}^{IJ}} 
\left( \frac{ff\tilde{h}\tilde{h}\tilde{h}}{(Tr\, \tilde{h})^3}\right)
\\ \notag
+\sum_{p=2}^{n}\sum_{q=2}^{n} \left( F''_{pq} \frac{\partial}{\partial \tilde{h}^{KL}}
\left( \frac{Tr\, \tilde{h}^q}{(Tr\, \tilde{h})^q}\right)+ F''_{p(n+1)}\frac{\partial}{\partial \tilde{h}^{KL}}\left(\frac{ff\tilde{h}\tilde{h}\tilde{h}}{(Tr\, \tilde{h})^3} \right) \right) \frac{\partial}{\partial \tilde{h}^{IJ}}
\left( \frac{Tr\, \tilde{h}^p}{(Tr\, \tilde{h})^p} \right) \\ \notag  
+\sum_{p=2}^{n} \left( F''_{(n+1)p} \frac{\partial}{\partial \tilde{h}^{KL}}
\left( \frac{Tr\, \tilde{h}^p}{(Tr\, \tilde{h})^p}\right)+ F''_{(n+1)(n+1)}\frac{\partial}{\partial \tilde{h}^{KL}}\left(\frac{ff\tilde{h}\tilde{h}\tilde{h}}{(Tr\, \tilde{h})^3} \right) \right) 
\frac{\partial}{\partial \tilde{h}^{IJ}}\left(\frac{ff\tilde{h}\tilde{h}\tilde{h}}{(Tr\, \tilde{h})^3} \right)\, ,
\end{align}
where $F''_{pq}$ stands for the derivative of $F'_p$ with respect to its $q$ argument and similar for $F''_{p(n+1)}$ and $F''_{(n+1)(n+1)}$. It is easy to show that
\begin{equation}
\frac{\partial^2 \, (ff\tilde{h}\tilde{h}\tilde{h})}{\partial \tilde{h}^{IJ}\partial \tilde{h}^{KL}}
=-6\, f^P_{\;\;\,K)(I}f_{J)(L}^{\;\;\;\;\;\;\,Q}\,\, \tilde{h}_{PQ} \, .
\end{equation}
Using the equations above, we obtain the following expressions
\begin{align}
\left. \frac{\partial V}{\partial \tilde{h}^{ab}} \right|_0=&0 \, ,
\\
\left. \frac{\partial V}{\partial \tilde{h}^{\alpha\beta}} \right|_0=&-\left( \kappa\, +\frac{2 \lambda}{3}\right) \, g_{\alpha\beta} \, ,
\\
\left. \frac{\partial^2 V}{\partial \tilde{h}^{b\beta}\partial \tilde{h}^{a\alpha}} \right|_0=&
\frac{\kappa}{4 \im}\, g_{ab}\, g_{\alpha\beta}
+\frac{\lambda}{6 \im} g_{cd} f^c_{ab} f^{d}_{\alpha\beta} \, ,
\end{align}
where we have set $(F)_0=0$ and defined
\be\label{lambda}
\lambda=\frac{ (F'_{n+1})_0}{8} \;\; . 
\ee
The parameter $\kappa$ is as before, see (\ref{kappa}), with the function $F(\cdot)$ of
one more invariant in place of $f(\cdot)$.

\subsection{Higgs sector masses}

In this subsection we show that the new parameter $\lambda$ introduced above
receives the interpretation of mass squared of the Higgs sector scalar fields. To
this end, let us work out the quadratic part of the action that comes from the potential,
concentrating only on the $\lambda$-dependent part. The $\kappa$-dependent
part was already taken care of by setting the Higgs sector perturbation two-forms
$b^\alpha_{\mu\nu}$ to be self-dual, and this is unchanged for our more
general potential. Dividing (\ref{bb-lin}) by two, using the self-duality of 
$b^\alpha_{\mu\nu}$ in the second term and simplifying we get:
\be
S^{(2)}_\lambda=-\frac{2\lambda}{3} \int 
\frac{1}{4} g_{cd} f^{c}_{ab} f^d_{\alpha\beta} 
(\Sigma^{a\,\mu\nu}_0 b^\alpha_{\mu\nu})
(\Sigma^{b\,\mu\nu}_0 b^\beta_{\mu\nu}) 
- g_{\alpha\beta}b^{\alpha\,\mu\nu} b^\beta_{\mu\nu}.
\ee
We now substitute in this expression the expansions (\ref{b-alpha}) for our two-forms
(in a specific gauge). It is not hard to see that only the term $f^z_{ab} f^z_{\alpha\beta}$
contributes and we get:
\be
S^{(2)}_\lambda=\lambda \int b^{4-} b^{5+} + b^{6+} b^{7-} + b^{4z} b^{5z} + b^{6z} b^{7z}
= - m^2_{Higgs} \langle {\bf b}^\dagger, {\bf b}\rangle,
\ee
where 
\be\label{m-higgs}
m^2_{Higgs} = - \lambda.
\ee
Thus, as all other physical parameters arising in our theory, the mass of the 
Higgs sector particles also comes from the defining potential.

\section{Discussion}

In view of the length of this paper it is probably appropriate to recap our logic and
emphasize the main results that we have obtained. Thus, we have started with a
generally-covariant gauge theory for a group $G$, with the action given by
(\ref{action-A}). At this stage all fields are complex and reality conditions are
later imposed to select the physical, real sector of the theory. We then perform
the Legendre transform and pass to the two-form field formulation (\ref{action-AB}).
Our phase space analysis in section \ref{sec:ham} is only needed to get a better
idea of what should be expected for the number of propagating DOF of the theory.
It does not form an essential part of our argument. The main analysis starts in
section \ref{sec:grav} where we analyze the simplest case $G={\rm SU}(2)$
and show how it describes the usual gravity in the limit when a certain parameter
of the potential is taken to be large, or, alternatively, for low energies. 
For a finite value of the parameter (or for Planckian energies) one gets
a modified gravity theory with two propagating DOF. However, as the low-energy limit
of our theory is still given by GR, we do not need to understand the nature
of this modification for purposes of this paper. 

We start with the analysis of the pure gravity case by describing how the Minkowski
spacetime looks like in the language of two-forms, see (\ref{s}). The action is then
expanded to quadratic order, and the field equations for the connection field are
solved for, with the solution given by (\ref{a-1}). After the solution is substituted into
the action one gets the linearized kinetic term (\ref{S-bf-lin}) as a functional of
only the two-form perturbation. This is supplemented
with the potential term part (\ref{lapsu2}). After the parameter $g$ is taken to infinity
one gets GR written in terms of two-forms, with a very compact linearized action
(\ref{S-bf-lin}). This action is considerably simpler than the one in terms of the metric perturbation,
and the relation between the two arises via (\ref{b-h}). We also perform the
Hamiltonian analysis of the linearized theory, to show how the usual two polarizations of the 
graviton arise in this language. In the $g\to\infty$ limit
this analysis reproduces Ashtekar's Hamiltonian formulation of GR, in its
linearized version. The main purpose of this analysis is to select the reality
conditions for the gravitational sector. These are particularly clear in
the Hamiltonian formulation, and later in the paper the same strategy of deducing
the reality conditions from the form of the Hamiltonian is used for other fields.
In this section we only discuss the rather simple reality conditions appropriate
in the GR limit $g\to\infty$. The finite $g$ case reality conditions are deduced in
the Appendix, for completeness.

Once the ${\rm SU}(2)$ case is understood we enlarge the gauge group to $G={\rm SU}(3)$.
We take the same set of two-forms (\ref{s}) for the background, which thus selects
in the ${\mathfrak su}(3)$ Lie algebra a preferred gravitational ${\mathfrak su}(2)$ 
subalgebra. The analysis of the gravitational part is unchanged, but we have carried
it out once more using a different basis in the Lie algebra (root basis), in preparation for
the analysis of the non-gravitational sectors. These split into a part that commutes
with ${\mathfrak su}(2)$ and that will later be identified with the YM sector, and the
part that does not commute with ${\mathfrak su}(2)$ and becomes the Higgs sector. 

Let us start with the Higgs sector. As in the case of gravity, we first solve the equations
for the connections $a^\alpha_\mu$ in terms of the perturbation two-forms $b^\alpha_{\mu\nu}$
and then substitute the result back into the action. The resulting kinetic part of the 
action as a functional of the two-forms $b^\alpha_{\mu\nu}$ is given by
(\ref{S-bf-Higgs*}). There is also the part (\ref{S-bb-higgs}) coming from the potential.
Similarly to the case of gravity, the role of the potential part, in the low-energy
limit, is to set certain components of the two-form field $b^\alpha_{\mu\nu}$ to zero.
After this is done, the perturbation two-forms $b^\alpha_{\mu\nu}$ becomes self-dual
and can be expanded in the basis of self-dual two-forms (\ref{s}). The coefficients
in this expansion become our Higgs fields. They can be seen to be charged under
the gravitational ${\rm SU}(2)$ subgroup, comprising 
two irreducible representations of spin $1/2$ of ${\rm SU}(2)$. They also transform
non-trivially under the part of the gauge group that does not commute with 
${\rm SU}(2)$, and so they are not all physical. A convenient gauge is given 
by (\ref{b-alpha}). Finally, our Higgs fields are charged under the part of the
gauge group that commutes with ${\rm SU}(2)$, i.e. under the YM subgroup,
which in the case of $G={\rm SU}(3)$ is ${\rm U}(1)$. After a gauge is fixed,
one obtains a Lagrangian for the physical fields, and this is found to be
just the usual one for a set of 8 massless fields. We then determined
the reality conditions needed to make it into a real Lagrangian with positive
definite Hamiltonian. These can be read off from either the Lagrangian we
have obtained, or from the Hamiltonian formulation that is also developed.
The end result is a set of 4 complex (and at this stage massless) scalar fields with
the usual real Lagrangian (\ref{L-Higgs*}). These fields are later made massive
by considering a slightly more general set of defining potentials.

We then analyze the YM sector, both in the Lagrangian and Hamiltonian frameworks.
As usual in this paper, the Hamiltonian framework considerations are most useful
for determining the reality conditions that need to be imposed. After these are
deduced, the derivation of the Lagrangian becomes straightforward, with the result
given by (\ref{Lagr-YM}). The YM coupling constant arises
as (\ref{YM-coupl}), with the parameter $\kappa$ related to the first derivatives
of the potential function via (\ref{kappa}). 

We then discuss (cubic) interactions between the various sectors of our theory and confirm
that they are as expected for such fields. Namely, the interactions of all fields with
gravity are via their stress-energy tensor, and interactions of the Higgs sector
with the YM fields is via the Higgs conserved current. 

Finally, we consider potentials more general than has been the case before,
and show how the first derivative (\ref{lambda}) of the potential with respect to the new
invariant becomes (minus) the mass squared (\ref{m-higgs}) of the Higgs sector fields. 
The parameter $\lambda=-m_{Higgs}^2$ can be both positive and negative so we
have the possibility of the Higgs potential pointing both up and down, depending on the
form of the defining potential. For negative $m^2_{Higgs}$ and thus positive $\lambda$
the configuration ${\bf b}=0$ is unstable and a new vacuum to expand about should be chosen, 
as in the standard Higgs mechanism. This finishes our demonstration of the fact that 
the content of the theory expanded around the Minkowski spacetime background is as desired.

Let us now discuss whether the unification scheme described in this paper can be
deemed "natural" in the sense that it naturally produces "realistic" values
of the parameters such as masses and coupling constants. To this end let us see
what dimensionful parameters are present in our theory. When the action is written in the
form (\ref{action-A}) the integrand has the mass dimension 4 (assuming that the connection
has the mass dimension one), and there are no dimensionful parameters in the theory at
all. After the Legendre transform to (\ref{action-AB}) the two-form field has the mass
dimension 2, and there are still no dimensionful parameters. However, since a part
of this field is to be interpreted as the spacetime metric, it needs to be made dimensionless,
and this is when a dimensionful parameter is introduced into the story. 
Rescaling the two-form field to give it the mass dimension zero introduces a parameter
of the mass dimension 2 in front of the action (interpreted as $1/G$, where $G$ is the
Newton's constant), as well as makes the potential function to have the mass dimension 2. 
This introduces a length (or mass) scale into the theory, and it is clear that there
is only one natural mass scale given by $M_p$. 

Various parameters of the theory are then obtained as derivatives of the potential
function evaluated at the background, and these have mass dimension 2, or, in
the case of the YM coupling (\ref{YM-coupl}) as the product of the derivative
of the potential times $G$. It is thus clear that the natural values for mass parameters
arising in our theory are $M_p$, and for the dimensionless parameters such as
the coupling constant $g_{YM}\sim 1$. However, this is precisely the values
that are "realistic". Indeed, as our Higgs fields interact with the "visible" YM sector,
we need to explain why they are not observed. This is explained by their very high
mass that makes them essentially irrelevant for the low energy physics. Second, the
realistic values of the YM coupling constants of particle physics are order one,
and precisely such values are natural in our unification scheme. Overall, our
unification model is "realistic" in the sense that it reproduces everything that
could be desired from such a simple setup.

An important ingredient that is missing from our simple-minded model
is that of the usual symmetry breaking mechanism of particle physics. Such a breaking, if present,
would introduce additional mass scales into the theory and make it much richer. The
model considered in this paper in which the bakground only broke the $G$
symmetry down to the gravitational and YM ones did not break the YM gauge group.
However, it is clear that our model naturally allows for such further breaking of symmetry. 
Indeed, we could take the background to be more non-trivial
and give to some of our Higgs fields a non-trivial vacuum expectation value.
Since our Higgs fields interact with the YM sector in the standard way, the
effect of such a non-trivial VEV is also going to be standard - the YM symmetry
is going to be broken, with some of the gauge fields becoming massive.
It is then very interesting that in our scheme this standard particle physics symmetry breaking
mechanism receives a new interpretation. Indeed, a non-trivial VEV for
the Higgs is now on the same footing as a non-zero value for the metric. 
In other words, in our unification scenario the Higgs fields and the metric
are just different parts of a single two-form field multiplet $B^I_{\mu\nu}$. 
Details of, for example, Hamiltonian analysis of the gravitational and
Higgs sectors also confirm a very close analogy between the two.
Thus, in a sense, it is the Higgs fields and the metric that become truly
unified in our scenario. It is of considerable interest to study such more involved symmetry
breaking scenarios. The goal would be to see if a truly realistic unification that
puts together some GUT gauge group, a set of Higgs fields
required to break it to the gauge group of the standard model as well as 
gravity is possible. This question is, however, beyond the scope of
this paper. 

Yet another very important ingredient that is missing from our scenario
is fermions. These are usually unproblematic for any scenario that
operates in Minkowski spacetime. However, we start with a generally
covariant theory with no metric in it, so it is not at all clear how and if fermions
can be added. At the moment, this is probably the most serious objection against
our scenario, but we remain hopeful that fermions can 
be described in our framework. The only possibility for this seems to be
to further enlarge the connection field in (\ref{action-A}) and make it
"fermionic". This might also require a "generalized" connection
that is no longer a one-form, as fermions that we would like to obtain
are not forms. We leave investigation of all these difficult but
very interesting questions to further research.

Finally, let us briefly touch on the question of quantization. The theory we
have considered was classical, but, of course, it has to be quantized. It is then
clear that our action (\ref{action-A}) is non-renormalizable in the usual sense
of the word. Indeed, expanding the theory around Minkowski spacetime
we have obtained a Lagrangian consisting of some renormalizable pieces --
in the Higgs and YM sectors -- as well as gravity with its non-renormalizable
interactions. However, there are also higher order interactions that are non-renormalizable,
and the full action is given by an expansion containing an infinite number
of non-renormalizable terms. Thus, the full theory is non-renormalizable. 
This is, of course, as expected, for we cannot hope to bring together a non-renormalizable
theory (gravity) with renormalizable other interactions in a renormalizable unified
theory. At best, we can hope for a non-renormalizable unified theory, and this
is what is happening in our scenario.

At the same time, what our starting action (\ref{action-A}) describes is just the most general
generally-covariant gauge theory. For this reason it can be expected that the class
of theories (\ref{action-A}) obtained by considering all possible potentials $f(\cdot)$
is closed under renormalization. Indeed, all terms that could arise as counterterms
are already included into (\ref{action-A}) and so the only effect of renormalization
should be in the renormalizing the defining function $f(\cdot)$. This expectation is consistent
with the outcome of our analysis. Indeed, we have seen that, for instance,
the YM theory gauge coupling is just a certain parameter of the potential defining
the theory. This parameter is known to flow with energy, and from the perspective
of our scheme this corresponds to a flow in the potential function. If the sole
effect of renormalization is a flow in the space of potentials, the non-renormalizability
of our theory ceases to be much of a problem. Indeed, it is then possible to hope
for a non-problematic UV fixed point corresponding to some very special potential
that would thus provide a UV completion of our theory. In this context it is interesting to remark that, 
since the gauge coupling is known to flow to zero value in the UV (asymptotic freedom),
and such coupling in our scheme is on the same footing with e.g. the parameter
$g$ describing the strength of gravity modifications, it is natural to expect that
$g$ flows to zero in the UV as well. However, it is not hard to see that this corresponds
to the defining potential $V(\cdot)$ flowing towards the one of the topological BF theory.
Thus, at least prior to any concrete analysis, it seems that the sought UV completion 
may be given by the topological BF theory, something that in the 
past has been suggested in the literature in other contexts. All in all, the absence of the usual 
"finite number of counterterms" renormalizability of our theory may not 
be a problem as the theory may possibly be renormalizable in the sense of Weinberg
\cite{Weinberg:1978kz} as containing all possible counterterms, see also
\cite{Weinberg:2009bg} for a more modern exposition of the notion of 
"effective renormalizability".

To summarize, there are many open problems of our scenario, notably questions of whether a
realistic grand unification is possible, whether fermions can be described
in the same framework, and whether the expectation of effective renormalizability
is realized. However, it appears to us that in spite of all the open problems the  
scenario described already suggests some very interesting new interpretations and
and is thus worthy of further exploration.

\bigskip
\noindent{\bf Acknowledgements.} ATG was supported by a Mathematical Sciences
Research Scholarship and KK by an EPSRC Advanced Fellowship. 

\section*{Appendix: Reality conditions for modified gravity}

The "correct" reality conditions for the full modified gravity theory can be worked out from the 
condition $B^i\wedge (B^j)^*=0$. In linearized theory this becomes:
\be
\Sigma^a\wedge (b^b)^* =\bar{\Sigma}^b \wedge b^a, \qquad {\rm or} \qquad
\Sigma^{a\,\mu\nu} (b^b_{\mu\nu})^* + \bar{\Sigma}^{b\,\mu\nu} b^a_{\mu\nu}=0,
\ee
where $(b^a)^*$ is the complex conjugate two-form perturbation and $\bar{\Sigma}$ is
given by (\ref{sb}). We now rewrite this reality condition using the space plus time split.
We get:
\be\label{reality-su2}
\im (t^{ab}-(t^{ba})^*)+ 2(b^{ab}_{\,\,0} + (b^{ba}_{\,\,0})^*)=0.
\ee
To get this condition we have used $\bar{\Sigma}^a_{bc}=\epsilon^a_{\,\, bc}, 
\bar{\Sigma}^a_{0b}=\im\delta^a_b$ and recalled the definition (\ref{t-su2})
of the configurational variable. We should now analyze this condition
together with the already known solution (\ref{b0-su2}), (\ref{btf-su2})
for the components $b^{ab}_{\,\,0}$.

Let us first consider the trace and anti-symmetric parts of (\ref{reality-su2}). Then
in the tracefree symmetric gauge for $t^{ab}$ these conditions simply state that
the lapse and shift functions $N, N^a$ are real. This explains why the
factor of $\im$ was introduced in (\ref{b0-su2}) in front of the lapse. 

Consider now the symmetric tracefree part of (\ref{reality-su2}). The corresponding
components of $b^{ab}_{\,\,0}$ are known from (\ref{btf-su2}) and we arrive
at the following condition on the phase space variables:
\be\label{reality-2-su2}
\frac{1}{2g}{\rm Re}\left(\epsilon^{ef(a}\partial_e \pi^{b)}_f\right)_{tf}= {\rm Im}(t^{ab})_{tf}.
\ee
In the case $g\to\infty$ that corresponds to GR this implies that $(t^{ab})_{tf}$
is real, but in the modified case the situation is more interesting.

In addition to (\ref{reality-2-su2}) there is another condition that is obtained
by requiring that (\ref{reality-2-su2}) is preserved under the evolution. Thus,
we need to compute the Poisson bracket of (\ref{reality-2-su2}) with the
Hamiltonian and impose the resulting condition as well. The computation is
a bit technical, but at this phase space level there is no way to avoid it.
Indeed, even in the case of GR it is clear from the form of the Hamiltonian
(\ref{Ham-su2}) that the relevant condition cannot be that the momentum is
real, for the Hamiltonian would be complex due to the presence of the 
second term in the second line. The computation of the Poisson bracket
can be done as follows. First, we introduce the real and imaginary parts of
the phase space variables:
\be
t^{ab}=t^{ab}_1 + \im t^{ab}_2, \qquad \pi^{ab}=\pi^{ab}_1+\im \pi^{ab}_2.
\ee
Second, we substitute this decomposition into the action written in the Hamiltonian
form. The resulting action has real and imaginary parts. It is not hard to convince
oneself that any one of these two parts can be used as an action for the system,
the resulting equations are the same due to Riemann-Cauchy equations that
follow from the fact that the original action was holomorphic. We choose to work
with the real part of the action. The relevant Poisson brackets are easily seen to be
\be
\{ \pi^{ab}_1(x),t_{1\,cd}(y)\} = \delta^{(a}_c\delta^{b)}_d \delta^3(x-y), \qquad 
\{ \pi^{ab}_2,(x)t_{2\,cd}(y)\}  = -\delta^{(a}_c\delta^{b)}_d \delta^3(x-y),
\ee
with all the other ones being zero. The real part of the Hamiltonian 
(with the constraint part already imposed and dropped) reads:
\be\nonumber
{\cal H}^{real} = \frac{1}{2}(\pi^{ab}_1)^2 - \frac{1}{2}(\pi_2^{ab})^2
- \epsilon^{efa}\partial^e \pi^{bf}_1 t^{ab}_2 - \epsilon^{efa}\partial^e \pi^{bf}_2 t^{ab}_1
+\frac{1}{2g} (\partial^a \pi^{bc}_1)^2 
- \frac{1}{2g} (\partial^a \pi^{bc}_2)^2.
\ee
We can now compute the Poisson bracket with the reality 
condition (\ref{reality-2-su2}) that becomes:
\be\label{real-1-su2}
\frac{1}{2g} \epsilon^{efa} \partial_e \pi^{fb}_1=t_2^{ab}.
\ee
The Poisson bracket with the left-hand-side is:
\be
\{{\cal H}^{real}, \frac{1}{2g} \epsilon^{efa}\partial^e \pi^{bf}_1\}=-\frac{1}{2g}\Delta \pi_2^{ab}.
\ee
The Poisson bracket with the right-hand-side is:
\be
\{{\cal H}^{real}, t^{ab}_2\}=\pi_2^{ab}+\epsilon^{efa}\partial^e t_1^{bf}-\frac{1}{g}\Delta \pi_2^{ab}.
\ee
Thus, the sought conditions that guarantees the consistency of (\ref{real-1-su2}) is:
\be\label{real-2-su2}
\pi_2^{ab}+\epsilon^{efa}\partial^e t_1^{bf}-\frac{1}{2g}\Delta \pi_2^{ab}=0.
\ee
We now need to solve this for $\pi^{ab}_2$, which gives:
\be
\pi_2^{ab} = -\frac{\epsilon^{efa}\partial^e t_1^{bf}}{1-\Delta/2g},
\ee
where the denominator should be understood as a formal power series in powers of
$\Delta/g$. When $g\to\infty$ we reproduce the GR result reviewed in the beginning
of this subsection.

We now have to substitute this, as well as the expression (\ref{real-1-su2}) 
for $t^{ab}_2$ into the action. This is a simple exercise with the result being:
\be
S^{real}=\int dt \,d^3x\,\left(\pi_{GR}^{ab} \partial_0 t_{GR}^{ab}
- \frac{1}{2} \left( (\pi_{GR}^{ab})^2 +  (\partial^a t^{bc}_{GR})^2
\right)\right),
\ee
where we have defined:
\be\label{redef-su2}
\pi_{GR}^{ab}=\pi_1^{ab}, \qquad t_{GR}^{ab}= \frac{t_1^{ab}}{1-\Delta/2g}.
\ee
These are the phase space variables in terms of which the Hamiltonian
takes the standard GR form. This shows how an explicitly real formulation
with a positive definite Hamiltonian can be obtained. We also see
that for any finite value of $g$ the graviton is unmodified. 



\begin{thebibliography}{24}

\bibitem{Vafa:2009se}
  C.~Vafa,
  ``Geometry of Grand Unification,''
  arXiv:0911.3008 [math-ph].

\bibitem{Coleman:1967ad}
  S.~R.~Coleman and J.~Mandula,
  ``All Possible Symmetries of the S Matrix,''
  Phys.\ Rev.\  {\bf 159}, 1251 (1967).

\bibitem{Haag:1974qh}
  R.~Haag, J.~T.~Lopuszanski and M.~Sohnius,
  ``All Possible Generators Of Supersymmetries Of The S Matrix,''
  Nucl.\ Phys.\  B {\bf 88}, 257 (1975).

\bibitem{Percacci:2008zf}
  R.~Percacci,
  ``Mixing internal and spacetime transformations: some examples and
  counterexamples,''
  J.\ Phys.\ A  {\bf 41}, 335403 (2008)
  [arXiv:0803.0303 [hep-th]].


\bibitem{Holst:1995pc}
  S.~Holst,
  ``Barbero's Hamiltonian derived from a generalized Hilbert-Palatini action,''
  Phys.\ Rev.\  D {\bf 53}, 5966 (1996)
  [arXiv:gr-qc/9511026].


\bibitem{Percacci:1984ai}
  R.~Percacci,
  ``Spontaneous Soldering,''
  Phys.\ Lett.\  B {\bf 144}, 37 (1984).
  
\bibitem{Percacci:1990wy}
  R.~Percacci,
  ``The Higgs Phenomenon in Quantum Gravity,''
  Nucl.\ Phys.\  B {\bf 353}, 271 (1991)
  [arXiv:0712.3545 [hep-th]].
  
\bibitem{Nesti:2007jz}
  F.~Nesti,
  ``Standard Model and Gravity from Spinors,''
  Eur.\ Phys.\ J.\  C {\bf 59}, 723 (2009)
  [arXiv:0706.3304 [hep-th]].

\bibitem{Nesti:2007ka}
  F.~Nesti and R.~Percacci,
  ``Graviweak Unification,''
  J.\ Phys.\ A  {\bf 41}, 075405 (2008)
  [arXiv:0706.3307 [hep-th]].
  
\bibitem{Nesti:2009kk}
  F.~Nesti and R.~Percacci,
  ``Chirality in unified theories of gravity,''
  arXiv:0909.4537 [hep-th].

\bibitem{Peldan:1992iw}
  P.~Peldan,
  ``Ashtekar's variables for arbitrary gauge group,''
  Phys.\ Rev.\  D {\bf 46}, 2279 (1992)
  [arXiv:hep-th/9204069].

\bibitem{Peldan:1992mp}
  P.~Peldan,
  ``Unification of gravity and Yang-Mills theory in (2+1)-dimensions,''
  Nucl.\ Phys.\  B {\bf 395}, 239 (1993)
  [arXiv:gr-qc/9211014].

\bibitem{Chakraborty:1994vx}
  S.~Chakraborty and P.~Peldan,
  ``Towards a unification of gravity and Yang-Mills theory,''
  Phys.\ Rev.\ Lett.\  {\bf 73}, 1195 (1994)
  [arXiv:gr-qc/9401028].

\bibitem{Chakraborty:1994yr}
  S.~Chakraborty and P.~Peldan,
  ``Gravity and Yang-Mills theory: Two faces of the same theory?,''
  Int.\ J.\ Mod.\ Phys.\  D {\bf 3}, 695 (1994)
  [arXiv:gr-qc/9403002].
  
\bibitem{Ashtekar:1987gu}
  A.~Ashtekar,
  ``New Hamiltonian Formulation of General Relativity,''
  Phys.\ Rev.\  D {\bf 36}, 1587 (1987).
  
\bibitem{Smolin:2007rx}
  L.~Smolin,
  ``The Plebanski action extended to a unification of gravity and Yang-Mills theory,''
  arXiv:0712.0977 [hep-th].
  
\bibitem{Krasnov:2006du}
  K.~Krasnov,
  ``Renormalizable Non-Metric Quantum Gravity?,''
  arXiv:hep-th/0611182.

\bibitem{Bengtsson:1990qg}
  I.~Bengtsson,
  ``The Cosmological constants,''
  Phys.\ Lett.\  B {\bf 254}, 55 (1991).
  
\bibitem{Alexandrov:2008fs}
  S.~Alexandrov and K.~Krasnov,
  ``Hamiltonian Analysis of non-chiral Plebanski Theory and its Generalizations,''
  Class.\ Quant.\ Grav.\  {\bf 26}, 055005 (2009)
  [arXiv:0809.4763 [gr-qc]].
  
\bibitem{Krasnov:2009ip}
  K.~Krasnov,
  ``Metric Lagrangians with two propagating degrees of freedom,''
  arXiv:0910.4028 [gr-qc].

\bibitem{Capovilla:1989ac}
  R.~Capovilla, T.~Jacobson and J.~Dell,
  ``General Relativity Without the Metric,''
  Phys.\ Rev.\ Lett.\  {\bf 63}, 2325 (1989).

\bibitem{Krasnov:2008fm}
  K.~Krasnov,
  ``Plebanski gravity without the simplicity constraints,''
  Class.\ Quant.\ Grav.\  {\bf 26}, 055002 (2009)
  [arXiv:0811.3147 [gr-qc]].

\bibitem{Krasnov:2007cq}
  K.~Krasnov,
  ``On deformations of Ashtekar's constraint algebra,''
  Phys.\ Rev.\ Lett.\  {\bf 100}, 081102 (2008)
  [arXiv:0711.0090 [gr-qc]].

\bibitem{Freidel:2008ku}
  L.~Freidel,
  ``Modified gravity without new degrees of freedom,''
  arXiv:0812.3200 [gr-qc].

\bibitem{Urbantke:1984eb}
  H.~Urbantke,
  ``On Integrability Properties Of SU(2) Yang-Mills Fields. I. Infinitesimal Part,''
  J.\ Math.\ Phys.\ {\bf 25}, 2321 (1984).
  
\bibitem{geor} H.~Georgi, "Lie Algebras in Particle Physics", 2nd Edition 
Westview Press, U.S., 1991.

\bibitem{cahn} R.~Cahn, "Semi-Simple Lie Algebras and Their Representation", 
The Benjamin/ Cummings Publishing Company, U.S., 1984.

\bibitem{Weinberg:1978kz}
  S.~Weinberg,
  ``Phenomenological Lagrangians,''
  Physica A {\bf 96}, 327 (1979).
  
\bibitem{Weinberg:2009bg}
  S.~Weinberg,
  ``Effective Field Theory, Past and Future,''
  arXiv:0908.1964 [hep-th].

\end{thebibliography}
\end{document}